%% file: main.tex
\documentclass{jfm}

\usepackage{graphicx}
\usepackage{newtxtext}
\usepackage{newtxmath}
\usepackage{bm}
\usepackage{natbib}
\usepackage{hyperref}
\usepackage{longtable}
\usepackage{romannum}
\usepackage{setspace}
\usepackage{lipsum}
\hypersetup{
    colorlinks = true,
    urlcolor   = blue,
    citecolor  = black,
    linkcolor=black,
}
\usepackage[table]{xcolor}

\newcommand*{\nameadjunct}{\relax}
\makeatletter
\renewcommand*{\NAT@nmfmt}[1]{\NAT@up #1\nameadjunct}
\makeatother

\newcommand*{\possessivecite}[2][]{%
  \begingroup
  \renewcommand*{\nameadjunct}{'s}%
  \citet[#1]{#2}%
  \endgroup
}

\usepackage{cleveref}

\newcommand{\RomanNumeralCaps}[1]
\linenumbers
\usepackage{multirow}

\usepackage{longtable}
\usepackage{romannum}
\usepackage{lipsum}
\usepackage{hyperref}
\hypersetup{
    colorlinks = true,
    urlcolor   = blue,
    citecolor  = blue,
    linkcolor=blue,
}
\usepackage{xpatch}

\makeatletter
\xpatchcmd\NAT@citex
 {%
  \@citea\NAT@hyper@{%
    \NAT@nmfmt{\NAT@nm}%
    \hyper@natlinkbreak{\NAT@aysep\NAT@spacechar}{\@citeb\@extra@b@citeb}%
    \NAT@date
  }%
 }
 {%
  \@citea
  \NAT@nmfmt{\NAT@nm}%
  \NAT@aysep\NAT@spacechar
  \NAT@hyper@{\NAT@date}%
 }
 {}{}
\xpatchcmd\NAT@citex
 {%
  \@citea\NAT@hyper@{%
    \NAT@nmfmt{\NAT@nm}%
    \hyper@natlinkbreak{\NAT@spacechar\NAT@@open\if*#1*\else#1\NAT@spacechar\fi}%
    {\@citeb\@extra@b@citeb}%
    \NAT@date
  }%
 }
 {
  \@citea
    \NAT@nmfmt{\NAT@nm}%
    \NAT@spacechar\NAT@@open\if*#1*\else#1\NAT@spacechar\fi
    \NAT@hyper@{\NAT@date}%
 }
 {}{}
\makeatother
  
\usepackage{array}
         
\usepackage{lipsum}

\usepackage{pgfplots}
\pgfplotsset{compat = newest}
\usepgfplotslibrary{fillbetween}
\usepgfplotslibrary{groupplots}
\usepackage{pgfplotstable}
\usepgfplotslibrary{external}
\usetikzlibrary{positioning,arrows.meta,shapes}
\pgfplotsset{compat=1.15}

\DeclareMathOperator\erf{erf}
\makeatletter
\newcommand{\Vast}{\bBigg@{5.5}}
\newcommand{\vast}{\bBigg@{5}}
\makeatother
\usepackage{comment}
\usepackage{tabularx}
\usepackage{booktabs}

\title{Turbulent free-surface in self-aerated flows: Superposition of entrapped and entrained air}

\author{Matthias Kramer\aff{1}
  \corresp{\email{m.kramer@unsw.edu.au}}}

\affiliation{\aff{1}UNSW Canberra, School of Engineering and Technology (SET), Canberra,
ACT 2610, Australia}
\begin{document}
\pagenumbering{arabic}
\maketitle

\begin{abstract}
The characterisation and the modelling of air concentration distributions in self-aerated free-surface flows has been subject to sustained research interest since the 1970s. Recently, a novel two-state formulation of the structure of a self-aerated flow was proposed by Kramer \& Valero [2023 J. Fluid Mech. 966, A37], which physically explains the air concentration through the weak interaction of two canonical flow momentum layers, comprising a Turbulent Boundary Layer (TBL) and a Turbulent Wavy Layer (TWL). The TWL was modelled using a Gaussian error function, assuming that the most dominant contribution are wave troughs. Here, it is shown that  air bubbles form an integral part of the TWL, and its formulation is expanded by adopting a superposition principle of entrapped air (waves) and entrained air (bubbles). Combining the superposition principle with the two-state formulation, an expression for the depth-averaged (mean) air concentration is derived, which allows to quantify the contribution of different physical mechanisms to the mean air concentration. Overall, the presented concepts help to uncover new flow physics, thereby contributing fundamentally to our understanding of self-aerated flows.
\end{abstract}

\begin{keywords}
channel flow, turbulent flows, bubble dynamics
\end{keywords}

\section{Introduction}
\label{sec:intro}
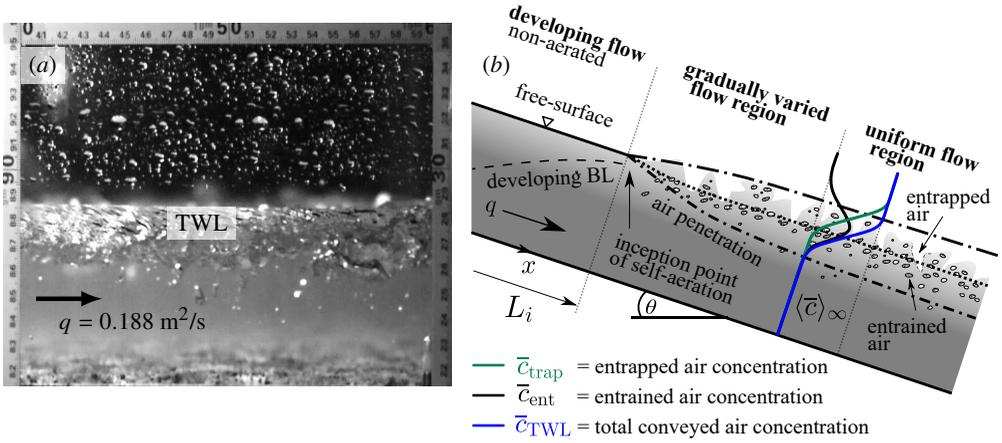
\begin{figure}
\centering   
\input{Figures/fig1.tex}
\vspace{-0.4cm}
\caption{Self-aeration in high Froude-number flows (\textit{a}) TWL of a flow over a micro-rough channel bed at the Water Research Laboratory (WRL), UNSW Sydney, Australia; specific water flow rate $q = 0.188$ m$^2$/s, chute angle $\theta = 10.8^\circ$, streamwise distance from invert $x = 6.3$ m to $6.5$ m; image courtesy of Armaghan Severi, adapted with permission (\textit{b}) schematic of the TWL of a self-aerated flow down a smooth chute, including a differentiation between entrapped, entrained and total conveyed air} 
\label{Fig1}
\end{figure}

Self-aeration is a fascinating flow phenomenon that is frequently observed in high Froude-number open-channel flows (Fig. \ref{Fig1}). Such flows are characterised by strong turbulence and neither surface tension nor gravity are able to maintain surface cohesion \citep{Brocchini2001JFM1}, causing entrainment of air bubbles into the flow column. These bubbles subsequently break down into a wide range of bubble sizes \citep{Lamarre1991,Deike2016,Deane2002,Chan2021}, and eventually penetrate towards the channel bottom through turbulent diffusion. It is known that entrained air can significantly alter flow properties, thereby leading to flow bulking, drag reduction, cavitation protection, and enhanced gas transfer \citep{Straub1958,Falvey1990,Gulliver1990,Kramer2021}. 

As such, the characterization and the modelling of air concentration distributions has been subject to sustained research interest over the last decades. Different groups of researchers have conceptualised the air concentration using single-layer \citep{Rao1971,Wood1991,Chanson2001DIFF,Zhang2017,Valero2016} or double-layer approaches  \citep{Straub1958,Killen1968,Wei2022a,Wei2022b}. Based on visual observations, various physical processes have been identified in self-aerated flows, comprising generation of free-surface waves, surface disruption, air entrainment, turbulent diffusion of air bubbles, and ejection of droplets. It is argued that single-layer approaches are unable to  represent these different flow processes, while good data-driven agreement between single-layer models and measurements has been achieved, which is however at the expense of empirically fitted coefficients. Recently, \cite{Kramer2023JFM} presented a physically based two-state convolution formulation for the air concentration that is built upon a Turbulent Boundary Layer (TBL) and a Turbulent Wavy Layer (TWL).

It is important to note that none of the previous single-layer or double-layer air concentration conceptualizations have taken into account the contribution of entrapped and entrained air, as depicted in Fig. \ref{Fig1}\textit{b}, which is introduced in the following. In a seminal series of experiments, \cite{Killen1968} investigated surface characteristics of self-aerated flows by deploying a common phase-detection probe as well as a larger-sized conduction probe that dipped in and out of the surface roughness/waves, hereafter referred to as \textit{dipping probe}. \cite{Wilhelms2005} re-analyzed the data of \cite{Killen1968} and pointed out that the dipping probe measured entrapped air, transported between wave crests and troughs, whereas the common phase-detection probe measured a combination of entrapped and entrained air, termed total conveyed air. Although \cite{Wilhelms2005} articulated the need for two measurements, one for entrapped air and one for total conveyed air, no other researchers have deployed a dipping probe since, showing the uniqueness of \possessivecite{Killen1968} data set.

\textcolor{black}{The key novelty of the present work is the introduction of a superposition principle, which explicitly accounts for entrapped air (waves) and entrained air (bubbles), allowing to quantify the importance of different physical mechanisms to the mean air concentration. In the following,} the two-state formulation of \cite{Kramer2023JFM} is briefly summarised ($\S$ \ref{sec:twostate}).  Thereafter, the superposition principle for the air concentration of the TWL is proposed, demonstrating that entrapped air and entrained air follow a Gaussian error function and a normal distribution, respectively ($\S$  \ref{sec:superposition}). The superposition principle is then combined with the two-state convolution in $\S$ \ref{sec:combination}, providing the most complete and physically consistent description of the air concentration distribution to date. A bed-normal integration of this expanded formulation allows to differentiate between three different physical mechanisms that contribute to the mean air concentration, comprising 
entrapped air within the TWL, entrained air within the TWL, and entrained air within the TBL ($\S$  \ref{sec:combination}). The different parameters of the superposition principle as well as the application of the new formulation are assessed against \possessivecite{Killen1968} data set in $\S$ \ref{sec:results}, followed by a discussion on model applicability and other  limitations ($\mbox{$\S$ \ref{sec:discussion}}$).

\section{Methods}
\subsection{Two-state convolution}
\label{sec:twostate}
This section provides a brief summary of the governing equations of the two-state convolution model, while more details are presented in \cite{Kramer2023JFM}. The air concentration of the TBL ($\overline{c}_\text{TBL}$) is reflected through a solution of the advection diffusion equation for air in water, whereas the air concentration of the TWL  ($\overline{c}_\text{TWL}$), \textcolor{black}{encompassing bubbles and waves, was found to follow a Gaussian error function} \citep{Kramer2023JFM}
\begin{equation}
\overline{c}_\text{TBL} = 
\begin{cases}
\textcolor{black}{\overline{c}_{\delta/2}}  \left(\textcolor{black}{\frac{y}{\delta-y}} \right)^{\beta}, & y \leq \delta/2,
\label{eq:voidfraction1}\\
\vphantom{\left(\frac{\frac{\delta}{y_\star}-1}{\frac{\delta}{y}-1} \right)^{\frac{\overline{v}_r S_c}{\kappa u_*}}}
\overline{c}_{\delta/2} \, \exp \left(\frac{4\beta}{\delta} \left(y - \frac{\delta}{2} \right)   \right), \quad & y > \delta/2, \\
\end{cases}
\end{equation}
\begin{equation}
\overline{c}_{\text{TWL}} = \frac{1}{2} \, \left( 1 + \erf  \left( \frac{y - y_{50}}{ \sqrt{2} \, \mathcal{H} } \right) \right),
\label{eq:voidfraction2}
\end{equation}
where $\overline{c}_{\delta/2}$ is the air concentration at half the boundary layer thickness ($\delta$), $y$ is the bed-normal coordinate, $\beta =  \overline{v}_r S_c /\kappa u_*$ is the Rouse number, $\overline{v}_r$ is the bed-normal bubble rise velocity, $\kappa$ is the van  Karman constant,  $u_*$ is the friction velocity, and $S_c$ is the turbulent Schmidt number, defined as the ratio of eddy viscosity and turbulent mass diffusivity. Further, $\mathcal{H}$ is a characteristic length-scale that is proportional to the thickness of the TWL, $\erf$ is the Gaussian error function, and $y_{50}$ is the mixture flow depth where the total conveyed air concentration is $\overline{c} = 0.5$; note that other mixture flow depths are represented in the same manner, e.g., $y_{90} = y(\overline{c}=0.9)$. The two-state model assumes a fluctuating interface that separates the TBL and the TWL, and a convolution of the two states with a Gaussian interface probability led to the following expression for the mean air concentration \citep{Krug2017,Kramer2023JFM}
\begin{equation}
\overline{c} =   \overline{c}_{\text{TBL}}(1-\Gamma) + \overline{c}_{\text{TWL}} \Gamma, 
\label{eq:voidfractionfinal}
\end{equation}
with
\begin{equation}
\Gamma(y; y_\star, \sigma_\star) = \frac{1}{2} \left( 1+\erf \left(\frac{y - y_\star }{ \sqrt{2} \sigma_\star}  \right)  \right),
\label{eq:gaussianerr}
\end{equation}
where $y_\star$ is the \textcolor{black}{time-averaged} interface position and $\sigma_\star$ is its standard deviation. 
\textcolor{black}{It is noted that the two-state formulation [Eq. (\ref{eq:voidfractionfinal})] has been successfully validated against more than 500 air concentration data-sets from literature,  hinting at universal applicability. For more information on the development of Eqns. (\ref{eq:voidfraction1}) to (\ref{eq:gaussianerr}), as well as on the definition and determination of associated physical model parameters, the reader is referred to \citet{Kramer2023JFM}.}

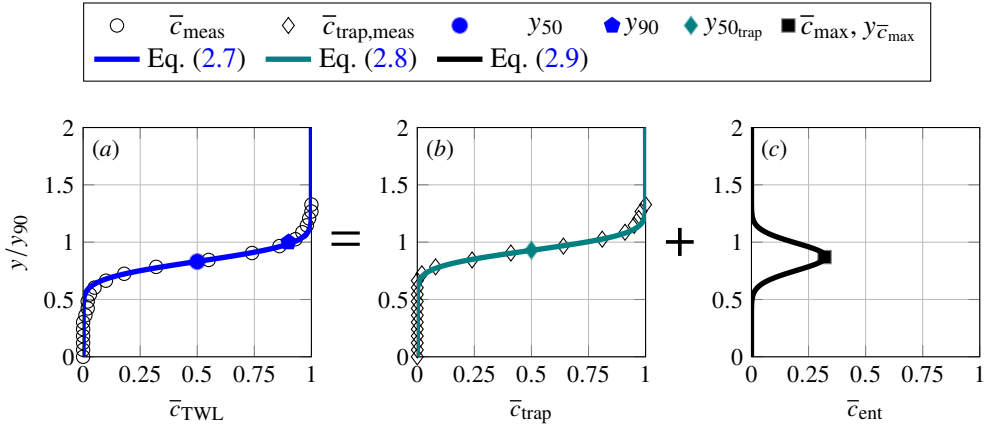
\begin{figure}
\centering
\input{Figures/fig2.tex}
\caption{Representation of \possessivecite{Killen1968}  measurements in a self-aerated flow with $q = 0.78$ m$^2/s$, $\theta = 30^\circ$, and $x = 7.4$ m; meas = measured (\textit{a}) total conveyed air concentration, measured with a common phase-detection probe  (\textit{b}) entrapped air concentration, measured with a  dipping probe (\textit{c}) entrained air concentration, determined through the superposition principle} 
\label{fig:airconcentration} 
\end{figure}

\subsection{Superposition principle (TWL)}
\label{sec:superposition}
Herein, it is hypothesized that the air concentration of the TWL results from a superposition of waves and entrained air bubbles, which was similarly proposed by \cite{Wilhelms2005} for the mean air concentration. To formulate this principle, the focus is set on a flow situation where aeration is confined to the wavy layer, similar to Fig. \ref{Fig1}. The entrapped air concentration of the TWL can be interpreted as the probability of encountering entrapped air at a certain location within the flow. In a time-averaged sense, this probability can be expressed as $p(\overline{c}_\text{trap}) =  \mathcal{V}_\text{trap}/\mathcal{V}_\text{tot}$, where $\mathcal{V}_\text{trap} =$  volume of entrapped air and $\mathcal{V}_\text{tot} = \mathcal{V}_\text{trap} + \mathcal{V}_\text{ent} + \mathcal{V}_\text{W} =$ total volume of the mixture, including the volume of entrained air ($\mathcal{V}_\text{ent}$) and the volume of water ($\mathcal{V}_\text{W}$). The probability of encountering an entrained air bubble within a wave is $p(\overline{c}^*_\text{ent} \mid \overline{c}_\text{trap}) = \mathcal{V}_\text{ent}/(\mathcal{V}_\text{ent} + \mathcal{V}_\text{W})$, which is a conditional probability, given that a wave/water phase is present. Considering the two complementary events $\overline{c}_\text{trap}$ and $(1 - \overline{c}_\text{trap})$, the expression for the total conveyed air concentration of the TWL reads 
\begin{equation}
\overline{c}_\text{TWL} = \overline{c}_\text{trap} + \left( 1 -\overline{c}_\text{trap}\right)\, \overline{c}^*_\text{ent}.
\label{eq:sup1}
\end{equation}

It is recognised that
\begin{equation}
\left( 1 -\overline{c}_\text{trap}\right) \,  \overline{c}^*_\text{ent} =  \frac{\mathcal{V}_\text{ent} + \mathcal{V}_\text{W}}{\mathcal{V}_\text{tot}} \, \frac{\mathcal{V}_\text{ent}}{\mathcal{V}_\text{ent} + \mathcal{V}_\text{W}} = \frac{\mathcal{V}_\text{ent}}{\mathcal{V}_\text{tot}} =  \overline{c}_\text{ent},
\label{eq:sup2}
\end{equation}
where $\overline{c}_\text{ent}$ is the entrained air concentration, defined as the volume of entrained air bubbles related to the total mixture volume. Combining Eqns. (\ref{eq:sup1}) and (\ref{eq:sup2}) leads to the final superposition equation for the TWL (Fig. \ref{fig:airconcentration})
\begin{equation}
\overline{c}_\text{TWL} = \overline{c}_\text{trap} +   \overline{c}_\text{ent}. 
\label{eq:superposition}
\end{equation}

It is noted that the total conveyed air concentration $\overline{c}_\text{TWL}$ is described by Eq. (\ref{eq:voidfraction2}),  c.f. Fig. \ref{fig:airconcentration}\textit{a}. \cite{Valero2016} discussed that the entrapped air concentration $\overline{c}_\text{trap}$ \textcolor{black}{follows an analytical solution of the air-water surface height distribution}, which is (also) reflected by a Gaussian error function (see Fig. \ref{fig:airconcentration}\textit{b}) 
\begin{equation}
\overline{c}_\text{trap}
= \frac{1}{2} \left(1 + \erf  \left( \frac{y - y_{50_\text{trap}}}{\sqrt{2} \mathcal{H}_\text{trap}} \right)  \right),
\label{entrappedair}
\end{equation}
where $y_{50_\text{trap}}$ corresponds to the mean water level, and $\mathcal{H}_\text{trap}$ is the root-mean-square wave height. Re-arranging Eq. (\ref{eq:superposition}), an analytical solution for the entrained air concentration can be written as
\begin{eqnarray}
\overline{c}_\text{ent} = \overline{c}_\text{TWL} - \overline{c}_\text{trap} =\frac{1}{2}  \left( \erf  \left( \frac{y - y_{50}}{\sqrt{2} \mathcal{H}} \right)  -  \erf  \left( \frac{y - y_{50_\text{trap}}}{ \sqrt{2} \, \mathcal{H}_\text{trap} } \right) \right).  
\label{entrainedair}
\end{eqnarray}

Figure \ref{fig:airconcentration}\textit{c} shows that the entrained air concentration corresponds to the difference of two cumulative Gaussians [Eq. (\ref{entrainedair})], which in turn reflects a  Gaussian distribution. Further parameters of interest are the peak entrained air concentration
$\overline{c}_\text{max}$ and its corresponding position $y_{\overline{c}_\text{max}}$, which were added to Fig. \ref{fig:airconcentration}\textit{c} for completeness. 
It is emphasized that these two parameters are not necessarily required, as the profile of entrained air (TWL) is mathematically defined by Eq. (\ref{entrainedair}).

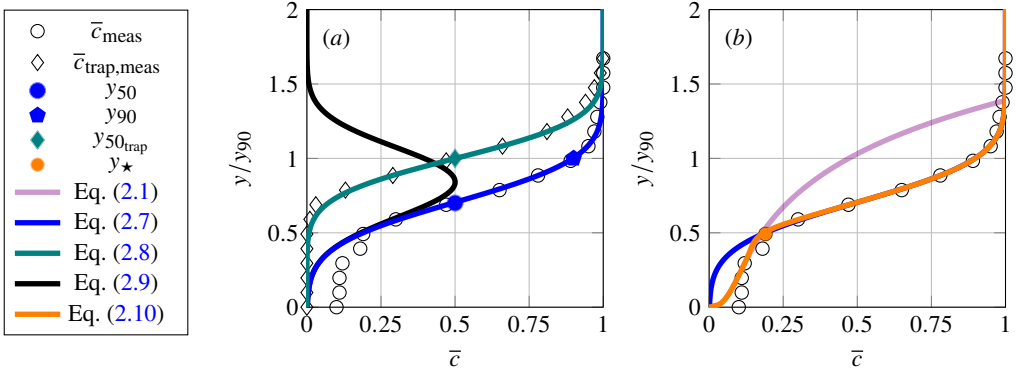
\begin{figure}
\centering
\input{Figures/fig3.tex}
\caption{Representation of \possessivecite{Killen1968} measurements in a self-aerated flow with $q = 0.39$ m$^2/s$, $\theta = 52.5^\circ$, $x = 3.7$ m (\textit{a}) superposition principle (\textit{b}) two-state air concentration convolution} 
\label{fig3}
\end{figure}

\subsection{Combining both approaches}
\label{sec:combination}
In $\S$ \ref{sec:superposition}, the superposition principle was formulated for flow situations where aeration is confined to the wavy layer (pure TWL, compare Figs. \ref{Fig1} and \ref{fig:airconcentration}). In practice, air bubbles are often diffused deeper into the flow column, one example being shown in Fig. \ref{fig3}, where it is seen that the superposition principle still holds for the TWL (Fig. \ref{fig3}\textit{a}). In order to explicitly account for the contribution of entrapped and entrained air within the TWL, the two-state convolution [Eq. (\ref{eq:voidfractionfinal})] is combined with the superposition principle [Eq. (\ref{eq:superposition})]
\begin{equation}
\overline{c} = \left(\overline{c}_\text{trap} +   \overline{c}_\text{ent}\right)  \Gamma +  \overline{c}_{\text{TBL}}(1-\Gamma), 
\label{eq:voidfractionfinal1}
\end{equation}
which describes the complete air concentration profile, see Fig. \ref{fig3}\textit{b}. Further, Eq. (\ref{eq:voidfractionfinal1}) can be integrated between the channel invert and $y_{90}$, yielding an expression for the depth-averaged (mean) air-concentration
\begin{eqnarray}
 \label{eq:meanair0}
 \langle \overline{c} \rangle &=& \frac{1}{y_{90}}  \int_{y=0}^{y_{90}} \overline{c} \, \text{d}y \\
 &=& 
 \underbrace{\frac{1}{y_{90}} \int_{y=0}^{y_{90}}   \overline{c}_\text{trap}  \Gamma  \, \text{d}y}_{\langle \overline{c} \rangle_{\text{TWL}_\text{trap}} } +
 \underbrace{\frac{1}{y_{90}}  \int_{y=0}^{y_{90}}\overline{c}_\text{ent}    \Gamma \, \text{d}y}_{\langle \overline{c}\rangle_{\text{TWL}_\text{ent}}} + 
 \underbrace{\frac{1}{y_{90}}\int_{y=0}^{y_{90}} \overline{c}_{\text{TBL}}(1-\Gamma) \, \text{d}y}_{\langle \overline{c} \rangle_\text{TBL}}.
\label{eq:meanair}
\end{eqnarray}

Equation (\ref{eq:voidfractionfinal1}) represents the most complete and physically consistent description of the air concentration distribution in self-aerated flows to date. Its integrated form [Eq. (\ref{eq:meanair})] allows to differentiate between three different physical mechanisms contributing to the mean air concentration, comprising (i) entrapment of air due to free-surface deformations, (ii) entrainment of air due to turbulent forces exceeding gravity and surface tension forces, and (iii) turbulent diffusion of air bubbles into the TBL, represented through $\langle \overline{c} \rangle_{\text{TWL}_\text{trap}}$, $\langle \overline{c} \rangle_{\text{TWL}_\text{ent}}$, and 
 $\langle \overline{c} \rangle_\text{TBL}$, respectively.

\section{Results}
\label{sec:results}
The application of the superposition principle requires two different measurements, one for entrapped air and one for total conveyed air. Commonly, the total conveyed air has been measured using intrusive phase-detection probes, e.g., \citet{Straub1958,Chanson2001DIFF,Bung2009,Severi2018}, whereas entrapped air has rarely been measured, one exception being the smooth chute data from \citet[20 profiles]{Killen1968}, to which the expanded formulation of the two-state model is applied. \textcolor{black}{This re-analysis of all 20 profiles is presented in Appendix \ref{appendixA},} and 
more details of the original measurements are provided in Table \ref{tab:data}. \textcolor{black}{Here, the local Froude-number is defined as  $Fr = q / (g d_\text{eq}^3)^{1/2}$, with $g$ being the gravitational acceleration and $d_\text{eq} = \int_{y=0}^{y_{90}} (1-\overline{c}) \, \text{d}y$ the equivalent clear water flow depth.} 

\begin{longtable}{l c c c c c c c c}
\caption{Experimental flow conditions of \citet{Killen1968}; all re-analysed profiles extracted from \citet[Tables B1 to B4]{Wilhelms1994}; \textcolor{black}{note that the profile number corresponds to Appendix \ref{appendixA}}}
\label{tab:data}\\
\toprule
Reference & chute type & profile & $q$ & $\langle \overline{c} \rangle $ & \textcolor{black}{$Fr$} & $\theta$  &$k_s$   \\
(-) & (-) & (-) & (m$^2$/s) & (-) & \textcolor{black}{(-)} &($^\circ$) & 
 (mm) &  \\
\midrule
\citet{Killen1968} & smooth & 1 to 5 & 0.39 & 0.20 to 0.33 & \textcolor{black}{9.4 to 11.8} & 30  & 0.71   \\
\citet{Killen1968} & smooth & 6 to 9 & 0.78 & 0.15 to 0.25 & \textcolor{black}{10.3 to 12.2} & 30 & 0.71    \\
\citet{Killen1968} & smooth &  10 to 17 & 0.39 &  0.19 to 0.55 & \textcolor{black}{14.5 to 21.1} & 52.5 & 0.71  \\
\citet{Killen1968} & smooth &  18 to 20 & 0.20 & 0.35 to 0.42 & \textcolor{black}{6.8 to 8.5} &  30 & 0.71   \\
\bottomrule
\multicolumn{8}{l}{$k_s$ = roughness height; chute length $= 15.25$ m;  chute width $= 0.46$ m} 
\end{longtable}

The two free parameters $\mathcal{H}$ and $\mathcal{H}_\text{trap}$ of the superposition principle were obtained through least squares fitting. Here, $\mathcal{H}$ was obtained by minimizing the sum of squared differences between measurements and modelled air
concentrations within the upper flow region, while the full profile was used for determination of $\mathcal{H}_\text{trap}$, as depicted in Fig. \ref{fig3}\textit{a}. The flow depths $y_{50}$ and $y_{50_\text{trap}}$ could be directly extracted from \possessivecite{Killen1968} data, and were therefore regarded as fixed. Other free and fixed parameters of the two-state convolution ($\beta$, $y_\star$, $\sigma_\star$, $\overline{c}_{\delta/2}$, and $\delta$) had already been determined by \cite{Kramer2023JFM}, and are therefore not discussed hereafter. In the following, the results of the re-analysis of \mbox{\possessivecite{Killen1968}} measurements are presented.

\subsection{Physical parameters of the TWL}
\label{sec:resultsparams}
\begin{figure}
\centering
\input{Figures/fig4.tex}
\vspace{-0.4cm}
\caption{Physical parameters of the TWL for the data of \citet{Killen1968} (\textit{a}) length scale $\mathcal{H}$ and root-mean-square wave height $\mathcal{H}_\text{trap}$ versus mean air concentration (\textit{b}) three-sigma rule applied to evaluate $\mathcal{H}$ (\textit{c}) characteristic depths $y_{50}$ and $y_{50_\text{trap}}$ versus mean air concentration
\vspace{-0.2cm} }
\label{fig:params}
\end{figure}
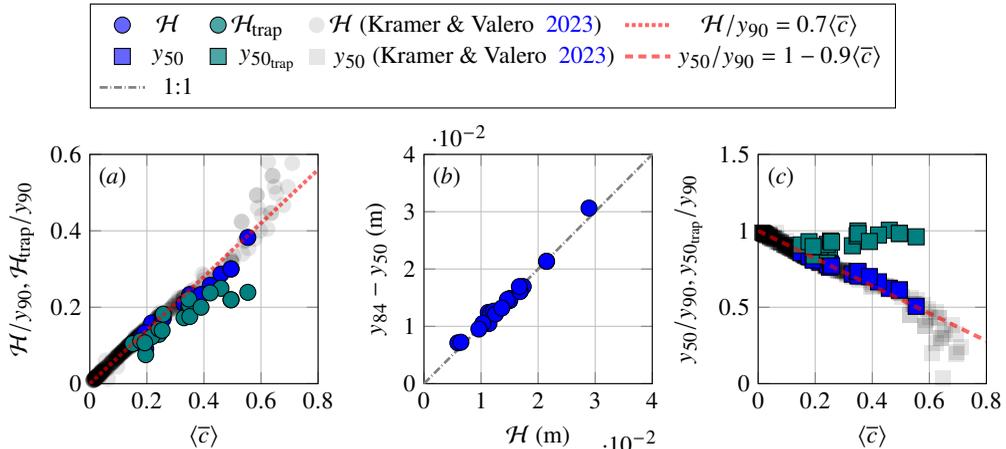

Figure \ref{fig:params}\textit{a} shows the length-scale of the of the TWL ($\mathcal{H}$) as well as the root-mean-square wave height ($\mathcal{H}_\text{trap}$), both normalised with $y_{90}$ and plotted against the mean air concentration. Similar to the data of \cite{Kramer2023JFM}, there was a linear dependence between $\mathcal{H}$ and $\langle \overline{c} \rangle$. The present analysis reveals that the length scale of the TWL ($\mathcal{H}$) and the root-mean-square wave height ($\mathcal{H}_\text{trap}$) \textcolor{black}{showed some similar trends} (Fig. \ref{fig:params}\textit{a}), which implies that $\mathcal{H}$ \textcolor{black}{can provide a rough} indication for $\mathcal{H}_\text{trap}$, \textcolor{black}{while it is acknowledged that $\mathcal{H}$ may not capture the full complexity of the air-water interface geometry}. Further, the empirical three-sigma rule was applied to show that $\mathcal{H}$ corresponded to the difference between the characteristic flow depths $y_{84}$ and $y_{50}$ (Fig. \ref{fig:params}\textit{b}), which was also applicable to $\mathcal{H}_\text{trap}$ (not shown). It is worthwhile to mention that roughness effects as well as the streamwise dependence of model parameters are implicitly accounted for in $\langle \overline{c} \rangle$.

As discussed in \cite{Kramer2023JFM}, the normalised flow depth $y_{50}$ was linearly related to the mean air concentration (Fig. \ref{fig:params}\textit{c}). In contrast, the mean water depth $y_{50_\text{trap}}$ was found to be less dependent on $\langle \overline{c} \rangle$, and $y_{50_\text{trap}}/y_{90}$ became constant for $\langle \overline{c} \rangle > 0.4$ (Fig. \ref{fig:params}\textit{c}). The difference between $y_{50}$ and $y_{50_\text{trap}}$ was indicative for the 
the downwards shift of the Gaussian error function for entrapped air, compare Fig. \ref{fig3}\textit{a}. 

\subsection{Mean air concentration decomposition}
\label{sec:resultsmeanair}
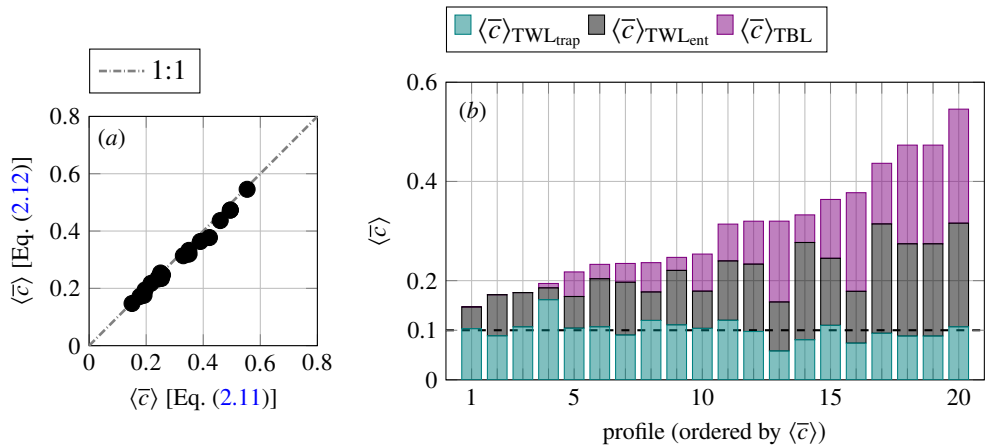
\begin{figure}
\centering
\input{Figures/fig5}
\caption{Mean air concentrations derived from \possessivecite{Killen1968} data   (\textit{a}) measured  mean air concentrations versus Eq. (\ref{eq:meanair})
(\textit{b}) contribution of different physical mechanisms to the mean air concentration as per Eq. (\ref{eq:meanair})}
\label{fig:cmean}
\end{figure}

Figure \ref{fig:cmean}\textit{a} confirms that predicted mean air concentrations [Eq. (\ref{eq:meanair})] were in good agreement with measured mean air concentrations, the latter directly evaluated from phase-detection intrusive measurements using Eq. (\ref{eq:meanair0}). Note that Eq. (\ref{eq:meanair}) was numerically integrated, incorporating the analytical solutions for the $\overline{c}_\text{TBL}$ [Eq. (\ref{eq:voidfraction1})],  $\overline{c}_\text{trap}$ [Eq. (\ref{entrappedair})],  $\overline{c}_\text{ent}$ [Eq. (\ref{entrainedair})], as well as $\Gamma$ [Eq. (\ref{eq:gaussianerr})].

Equation (\ref{eq:meanair}) allows to differentiate between three different physical mechanisms contributing to the mean air concentration. It was found that $\langle \overline{c} \rangle_\text{TBL}$ and $\langle \overline{c}\rangle_{\text{TWL}_\text{ent}}$ increased with increasing $\langle \overline{c} \rangle$, whereas the entrapped air concentration of the TWL was approximately constant, at $\langle \overline{c} \rangle_{\text{TWL}_\text{trap}} \approx 0.1$ (Fig. \ref{fig:cmean}\textit{b}; \textcolor{black}{profiles ordered by $\langle \overline{c} \rangle$}), \textcolor{black}{which hints at the fact that the geometry of the (mean) air-water interface of the TWL varied only slightly in \possessivecite{Killen1968} experiments.} Note that a constant entrapped mean air concentration was previously reported by \citet{Wilhelms2005}, \textcolor{black}{which is further corroborated by recent computations of the free-surface roughness wavelength
distribution in supercritical flows \citep[Fig. 12]{Valero2018}}. More generally, the deformation of the free-surface in shallow turbulent flows, and therefore the entrapped air concentration, is known to be driven by various processes, including the interaction of turbulent coherent structures with the water surface, resonant wave growth, and effects of  bed topography \citep{Brocchini2001JFM1,Valero2016,Muraro2021}. While effects of these different processes on entrapped and entrained air concentrations have not been studied in the past, the current decomposition of the mean air concentration provides a versatile framework that  allows to assess the contribution of individual physical mechanisms. 

\subsection{Streamwise self-aeration development and equilibrium state}
\label{sec:Cdevelopment}

\textcolor{black}{In Fig. \ref{fig:cmean}\textit{b}, 
the profiles of \possessivecite{Killen1968} four measurement series (Table \ref{tab:data}) were ordered by increasing $\langle \overline{c} \rangle$, which is appropriate to exemplify general trends of $\langle\overline{c} \rangle_{\text{TWL}_\text{trap}}$, $\langle\overline{c}\rangle_{\text{TWL}_\text{ent}}$, and $\langle\overline{c}\rangle_\text{TBL}$ with respect to $\langle \overline{c} \rangle$. To provide more insights on $\langle \overline{c} \rangle$ and its controlling parameters, it is important to point out different distinct regions in self-aerated flows, including the non-aerated developing flow region, the aerated gradually varied flow (GVF) region, and the aerated uniform flow (UF) region (Fig. \ref{Fig1}\textit{b}), which have been described in literature, e.g., \cite{Wood1991,Chanson1996}, amongst others. In the GVF region (Fig. \ref{Fig1}\textit{b}), the mean air concentration $\langle \overline{c} \rangle$ depends on the streamwise location with respect to the inception point of air entrainment ($L_i$, Fig. \ref{Fig1}\textit{b}), as well as on the slope $\theta$ and (similarly) on the Froude-number. In contrast, the mean air concentration
in the UF region, termed equilibrium mean air concentration $\langle \overline{c}\rangle_\infty$ (Fig. \ref{Fig1}\textit{b}), is known to be solely a function of $\theta$ (or $Fr$) \citep{Hager91,Matos1995}.}

\begin{figure}
\centering
\input{Figures/fig6.tex}
\caption{Equilibrium air concentration and streamwise self-aeration development  (\textit{a}) equilibrium air concentrations $\langle \overline{c} \rangle_\infty$ and $\langle \overline{c} \rangle_{\text{TBL} \infty}$ versus Froude-numbers for the data of \protect \citet{Straub1958}, compared with non-equilibrium concentrations from \protect \citet{Killen1968} (\textit{b},\textit{c}) evolution of depth-averaged (mean) air concentrations in \protect \possessivecite{Killen1968} high Froude-number flows; GVF = gradually varied flow, UF = uniform flow}
\label{fig:Cdevelopment}
\end{figure}
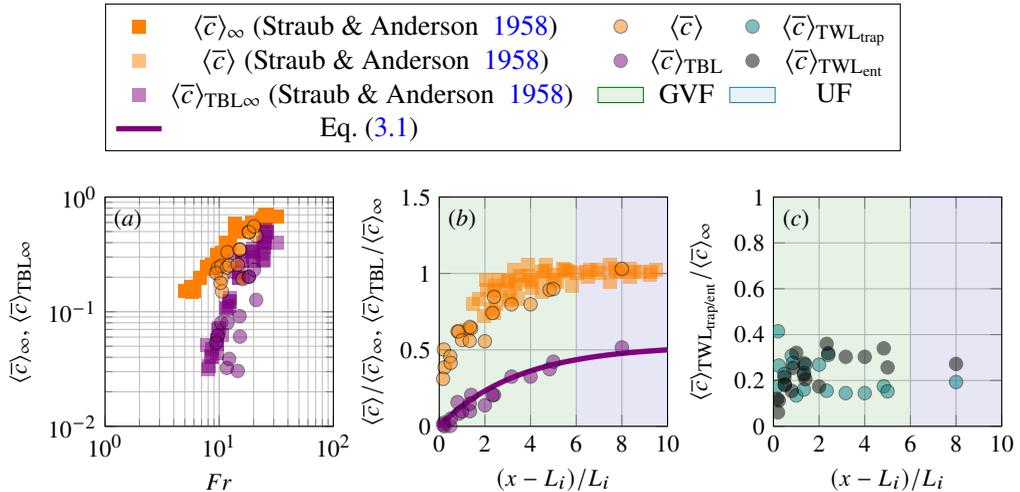

\textcolor{black}{Figure \ref{fig:Cdevelopment}\textit{a} compares equilibrium air concentrations from \citet{Straub1958} with the present re-analysis, showing that \possessivecite{Killen1968} measuremets were taken in the GVF region. Following \citet{Wei2022a}, \possessivecite{Killen1968}  mean air concentrations are normalised with their equilibrium value, approximated as $\langle \overline{c} \rangle_\infty = 0.75 \sin \theta^{0.75}$ \citep{Hager91}, and plotted against the dimensionless streamwise coordinate $(x-L_i)/L_i$ (Figs. \ref{fig:Cdevelopment}\textit{b,c}). This normalisation shows a good collapse of the four different measurement series, thereby demonstrating how the two-state superposition model can be used to finely differentiate between different physical processes in the streamwise decomposition of $\langle \overline{c} \rangle$. The evolution of $\langle \overline{c} \rangle_\text{TBL}$ displays asymptotic behaviour towards equilibrium and is well described by an analytical solution of the continuity equation for air in water (Appendix \ref{appendixB})
\begin{equation}
\langle \overline{c} \rangle_\text{TBL} =  \langle \overline{c} \rangle_{\text{TBL}\infty} \left( 1 - \exp \left(-\frac{u_r \cos \theta}{q} \left(x - L_i \right) \right) \right),
\label{eqCdevelopment}
\end{equation}
where $\langle \overline{c} \rangle_{\text{TBL} \infty}$ is the equilibrium mean air concentration of the TBL, and $u_r$ is the depth-averaged bubble rise velocity. Here, Eq. (\ref{eqCdevelopment}) was evaluated for \possessivecite{Killen1968} data on the 52.5 degree slope, and good agreement with measurements was achieved using $u_r = 0.1$ m/s and $\langle \overline{c} \rangle_{\text{TBL}\infty} = 0.53 \langle \overline{c} \rangle_{\infty}$ (Fig. \ref{fig:Cdevelopment}\textit{b}). The mean air concentration $\langle \overline{c} \rangle$ similarly approaches equilibrium, and an additional comparison with data from \citet{Straub1958} reveals that the length of the GVF region is approximately 4 to 6 times $L_i$ (Fig. \ref{fig:Cdevelopment}\textit{b}). Further, Fig. \ref{fig:Cdevelopment}\textit{c} shows that the 
trends in $\langle \overline{c} \rangle_{\text{TWL}_\text{trap}}$ and $\langle \overline{c} \rangle_{\text{TWL}_\text{ent}}$ are opposite for $(x-L_i)/L_i \lessapprox 2$, which is consistent with experimental observations of free-surface roughness/waves, but no entrained air, upstream of the inception point of free-surface aeration \citep{Felder2023}. In the GVF region, around $(x-L_i)/L_i \gtrapprox 2$, the contributions of $\langle \overline{c} \rangle_{\text{TWL}_\text{trap}}$ and $\langle \overline{c} \rangle_{\text{TWL}_\text{ent}}$ become approximately constant, which suggests that equilibrium for the TWL is achieved further upstream than for the TBL. Additional research is required to confirm these findings.}

\section{Discussion: Model applicability and limitations}
\label{sec:discussion}
\begin{figure}
\centering
\input{Figures/fig7}
\caption{Applicability of the proposed equations for smooth chute flows: variation of $\overline{c}_0$ versus $\langle \overline{c} \rangle$}
\label{fig:bottom}
\end{figure}
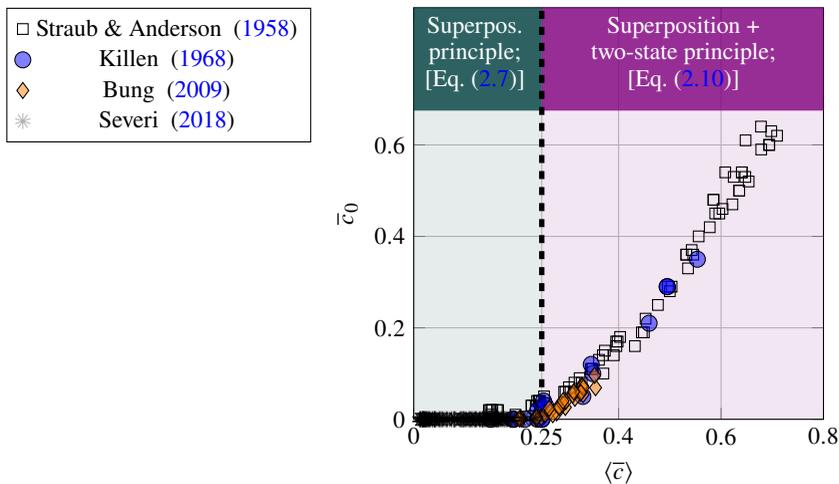

In $\S$ \ref{sec:superposition}, the superposition principle [Eq. (\ref{eq:superposition})] was formulated for flow situations where aeration is confined to the wavy layer of a supercritical free-surface flow, i.e., a pure TWL, and it was combined with the two-state convolution [Eq. (\ref{eq:voidfractionfinal1})]  to account for flows where the air bubble diffusion layer protrudes to the channel bottom, see $\S$ \ref{sec:combination}. Therefore, the bottom air concentration $\overline{c}_0$, defined as the air concentration in the vicinity of the solid invert \citep{Hager91,Kramer2021}, is a natural choice to exemplify the application range of proposed equations. Figure \ref{fig:bottom} shows a plot of $\overline{c}_0$ versus $\langle \overline{c} \rangle$, illustrating that Eq. (\ref{eq:superposition}) is valid for $\langle \overline{c} \rangle \lessapprox  0.25$, while Eq. (\ref{eq:voidfractionfinal1}) is valid for $\langle \overline{c} \rangle \gtrapprox  0.25$. It is noteworthy mentioning that the total conveyed air concentration is fully described by the two-state convolution [Eq. (\ref{eq:voidfractionfinal})] of \cite{Kramer2023JFM}, while the superposition principle provides additional physical insights into the structure of the TWL, given that additional measurements of entrapped air are made. 

\textcolor{black}{The model parameters of the extended two-state superposition principle [Eq. (\ref{eq:voidfractionfinal1})] are the Rouse number ($\beta$), the boundary layer thickness ($\delta$), the air concentration at half the boundary layer thickness ($\overline{c}_{\delta/2}$), the transition/interface parameters $y_\star$ and $\sigma_\star$, mixture flow depths $y_{50}$ and $y_{{50}_\text{trap}}$, as well as the length scale of the TWL ($\mathcal{H}$) and the root-mean-square wave height ($\mathcal{H}_\text{trap})$. Of these parameters, $y_{50}$, $y_{50_\text{trap}}$, $\overline{c}_{\delta/2}$, and $\delta$ were directly extracted from measurements, whereas $\beta$, $y_\star$, $\sigma_\star$, $\mathcal{H}$, and $\mathcal{H}_\text{trap}$ were obtained through fitting. It is acknowledged that the predictive capability for some parameters is currently limited, which is however deemed acceptable, as the aim of the present model is to establish a physically based description of the air concentration distribution, with physical parameters responding to the flow. Further details on $\mathcal{H}, \mathcal{H}_\text{trap}$, $y_{50}$, and  $y_{{50}_\text{trap}}$ are presented in Fig. \ref{fig:params}, while the interface parameters and the Rouse number range between $\beta = 0.05$ to 1.2, $y_\star/\delta = 0.6$ to 0.9, and $\sigma_\star/\delta = 0.1$ to 0.2 \citep{Kramer2023JFM}.}

As mentioned before, the application of the superposition principle requires two separate measurements, one for entrapped air and one for total conveyed air. \citet{Killen1968} used a common intrusive phase-detection probe for the measurement of total conveyed air, while a larger-sized dipping probe was used for the measurement of entrapped air. It is emphasized that these measurements are unique, and no other researchers have deployed a comparable setup since. \textcolor{black}{Future measurement of entrapped air, either through a measurement setup similar to that of \citet{Killen1968} or via non-intrusive measurement techniques, such as acoustic displacement meters \citep{Cui2022} or laser time-of-flight or triangulation sensors, are of high relevance to increase our fundamental physical understanding of air-water flow processes, which is anticipated to lead to an improvement/revision of some existing modelling approaches, e.g., for air-water mass transfer in supercritical flows \citep{Bung2018,Kramer2020}}.

\textcolor{black}{Lastly, it is stressed that the two-state superposition model has been developed for statistically steady self-aerated flows on steep slopes in prismatic rectangular channels. However, the model can readily be adapted to characterize air concentration distributions of other statistically steady self-aerated flows in prismatic geometries, e.g., hydraulic jumps. The application to unsteady aerated flows, such as breaking waves, is more involved and requires the hundredfold repetition of experiments, followed by an application of ensemble-averaging techniques, see \cite{Blenkinsopp2007,Wuthrich2022}.}

\section{Conclusion}
In this work, a novel superposition principle for entrapped and entrained air within the TWL of a supercritical open-channel flow is presented. The corresponding air concentration distributions for entrapped air and total conveyed air both follow a Gaussian error function, while entrained air is characterized by a Gaussian normal distribution. The free parameters of the mathematical formulation are the root-mean-square wave height and the length scale of the TWL, which are shown to be of similar magnitude and dependent on the mean air concentration. Subsequently, the superposition principle is combined with the two-state convolution of \citet{Kramer2023JFM}, representing the most complete and physical description of the air concentration distribution to date. A bed-normal integration of this combined equation allows to differentiate between three different physical mechanisms that contribute to the mean air concentration, comprising entrapment of air due to free-surface deformations, entrainment of air due to turbulent forces, and turbulent diffusion of air bubbles into the TBL. \textcolor{black}{The subsequent analysis of  the streamwise development of these mechanisms suggests that the equilibrium for the TWL is achieved further upstream than for the TBL. While further research is required to confirm this finding, the presented application nicely demonstrates how the two-state superposition model can be used to uncover new flow physics in self-aerated flows. It is acknowledged that only a limited data set was analysed herein, which is because the quantification of entrapped air requires specific flow measurement instrumentation, i.e., a dipping probe.} 

Overall, it is anticipated that the presented theory holds for a wide range of high Froude-number self-aerated flows, \textcolor{black}{encompassing the range tested by \citet[$5  \lessapprox Fr \lessapprox 32$]{Straub1958}.} A meaningful extension of this work would  comprise a thorough development/testing of new sensors for the non-intrusive measurement of entrapped air, as well as the development of advanced phase-detection signal processing techniques that allow to discriminate between entrapped and entrained air. These developments are to be followed by detailed investigations on the functional dependence between model parameters and flow/geometric properties, including bottom-surface roughness, friction velocity, flow depth, as well as other statistical measures of bulk flow and turbulence. A better understanding of the underlying physics of self-aerated flows will enable the formulation and implementation of more physically consistent numerical models for air entrainment and transport.

\backsection[Acknowledgements]{I would like to thank Dr Daniel Valero (Karlsruhe Institute of Technology, KIT) for fruitful discussions, Prof Daniel Bung (FH Aachen) for proof-reading and sharing of data sets, and lastly  Dr Armaghan Severi (Manly Hydraulics Laboratory) for sharing her images.}

\backsection[Funding]{This research received no specific grant from any funding agency, commercial or not-for-profit sectors.}

\backsection[Declaration of interests]{The author reports no conflict of interest.}

\backsection[Data availability statement]{All data, models, or code that support the findings of this study are available from the corresponding author upon reasonable request.}

\backsection[Author ORCID]{M. Kramer, https://orcid.org/0000-0001-5673-2751}

\bibliographystyle{jfm}

\appendix
\section{Re-analysis of \possessivecite{Killen1968} measurements}
\label{appendixA}
\textcolor{black}{This appendix presents the application of the combined superposition two-state formulation to 20 concentration profiles of \possessivecite{Killen1968} data set, with corresponding flow conditions indicated in Table \ref{tab:data}. Each measured profile with $\langle \overline{c} \rangle \gtrapprox 0.25$  is represented by two sub-figures. In the first sub-figure (index \textit{a}), the superposition principle is plotted with its corresponding Eqns. [(\ref{entrappedair}), (\ref{entrainedair}), (\ref{eq:superposition})], together with the profile number, chute angle ($\theta$), specific discharge ($q$), and streamwise distance ($x$ in m) from the upstream crest. Each second sub-figure (index \textit{b}) contains  plots of the two-state formulation [Eqns. (\ref{eq:voidfraction1}), (\ref{eq:superposition}), (\ref{eq:voidfractionfinal1})], including the mean air concentration. For $\langle \overline{c} \rangle \lessapprox 0.25$, the air concentration distribution is characterised by the superposition principle alone, and only the first sub-figure is plotted (index dropped). The numbering of the profiles increases with streamwise distance for each test series, as per Table \ref{tab:data}, and the background of every first profile of the four series is shaded in gray.}

\input{Figures/figAppendixKillen}

\section{Streamwise evolution of $\langle \overline{c} \rangle_\text{TBL}$}
\label{appendixB}
\textcolor{black}{To develop an equation for the streamwise evolution of $\langle \overline{c} \rangle_\text{TBL}$, the continuity equation for entrained air within the TBL is written \citep{Wood1985} 
\begin{equation}
\frac{\text{d} (q_{a_\text{TBL}})}{\text{d}x} =  v_e - \langle \overline{c} \rangle_\text{TBL} \, u_r \cos \theta, 
\label{eq:Cdevelopment1}
\end{equation}
where $q_{a_\text{TBL}}$ is the specific air flow rate of the TBL per unit width, $v_e$ is the entrainment velocity of air into the TBL, and $u_r$ $\cos \theta$ represents detrainment of air, with $u_r$ being a depth-averaged rise velocity of air bubbles. It is noted that previous researchers used the total air flow rate $q_a$ instead of $q_{a_\text{TBL}}$, which however is thought to be incorrect, as the volume of entrapped air, for example in the developing non-aerated region, is not balanced by rising air bubbles. Similar to \citet{Wood1985}, it is now assumed $q_{a_\text{TBL}}/q \approx \langle \overline{c} \rangle_\text{TBL}$; note that this assumption represents a simplification, and more elaborate relationships for $q_{a_\text{TBL}}/q$ may be used, see \citet{Wood1991,Chanson1996}, which however would not lead to an explicit solution. Substitution of $q_{a_\text{TBL}}/q \approx \langle \overline{c} \rangle_\text{TBL}$ into Eq. (\ref{eq:Cdevelopment1}) leads to
\begin{equation}
q \frac{\text{d} \langle \overline{c} \rangle_\text{TBL}}{\text{d}x} =  v_e - \langle \overline{c} \rangle_\text{TBL} \, u_{r} \cos \theta.
\label{eq:Cdevelopment2}
\end{equation}}

\textcolor{black}{In the uniform flow region (Fig. \ref{Fig1}), streamwise gradients vanish, implying that Eq. (\ref{eq:Cdevelopment2}) simplifies to
\begin{equation}
0 =  v_{e\infty} - \langle \overline{c} \rangle_{\text{TBL}\infty}  \, u_{r \infty} \cos \theta, 
\label{eq:Cdevelopment3}
\end{equation}
where $v_{e \infty}, \langle \overline{c} \rangle_{\text{TBL} \infty}$, and $u_{r \infty}$ are the entrainment velocity, mean air concentration, and bubble rise velocity in the uniform flow region. Equation (\ref{eq:Cdevelopment3}) is now subtracted from Eq. (\ref{eq:Cdevelopment2}), further assuming $v_e \approx v_{e \infty}$ and $u_r \approx u_{r \infty}$
\begin{equation}
q  \frac{\text{d} \langle \overline{c} \rangle_\text{TBL}}{\text{d}x} =   u_r \cos \theta \left(\langle \overline{c} \rangle_{\text{TBL}\infty}  - \langle \overline{c} \rangle_{\text{TBL}} \right).
\end{equation}}

\textcolor{black}{Separating variables
\begin{equation}
 \frac{1}{\langle \overline{c} \rangle_{\text{TBL}\infty}  - \langle \overline{c} \rangle_{\text{TBL}}} \, \text{d} \langle \overline{c} \rangle_{\text{TBL}} =  \frac{u_r \cos \theta}{q}  \text{d}x, 
\end{equation}
and integrating between the inception point of air entrainment ($x = L_i$) and an arbitrary downstream location
\begin{equation}
\int_{0}^{\langle \overline{c} \rangle_\text{TBL}} \frac{1}{\langle \overline{c} \rangle_{\text{TBL}\infty}  - \langle \overline{c} \rangle_{\text{TBL}}} \text{d} \langle \overline{c} \rangle_{\text{TBL}} = \frac{u_r \cos \theta}{q} \int_{x=L_i}^{x} \text{d}x, 
 \label{eq:Cdevelopment4a}
\end{equation}
yields the following solution
\begin{equation}
 \ln \left( \frac{\langle \overline{c} \rangle_{\text{TBL}\infty}} {\langle \overline{c} \rangle_{\text{TBL}\infty} - \langle \overline{c} \rangle_\text{TBL}  } \right) = \frac{u_r \cos \theta}{q} \left( x - L_i \right), 
 \label{eq:Cdevelopment4}
\end{equation}
where the lower limit of the integral on the LHS (left-hand side) of Eq. (\ref{eq:Cdevelopment4a}) corresponds to the entrained air concentration at the inception point, which per definition $\langle \overline{c} \rangle_\text{TBL}(x=L_i) = 0$. Equation (\ref{eq:Cdevelopment4}) can be re-arranged/simplified to obtain the following analytical expression for the streamwise development of $\langle \overline{c} \rangle_\text{TBL}$
\begin{equation}
\langle \overline{c} \rangle_\text{TBL} =  \langle \overline{c} \rangle_{\text{TBL}\infty} \left( 1 - \exp \left(-\frac{u_r \cos \theta}{q} \left(x - L_i \right) \right) \right). \label{eq:Cdevelopment5}
\end{equation}}

This equation provides a simple method to characterise the increase of the mean air concentration of the TBL as function of the equilibrium air concentration ($\langle \overline{c} \rangle_{\text{TBL}\infty}$), depth-averaged bubble rise velocity ($u_r$), slope ($\theta$), specific water flow rate ($q$), and the streamwise distance from the inception point of air entrainment ($x-L_i$). Substituting $q = \langle \overline{u} \rangle_i d_i$, with $\langle \overline{u}\rangle_i$ and $d_i$ being the mean water velocity and the water depth at the inception point, \possessivecite[Eq. 4]{Wilhelms2005} empirical relationship can be recovered
\begin{equation}
\langle \overline{c} \rangle_\text{TBL} =  \langle \overline{c} \rangle_{\text{TBL}\infty} \left( 1 - \exp \left(-\frac{u_r \cos \theta}{\langle \overline{u}\rangle_i} \,  \frac{x - L_i}{d_i}  \right) \right) = \langle \overline{c} \rangle_{\text{TBL}\infty} \left( 1 - \exp \left(- \alpha  \frac{x - L_i}{d_i}  \right) \right), 
\label{eq:Cdevelopment6}
\end{equation}
thereby revealing that their coefficient $\alpha = (u_r \cos \theta) / \langle \overline{u}\rangle_i$ corresponds to a dimensionless bubble rise velocity. In order to solve Eqns. (\ref{eq:Cdevelopment5}) or (\ref{eq:Cdevelopment6}), the unknowns $u_r$ and $\langle \overline{c} \rangle_{\text{TBL}\infty}$ need to be determined. One can  perform some air-water flow measurements or, alternatively, use a best-fit approach.
\end{document}

%% file: Figures/fig1.tex
\begin{tikzpicture}

\node[anchor=north west] at  (-1,0)
{\includegraphics[width= 13.5 cm]{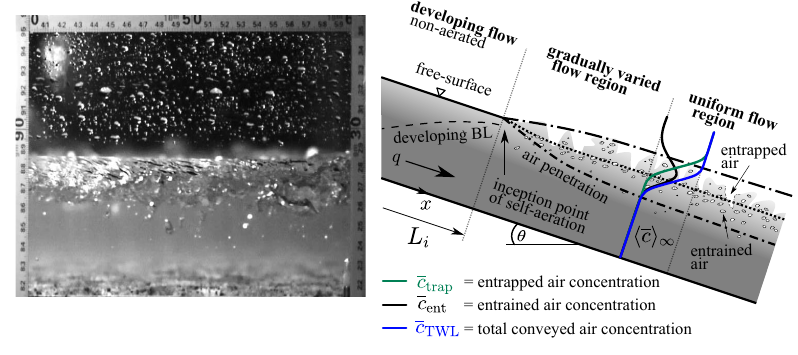} };  

\node[align=center,color=black,anchor=north west] at (0cm,-4cm) {$q = 0.188$ m$^2$/s};

\node (A) at (-0.3, -4){};
\node (B) at (0.8, -4){};

\node[align=center,fill=white,fill opacity=0.7,text opacity=1] at (-0.1,-0.93) {(\textit{a})};

\node[align=center,fill=white,fill opacity=0.7,text opacity=1] at (5.9,-0.93) {(\textit{b})};

\node[align=center,fill=white,fill opacity=0.65,text opacity=1] at (2,-3) {TWL};

% arrows
\draw[-latex, line width = 1.7] (A) edge (B);

\end{tikzpicture}

%% file: Figures/fig2.tex
\begin{tikzpicture}
\definecolor{color1}{rgb}{1,0.7,0.1}
\definecolor{color2}{rgb}{0.15,0.45,0.55}

 \begin{groupplot}[
     group style = {group size = 3 by 2,horizontal sep=1.4cm,vertical sep=1.4cm},
     width = 1\textwidth]

\nextgroupplot[height = 4.6cm,
     width = 4.6cm,
         xlabel={$\overline{c}_\text{TWL}$},
         ylabel={$y/y_{90}$},
         xtick={0, 0.25, 0.5 , 0.75, 1},
         ytick={0, 0.5, 1,  1.5,2 },
       %yticklabels={0, 0.5, \phantom{0 , 1,  1.5,2 },
     grid=both,
     xmin=0,
     xmax=1,
     ymin=0,
     ymax=2,
    legend columns=6,
   legend style={cells={align=left},anchor = south west,at={(0,1.2)},font=\normalsize},
   clip mode = individual,
   ]

\node[align=center] at (rel axis cs:1.15,0.5) {\huge{$\bf{=}$}};

\addplot [fill=black,mark=*,only marks,fill opacity=0,draw opacity=1, mark size=2.5,clip mode=individual]
table[row sep=crcr]{%
0 -1	\\
}; \addlegendentry{$\overline{c}_\text{meas} \, \,$};

\addplot [fill=black,mark=diamond,only marks,fill opacity=0,draw opacity=1, mark size=3,clip mode=individual]
table[row sep=crcr]{%
0 -1	\\
}; \addlegendentry{$\overline{c}_\text{trap,meas} \, \,$};

\addplot [fill=blue,draw=blue,mark=*,only marks,fill opacity=1,draw opacity=0.5, mark size=3,clip mode=individual]
table[row sep=crcr]{%
0 -1	\\
}; \addlegendentry{$y_{50}$};

\addplot [fill=blue,draw=blue,mark=pentagon*
,only marks,fill opacity=1,draw opacity=1, mark size=2.5,clip mode=individual,line width =1.3]
table[row sep=crcr]{%
0 -1	\\
}; \addlegendentry{$y_{90}$ \, };

\addplot [fill=teal,draw=teal,mark=diamond*,only marks,fill opacity=1,draw opacity=0.5, mark size=3.5,clip mode=individual]
table[row sep=crcr]{%
0 -1	\\
}; \addlegendentry{$y_{50_\text{trap}}$ \,};

\addplot [fill=black,draw=black,mark=square*,only marks,fill opacity=1,draw opacity=0.5, mark size=2.5,clip mode=individual]
table[row sep=crcr]{%
0 -1	\\
}; \addlegendentry{$\overline{c}_\text{max}, y_{\overline{c}_\text{max}}$};

\addplot[color=blue,mark=none,line width = 2,draw opacity=1] table[row sep=crcr]{%
0 -1	\\
}; \addlegendentry{Eq. (\ref{eq:superposition}) \,};

\addplot[color=teal,mark=none,line width = 2,draw opacity=1] 
table[row sep=crcr]{%
0 -1	\\
}; \addlegendentry{Eq. (\ref{entrappedair}) \,};

\addplot[color=black,mark=none,line width = 2,draw opacity=1] 
table[row sep=crcr]{%
0 -1	\\
}; \addlegendentry{Eq. (\ref{entrainedair}) \,};

\addplot [fill=black,mark=*,only marks,fill opacity=0,draw opacity=1, mark size=2.5,clip mode=individual]
table[row sep=crcr]{%
0	0	\\
0	0.060344828	\\
0	0.120689655	\\
0	0.181034483	\\
0	0.24137931	\\
0	0.301724138	\\
0.01	0.362068966	\\
0.02	0.422413793	\\
0.02	0.482758621	\\
0.03	0.543103448	\\
0.05	0.603448276	\\
0.1	0.663793103	\\
0.18	0.724137931	\\
0.32	0.784482759	\\
0.55	0.844827586	\\
0.74	0.905172414	\\
0.86	0.965517241	\\
0.93	1.025862069	\\
0.96	1.086206897	\\
0.98	1.146551724	\\
0.99	1.206896552	\\
1	1.267241379	\\
1	1.327586207	\\
};

\addplot[color=blue,mark=none,line width = 2,draw opacity=1] 
table[row sep=crcr]{%
1.67E-12	0	\\
2.97E-12	0.009783532	\\
5.27E-12	0.019567065	\\
9.27E-12	0.029350597	\\
1.62E-11	0.039134129	\\
2.81E-11	0.048917662	\\
4.86E-11	0.058701194	\\
8.32E-11	0.068484727	\\
1.42E-10	0.078268259	\\
2.40E-10	0.088051791	\\
4.03E-10	0.097835324	\\
6.73E-10	0.107618856	\\
1.12E-09	0.117402388	\\
1.84E-09	0.127185921	\\
3.01E-09	0.136969453	\\
4.90E-09	0.146752985	\\
7.91E-09	0.156536518	\\
1.27E-08	0.16632005	\\
2.03E-08	0.176103582	\\
3.21E-08	0.185887115	\\
5.06E-08	0.195670647	\\
7.91E-08	0.20545418	\\
1.23E-07	0.215237712	\\
1.90E-07	0.225021244	\\
2.91E-07	0.234804777	\\
4.43E-07	0.244588309	\\
6.71E-07	0.254371841	\\
1.01E-06	0.264155374	\\
1.51E-06	0.273938906	\\
2.24E-06	0.283722438	\\
3.31E-06	0.293505971	\\
4.85E-06	0.303289503	\\
7.06E-06	0.313073035	\\
1.02E-05	0.322856568	\\
1.47E-05	0.3326401	\\
2.10E-05	0.342423633	\\
2.98E-05	0.352207165	\\
4.21E-05	0.361990697	\\
5.90E-05	0.37177423	\\
8.21E-05	0.381557762	\\
0.000113621	0.391341294	\\
0.000156241	0.401124827	\\
0.000213489	0.410908359	\\
0.000289874	0.420691891	\\
0.000391109	0.430475424	\\
0.000524384	0.440258956	\\
0.000698665	0.450042488	\\
0.000925049	0.459826021	\\
0.001217144	0.469609553	\\
0.001591509	0.479393086	\\
0.002068106	0.489176618	\\
0.002670799	0.49896015	\\
0.003427859	0.508743683	\\
0.004372468	0.518527215	\\
0.005543212	0.528310747	\\
0.006984531	0.53809428	\\
0.008747101	0.547877812	\\
0.010888122	0.557661344	\\
0.013471475	0.567444877	\\
0.016567715	0.577228409	\\
0.020253872	0.587011942	\\
0.024613015	0.596795474	\\
0.029733566	0.606579006	\\
0.035708326	0.616362539	\\
0.042633208	0.626146071	\\
0.050605672	0.635929603	\\
0.05972286	0.645713136	\\
0.070079466	0.655496668	\\
0.081765367	0.6652802	\\
0.094863071	0.675063733	\\
0.109445045	0.684847265	\\
0.125571009	0.694630797	\\
0.143285279	0.70441433	\\
0.162614264	0.714197862	\\
0.183564214	0.723981395	\\
0.206119336	0.733764927	\\
0.23024035	0.743548459	\\
0.255863595	0.753331992	\\
0.282900746	0.763115524	\\
0.311239184	0.772899056	\\
0.340743061	0.782682589	\\
0.371255048	0.792466121	\\
0.402598735	0.802249653	\\
0.434581639	0.812033186	\\
0.466998719	0.821816718	\\
0.49963631	0.83160025	\\
0.532276336	0.841383783	\\
0.564700674	0.851167315	\\
0.596695514	0.860950848	\\
0.628055578	0.87073438	\\
0.658588065	0.880517912	\\
0.688116172	0.890301445	\\
0.716482115	0.900084977	\\
0.743549548	0.909868509	\\
0.76920532	0.919652042	\\
0.793360559	0.929435574	\\
0.815951056	0.939219106	\\
0.836936996	0.949002639	\\
0.856302075	0.958786171	\\
0.874052073	0.968569703	\\
0.890212974	0.978353236	\\
0.904828722	0.988136768	\\
0.91795872	0.997920301	\\
0.929675184	1.007703833	\\
0.940060425	1.017487365	\\
0.949204187	1.027270898	\\
0.95720108	1.03705443	\\
0.96414822	1.046837962	\\
0.970143077	1.056621495	\\
0.97528162	1.066405027	\\
0.979656732	1.076188559	\\
0.983356945	1.085972092	\\
0.986465455	1.095755624	\\
0.989059433	1.105539156	\\
0.991209581	1.115322689	\\
0.992979929	1.125106221	\\
0.994427824	1.134889754	\\
0.995604085	1.144673286	\\
0.996553287	1.154456818	\\
0.997314141	1.164240351	\\
0.997919947	1.174023883	\\
0.998399076	1.183807415	\\
0.998775485	1.193590948	\\
0.99906922	1.20337448	\\
0.999296908	1.213158012	\\
0.99947222	1.222941545	\\
0.999606303	1.232725077	\\
0.999708167	1.24250861	\\
0.999785037	1.252292142	\\
0.999842659	1.262075674	\\
0.999885562	1.271859207	\\
0.999917294	1.281642739	\\
0.999940607	1.291426271	\\
0.999957619	1.301209804	\\
0.999969951	1.310993336	\\
0.999978831	1.320776868	\\
0.999985181	1.330560401	\\
0.999989693	1.340343933	\\
0.999992877	1.350127465	\\
0.999995109	1.359910998	\\
0.999996663	1.36969453	\\
0.999997738	1.379478063	\\
0.999998477	1.389261595	\\
0.999998981	1.399045127	\\
0.999999322	1.40882866	\\
0.999999552	1.418612192	\\
0.999999706	1.428395724	\\
0.999999808	1.438179257	\\
0.999999876	1.447962789	\\
0.99999992	1.457746321	\\
0.999999949	1.467529854	\\
0.999999968	1.477313386	\\
0.99999998	1.487096918	\\
0.999999987	1.496880451	\\
0.999999992	1.506663983	\\
0.999999995	1.516447516	\\
0.999999997	1.526231048	\\
0.999999998	1.53601458	\\
0.999999999	1.545798113	\\
0.999999999	1.555581645	\\
1	1.565365177	\\
1	1.57514871	\\
1	1.584932242	\\
1	1.594715774	\\
1	1.604499307	\\
1	1.614282839	\\
1	2	\\
};

\node[align=center] at (rel axis cs:0.1,0.9) {(\textit{a})};

\addplot [fill=blue,draw=blue,mark=*,only marks,fill opacity=1,draw opacity=0.5, mark size=3,clip mode=individual]
table[row sep=crcr]{%
0.5 0.83	\\
};

\addplot [fill=blue,draw=blue,mark=pentagon*
,only marks,fill opacity=1,draw opacity=1, mark size=2.5,clip mode=individual,line width =1.3]
table[row sep=crcr]{%
0.9  1\\
}; 

\nextgroupplot[height = 4.6cm,
     width = 4.6cm,
     xlabel={$\overline{c}_\text{trap}$},
     grid=both,
     xmin=0,
     xmax=1,
     ymin=0,
       xtick={0, 0.25, 0.5, 0.75,1},
     ymax=2,
    legend columns=4,
   legend style={cells={align=left},anchor = south west,at={(0,1.2)},font=\normalsize},
   clip mode=individual
   ]

\node[align=center] at (rel axis cs:0.1,0.9) {(\textit{b})};

\node[align=center] at (rel axis cs:1.15,0.5) {\huge{$\bf{+}$}};

\addplot [fill=black,mark=diamond,only marks,fill opacity=0,draw opacity=1, mark size=3,clip mode=individual]
table[row sep=crcr]{%
0	0	\\
0	0.060344828	\\
0	0.120689655	\\
0	0.181034483	\\
0	0.24137931	\\
0	0.301724138	\\
0	0.362068966	\\
0	0.422413793	\\
0	0.482758621	\\
0	0.543103448	\\
0	0.603448276	\\
0	0.663793103	\\
0.02	0.724137931	\\
0.08	0.784482759	\\
0.24	0.844827586	\\
0.41	0.905172414	\\
0.64	0.965517241	\\
0.81	1.025862069	\\
0.91	1.086206897	\\
0.95	1.146551724	\\
0.97	1.206896552	\\
0.98	1.267241379	\\
1	1.327586207	\\
};

\addplot[color=teal,mark=none,line width = 2,draw opacity=1] 
table[row sep=crcr]{%
5.55E-17	0	\\
1.11E-16	0.009783532	\\
2.22E-16	0.019567065	\\
5.00E-16	0.029350597	\\
9.44E-16	0.039134129	\\
1.94E-15	0.048917662	\\
3.89E-15	0.058701194	\\
7.72E-15	0.068484727	\\
1.53E-14	0.078268259	\\
2.98E-14	0.088051791	\\
5.79E-14	0.097835324	\\
1.12E-13	0.107618856	\\
2.13E-13	0.117402388	\\
4.05E-13	0.127185921	\\
7.64E-13	0.136969453	\\
1.43E-12	0.146752985	\\
2.65E-12	0.156536518	\\
4.89E-12	0.16632005	\\
8.93E-12	0.176103582	\\
1.62E-11	0.185887115	\\
2.92E-11	0.195670647	\\
5.22E-11	0.20545418	\\
9.27E-11	0.215237712	\\
1.63E-10	0.225021244	\\
2.86E-10	0.234804777	\\
4.96E-10	0.244588309	\\
8.54E-10	0.254371841	\\
1.46E-09	0.264155374	\\
2.48E-09	0.273938906	\\
4.17E-09	0.283722438	\\
6.98E-09	0.293505971	\\
1.16E-08	0.303289503	\\
1.91E-08	0.313073035	\\
3.12E-08	0.322856568	\\
5.07E-08	0.3326401	\\
8.16E-08	0.342423633	\\
1.31E-07	0.352207165	\\
2.07E-07	0.361990697	\\
3.26E-07	0.37177423	\\
5.11E-07	0.381557762	\\
7.93E-07	0.391341294	\\
1.22E-06	0.401124827	\\
1.87E-06	0.410908359	\\
2.84E-06	0.420691891	\\
4.28E-06	0.430475424	\\
6.41E-06	0.440258956	\\
9.52E-06	0.450042488	\\
1.40E-05	0.459826021	\\
2.06E-05	0.469609553	\\
2.99E-05	0.479393086	\\
4.31E-05	0.489176618	\\
6.18E-05	0.49896015	\\
8.79E-05	0.508743683	\\
0.000124079	0.518527215	\\
0.00017393	0.528310747	\\
0.000242059	0.53809428	\\
0.000334456	0.547877812	\\
0.000458816	0.557661344	\\
0.000624922	0.567444877	\\
0.000845103	0.577228409	\\
0.001134741	0.587011942	\\
0.001512854	0.596795474	\\
0.002002713	0.606579006	\\
0.002632521	0.616362539	\\
0.003436105	0.626146071	\\
0.004453618	0.635929603	\\
0.00573222	0.645713136	\\
0.007326691	0.655496668	\\
0.009299949	0.6652802	\\
0.011723419	0.675063733	\\
0.014677194	0.684847265	\\
0.01824995	0.694630797	\\
0.022538548	0.70441433	\\
0.027647286	0.714197862	\\
0.033686745	0.723981395	\\
0.040772213	0.733764927	\\
0.049021669	0.743548459	\\
0.058553319	0.753331992	\\
0.06948274	0.763115524	\\
0.081919644	0.772899056	\\
0.095964382	0.782682589	\\
0.11170425	0.792466121	\\
0.129209759	0.802249653	\\
0.148530979	0.812033186	\\
0.169694137	0.821816718	\\
0.192698627	0.83160025	\\
0.217514574	0.841383783	\\
0.244081132	0.851167315	\\
0.272305607	0.860950848	\\
0.302063529	0.87073438	\\
0.333199722	0.880517912	\\
0.365530381	0.890301445	\\
0.398846144	0.900084977	\\
0.432916067	0.909868509	\\
0.467492379	0.919652042	\\
0.502315875	0.929435574	\\
0.537121722	0.939219106	\\
0.571645491	0.949002639	\\
0.605629169	0.958786171	\\
0.638826922	0.968569703	\\
0.671010414	0.978353236	\\
0.70197347	0.988136768	\\
0.731535932	0.997920301	\\
0.759546605	1.007703833	\\
0.785885204	1.017487365	\\
0.810463282	1.027270898	\\
0.833224178	1.03705443	\\
0.854142023	1.046837962	\\
0.873219928	1.056621495	\\
0.890487472	1.066405027	\\
0.905997644	1.076188559	\\
0.919823399	1.085972092	\\
0.932053983	1.095755624	\\
0.9427912	1.105539156	\\
0.952145739	1.115322689	\\
0.960233704	1.125106221	\\
0.967173428	1.134889754	\\
0.973082665	1.144673286	\\
0.978076184	1.154456818	\\
0.982263812	1.164240351	\\
0.985748916	1.174023883	\\
0.988627305	1.183807415	\\
0.990986531	1.193590948	\\
0.992905533	1.20337448	\\
0.994454592	1.213158012	\\
0.995695518	1.222941545	\\
0.996682049	1.232725077	\\
0.997460375	1.24250861	\\
0.998069769	1.252292142	\\
0.99854327	1.262075674	\\
0.998908386	1.271859207	\\
0.999187785	1.281642739	\\
0.999399966	1.291426271	\\
0.999559876	1.301209804	\\
0.999679475	1.310993336	\\
0.999768246	1.320776868	\\
0.999833633	1.330560401	\\
0.999881431	1.340343933	\\
0.999916105	1.350127465	\\
0.999941067	1.359910998	\\
0.999958902	1.36969453	\\
0.999971547	1.379478063	\\
0.999980444	1.389261595	\\
0.999986657	1.399045127	\\
0.999990962	1.40882866	\\
0.999993923	1.418612192	\\
0.999995944	1.428395724	\\
0.999997312	1.438179257	\\
0.999998232	1.447962789	\\
0.999998846	1.457746321	\\
0.999999252	1.467529854	\\
0.999999519	1.477313386	\\
0.999999693	1.487096918	\\
0.999999805	1.496880451	\\
0.999999877	1.506663983	\\
0.999999923	1.516447516	\\
0.999999952	1.526231048	\\
0.999999971	1.53601458	\\
0.999999982	1.545798113	\\
0.999999989	1.555581645	\\
0.999999993	1.565365177	\\
0.999999996	1.57514871	\\
0.999999998	1.584932242	\\
0.999999999	1.594715774	\\
0.999999999	1.604499307	\\
1	1.614282839	\\
1	1.624066371	\\
1	2\\
};

\addplot [fill=teal,draw=teal,mark=diamond*,only marks,fill opacity=1,draw opacity=0.5, mark size=3.5,clip mode=individual]
table[row sep=crcr]{%
0.5 0.93	\\
};

\nextgroupplot[height = 4.6cm,
     width = 4.6cm,
         xlabel={$\overline{c}_\text{ent}$},
     grid=both,
     xmin=0,
     xmax=1,
       xtick={0, 0.25, 0.5, 0.75,1},
     ymin=0,
     ymax=2,
    legend columns=5,
   legend style={cells={align=left},anchor = south west,at={(0,1.2)},font=\normalsize},
   clip mode = individual,
   ]
   
\node[align=center] at (rel axis cs:0.1,0.9) {(\textit{c})};

\addplot[color=black,mark=none,line width = 2,draw opacity=1] 
table[row sep=crcr]{%
1.67E-12	0	\\
2.97E-12	0.009783532	\\
5.27E-12	0.019567065	\\
9.27E-12	0.029350597	\\
1.62E-11	0.039134129	\\
2.81E-11	0.048917662	\\
4.85E-11	0.058701194	\\
8.32E-11	0.068484727	\\
1.42E-10	0.078268259	\\
2.40E-10	0.088051791	\\
4.03E-10	0.097835324	\\
6.73E-10	0.107618856	\\
1.12E-09	0.117402388	\\
1.84E-09	0.127185921	\\
3.01E-09	0.136969453	\\
4.90E-09	0.146752985	\\
7.91E-09	0.156536518	\\
1.27E-08	0.16632005	\\
2.03E-08	0.176103582	\\
3.21E-08	0.185887115	\\
5.05E-08	0.195670647	\\
7.90E-08	0.20545418	\\
1.23E-07	0.215237712	\\
1.90E-07	0.225021244	\\
2.91E-07	0.234804777	\\
4.43E-07	0.244588309	\\
6.71E-07	0.254371841	\\
1.01E-06	0.264155374	\\
1.51E-06	0.273938906	\\
2.24E-06	0.283722438	\\
3.30E-06	0.293505971	\\
4.84E-06	0.303289503	\\
7.04E-06	0.313073035	\\
1.02E-05	0.322856568	\\
1.46E-05	0.3326401	\\
2.09E-05	0.342423633	\\
2.97E-05	0.352207165	\\
4.19E-05	0.361990697	\\
5.86E-05	0.37177423	\\
8.16E-05	0.381557762	\\
0.000112828	0.391341294	\\
0.000155019	0.401124827	\\
0.00021162	0.410908359	\\
0.000287034	0.420691891	\\
0.000386827	0.430475424	\\
0.000517975	0.440258956	\\
0.000689143	0.450042488	\\
0.000911004	0.459826021	\\
0.001196581	0.469609553	\\
0.001561618	0.479393086	\\
0.002024973	0.489176618	\\
0.002609009	0.49896015	\\
0.003339981	0.508743683	\\
0.004248389	0.518527215	\\
0.005369282	0.528310747	\\
0.006742472	0.53809428	\\
0.008412645	0.547877812	\\
0.010429306	0.557661344	\\
0.012846553	0.567444877	\\
0.015722613	0.577228409	\\
0.019119131	0.587011942	\\
0.023100162	0.596795474	\\
0.027730853	0.606579006	\\
0.033075805	0.616362539	\\
0.039197103	0.626146071	\\
0.046152054	0.635929603	\\
0.05399064	0.645713136	\\
0.062752775	0.655496668	\\
0.072465418	0.6652802	\\
0.083139652	0.675063733	\\
0.094767851	0.684847265	\\
0.10732106	0.694630797	\\
0.120746731	0.70441433	\\
0.134966977	0.714197862	\\
0.149877469	0.723981395	\\
0.165347123	0.733764927	\\
0.181218681	0.743548459	\\
0.197310276	0.753331992	\\
0.213418006	0.763115524	\\
0.229319539	0.772899056	\\
0.24477868	0.782682589	\\
0.259550798	0.792466121	\\
0.273388976	0.802249653	\\
0.28605066	0.812033186	\\
0.297304581	0.821816718	\\
0.306937683	0.83160025	\\
0.314761762	0.841383783	\\
0.320619542	0.851167315	\\
0.324389907	0.860950848	\\
0.32599205	0.87073438	\\
0.325388344	0.880517912	\\
0.322585791	0.890301445	\\
0.317635971	0.900084977	\\
0.310633481	0.909868509	\\
0.301712941	0.919652042	\\
0.291044685	0.929435574	\\
0.278829334	0.939219106	\\
0.265291504	0.949002639	\\
0.250672906	0.958786171	\\
0.235225151	0.968569703	\\
0.21920256	0.978353236	\\
0.202855252	0.988136768	\\
0.186422789	0.997920301	\\
0.170128579	1.007703833	\\
0.154175222	1.017487365	\\
0.138740904	1.027270898	\\
0.123976902	1.03705443	\\
0.110006197	1.046837962	\\
0.096923149	1.056621495	\\
0.084794148	1.066405027	\\
0.073659088	1.076188559	\\
0.063533546	1.085972092	\\
0.054411473	1.095755624	\\
0.046268233	1.105539156	\\
0.039063842	1.115322689	\\
0.032746225	1.125106221	\\
0.027254395	1.134889754	\\
0.02252142	1.144673286	\\
0.018477103	1.154456818	\\
0.015050329	1.164240351	\\
0.01217103	1.174023883	\\
0.00977177	1.183807415	\\
0.007788954	1.193590948	\\
0.006163687	1.20337448	\\
0.004842316	1.213158012	\\
0.003776702	1.222941545	\\
0.002924254	1.232725077	\\
0.002247792	1.24250861	\\
0.001715268	1.252292142	\\
0.001299389	1.262075674	\\
0.000977177	1.271859207	\\
0.000729509	1.281642739	\\
0.000540641	1.291426271	\\
0.000397743	1.301209804	\\
0.000290476	1.310993336	\\
0.000210585	1.320776868	\\
0.000151548	1.330560401	\\
0.000108263	1.340343933	\\
7.68E-05	1.350127465	\\
5.40E-05	1.359910998	\\
3.78E-05	1.36969453	\\
2.62E-05	1.379478063	\\
1.80E-05	1.389261595	\\
1.23E-05	1.399045127	\\
8.36E-06	1.40882866	\\
5.63E-06	1.418612192	\\
3.76E-06	1.428395724	\\
2.50E-06	1.438179257	\\
1.64E-06	1.447962789	\\
1.07E-06	1.457746321	\\
6.97E-07	1.467529854	\\
4.49E-07	1.477313386	\\
2.87E-07	1.487096918	\\
1.82E-07	1.496880451	\\
1.15E-07	1.506663983	\\
7.17E-08	1.516447516	\\
4.45E-08	1.526231048	\\
2.74E-08	1.53601458	\\
1.67E-08	1.545798113	\\
1.02E-08	1.555581645	\\
6.11E-09	1.565365177	\\
3.65E-09	1.57514871	\\
2.17E-09	1.584932242	\\
1.28E-09	1.594715774	\\
7.46E-10	1.604499307	\\
4.32E-10	1.614282839	\\
2.49E-10	1.624066371	\\
0	2	\\
};

\addplot [fill=black,draw=black,mark=square*,only marks,fill opacity=1,draw opacity=0.5, mark size=2.5,clip mode=individual]
table[row sep=crcr]{%
0.32 0.87	\\
};

\end{groupplot}

\end{tikzpicture}

%% file: Figures/fig3.tex
\pgfplotstableread{
series1 series2 series3 series4
0.100000000000000	0	0	0
0.110000000000000	0.0983606557377049	0	0.0983606557377049
0.110000000000000	0.196721311475410	0	0.196721311475410
0.120000000000000	0.295081967213115	0	0.295081967213115
0.180000000000000	0.393442622950820	0	0.393442622950820
0.190000000000000	0.491803278688525	0	0.491803278688525
0.300000000000000	0.590163934426229	0.0100000000000000	0.590163934426229
0.470000000000000	0.688524590163934	0.0300000000000000	0.688524590163934
0.650000000000000	0.786885245901639	0.130000000000000	0.786885245901639
0.780000000000000	0.885245901639344	0.290000000000000	0.885245901639344
0.890000000000000	0.983606557377049	0.470000000000000	0.983606557377049
0.950000000000000	1.08196721311475	0.660000000000000	1.08196721311475
0.970000000000000	1.18032786885246	0.810000000000000	1.18032786885246
0.980000000000000	1.27868852459016	0.880000000000000	1.27868852459016
0.990000000000000	1.37704918032787	0.940000000000000	1.37704918032787
1	1.47540983606557	0.970000000000000	1.47540983606557
1	1.57377049180328	0.990000000000000	1.57377049180328
1	1.67213114754098	1	1.67213114754098
}\dataone

\pgfplotstableread{
series1 series2 series3 series4 series5 series6 series7 series8
0.00124449286275530	0	5.37258511401628e-07	0	0.00124395560424390	0	0	1.62828508886210e-14
0.00155701356084331	0.0159469286215475	7.94801251646771e-07	0.0159469286215475	0.00155621875959167	0.0159469286215475	0.0328669496721849	0.0328669496704396
0.00193956939543649	0.0318938572430950	1.16894446017257e-06	0.0318938572430950	0.00193840045097632	0.0318938572430950	0.0432749772389666	0.0432749772295892
0.00240566851513857	0.0478407858646425	1.70919571584882e-06	0.0478407858646425	0.00240395931942272	0.0478407858646425	0.0510072406591362	0.0510072406162666
0.00297090364239061	0.0637877144861899	2.48459084467578e-06	0.0637877144861899	0.00296841905154593	0.0637877144861899	0.0574628140453403	0.0574628138667895
0.00365315972045327	0.0797346431077374	3.59075534533959e-06	0.0797346431077374	0.00364956896510793	0.0797346431077374	0.0631564871198889	0.0631564864278622
0.00447281849337300	0.0956815717292849	5.15925970717701e-06	0.0956815717292849	0.00446765923366582	0.0956815717292849	0.0683437060182891	0.0683437034989121
0.00545295412643032	0.111628500350832	7.36991541699927e-06	0.111628500350832	0.00544558421101332	0.111628500350832	0.0731728621345666	0.0731728534767149
0.00661951303312480	0.127575428972380	1.04668059593971e-05	0.127575428972380	0.00660904622716541	0.127575428972380	0.0777393043242309	0.0777392761582452
0.00800147018618541	0.143522357593927	1.47790166053485e-05	0.143522357593927	0.00798669116958006	0.143522357593927	0.0821088665643886	0.0821087796642584
0.00963095340127090	0.159469286215475	2.07472176664836e-05	0.159469286215475	0.00961020618360442	0.159469286215475	0.0863295872932015	0.0863293327238142
0.0115433264445321	0.175416214837022	2.89574659444458e-05	0.175416214837022	0.0115143689785877	0.175416214837022	0.0904381348311028	0.0904374262015567
0.0137772213836196	0.191363143458570	4.01838139864608e-05	0.191363143458570	0.0137370375696331	0.191363143458570	0.0944635974201731	0.0944617220811533
0.0163745104314184	0.207310072080117	5.54415494622429e-05	0.207310072080117	0.0163190688819561	0.207310072080117	0.0984298536083561	0.0984251338144958
0.0193802076763009	0.223257000701665	7.60531172858125e-05	0.223257000701665	0.0193041545590151	0.223257000701665	0.102357129540721	0.102345831115872
0.0228422916006508	0.239203929323212	0.000103728991109775	0.239203929323212	0.0227385626095410	0.239203929323212	0.106263067577392	0.106237341296627
0.0268114402012922	0.255150857944760	0.000140665940480289	0.255150857944760	0.0266707742608119	0.255150857944760	0.110163490027748	0.110107776518780
0.0313406718701449	0.271097786566307	0.000189665262897099	0.271097786566307	0.0311510066072478	0.271097786566307	0.114072967408743	0.113958237980315
0.0364848869849086	0.287044715187855	0.000254273589578302	0.287044715187855	0.0362306133953303	0.287044715187855	0.118005259349954	0.117780682028859
0.0423003073939163	0.302991643809402	0.000338948799222860	0.302991643809402	0.0419613585946935	0.302991643809402	0.121973672459566	0.121556028825424
0.0488438136320971	0.318938572430950	0.000449253351653800	0.318938572430950	0.0483945602804433	0.318938572430950	0.125991365352792	0.125254020729743
0.0561721827296490	0.334885501052497	0.000592076947172782	0.334885501052497	0.0555801057824762	0.334885501052497	0.130071622569952	0.128837090319430
0.0643412328019203	0.350832429674045	0.000775889791994722	0.350832429674045	0.0635653430099256	0.350832429674045	0.134228114085750	0.132270843694244
0.0734048841457986	0.366779358295592	0.00101102687191013	0.366779358295592	0.0723938572738885	0.366779358295592	0.138475154329239	0.135543110482763
0.0834141502012218	0.382726286917140	0.00131000247746027	0.382726286917140	0.0821041477237615	0.382726286917140	0.142827973440822	0.138691361428242
0.0944160753347149	0.398673215538687	0.00168785276555100	0.398673215538687	0.0927282225691639	0.398673215538687	0.147303013564054	0.141834657745099
0.106452639819772	0.414620144160235	0.00216250237866145	0.414620144160235	0.104290137441110	0.414620144160235	0.151915763526788	0.145199930795404
0.119559655472802	0.430567072781782	0.00275514908477831	0.430567072781782	0.116804506388023	0.430567072781782	0.156672960379669	0.149129913982862
0.133765677997947	0.446514001403330	0.00349065808108473	0.446514001403330	0.130275019916862	0.446514001403330	0.161579127434007	0.154079685199567
0.149090964049780	0.462460930024877	0.00439795507896057	0.462460930024877	0.144693008970819	0.462460930024877	0.166638929647257	0.160577747194012
0.165546502203715	0.478407858646425	0.00551040464086472	0.478407858646425	0.160036097562850	0.478407858646425	0.171857178058625	0.169156491217780
0.183133147315407	0.494354787267972	0.00686615758342019	0.494354787267972	0.176266989731987	0.494354787267972	0.177238834363577	0.180267305907327
0.201840887066810	0.510301715889520	0.00850844873568019	0.510301715889520	0.193332438331130	0.510301715889520	0.182789015631593	0.194201064682008
0.221648267787815	0.526248644511067	0.0104858241124930	0.526248644511067	0.211162443675322	0.526248644511067	0.188512999171658	0.211035823696631
0.242522003898937	0.542195573132615	0.0128522748155495	0.542195573132615	0.229669729083387	0.542195573132615	0.194416227550117	0.230626732980047
0.264416791577008	0.558142501754162	0.0156672539067659	0.558142501754162	0.248749537670242	0.558142501754162	0.200504313765656	0.252640549637265
0.287275342582331	0.574089430375710	0.0189955523096229	0.574089430375710	0.268279790272708	0.574089430375710	0.206783046586341	0.276623625121635
0.311028648726812	0.590036358997257	0.0229070106724517	0.590036358997257	0.288121638054361	0.590036358997257	0.213258396053786	0.302083141995529
0.335596481374038	0.605983287618805	0.0274760462363301	0.605983287618805	0.308120435137708	0.605983287618805	0.219936519159679	0.328559911447993
0.360888123845236	0.621930216240352	0.0327809772107860	0.621930216240352	0.328107146634450	0.621930216240352	0.226823765700065	0.355676935805083
0.386803327888591	0.637877144861900	0.0389031320374094	0.637877144861900	0.347900195851182	0.637877144861900	0.233926684312963	0.383157915588813
0.413233478700139	0.653824073483447	0.0459257372076273	0.653824073483447	0.367307741492511	0.653824073483447	0.241252028705037	0.410819566544368
0.440062946616563	0.669771002104995	0.0539325849055467	0.669771002104995	0.386130361711016	0.669771002104995	0.248806764073262	0.438547657211092
0.467170597782642	0.685717930726542	0.0630064904874099	0.685717930726542	0.404164107295232	0.685717930726542	0.256598073727682	0.466267958756524
0.494431431061500	0.701664859348089	0.0732275594087141	0.701664859348089	0.421203871652786	0.701664859348089	0.264633365921553	0.493920812093385
0.521718304408778	0.717611787969637	0.0846712932999627	0.717611787969637	0.437047011108815	0.717611787969637	0.272920280895372	0.521443844461605
0.548903711038932	0.733558716591184	0.0974065750229429	0.733558716591184	0.451497136015989	0.733558716591184	0.281466698141493	0.548763482653700
0.575861564093109	0.749505645212732	0.111493582198429	0.749505645212732	0.464367981894681	0.749505645212732	0.290280743896221	0.575793439866956
0.602468948241082	0.765452573834279	0.126981687331184	0.765452573834279	0.475487260909898	0.765452573834279	0.299370798866531	0.602437472252727
0.628607797726032	0.781399502455827	0.143907409708009	0.781399502455827	0.484700388018023	0.781399502455827	0.308745506198738	0.628593963638911
0.654166462745699	0.797346431077374	0.162292489173780	0.797346431077374	0.491873973571919	0.797346431077374	0.318413779696707	0.654160678062996
0.679041129657177	0.813293359698922	0.182142154225745	0.813293359698922	0.496898975431432	0.813293359698922	0.328384812297416	0.679038828117731
0.703137065147108	0.829240288320469	0.203443656233127	0.829240288320469	0.499693408913980	0.829240288320469	0.338668084811922	0.703136193763680
0.726369660033844	0.845187216942017	0.226165137744872	0.845187216942017	0.500204522288972	0.845187216942017	0.349273374940056	0.726369346065010
0.748665254540996	0.861134145563564	0.250254895709875	0.861134145563564	0.498410358831121	0.861134145563564	0.360210766567403	0.748665146875096
0.769961733458683	0.877081074185112	0.275641090096401	0.877081074185112	0.494320643362282	0.877081074185112	0.371490659353422	0.769961698318270
0.790208886336506	0.893028002806659	0.302231935142105	0.893028002806659	0.487976951194400	0.893028002806659	0.383123778619806	0.790208875419828
0.809368534479568	0.908974931428207	0.329916394757262	0.908974931428207	0.479452139722306	0.908974931428207	0.395121185548499	0.809368531251501
0.827414432808277	0.924921860049754	0.358565386071386	0.924921860049754	0.468849046736892	0.924921860049754	0.407494287699064	0.827414431899678
0.844331960380687	0.940868788671302	0.388033476524486	0.940868788671302	0.456298483856200	0.940868788671302	0.420254849855388	0.844331960137251
0.860117618381647	0.956815717292849	0.418161041122770	0.956815717292849	0.441956577258877	0.956815717292849	0.433415005212061	0.860117618319563
0.874778358514150	0.972762645914397	0.448776828417544	0.972762645914397	0.426001530096605	0.972762645914397	0.446987266911043	0.874778358499078
0.888330767886469	0.988709574535944	0.479700867333092	0.988709574535944	0.408629900553377	0.988709574535944	0.460984539939607	0.888330767882986
0.900800138620909	1.00465650315749	0.510747633011436	1.00465650315749	0.390052505609473	1.00465650315749	0.475420133400853	0.900800138620143
0.912219451507989	1.02060343177904	0.541729379092411	1.02060343177904	0.370490072415579	1.02060343177904	0.490307773168481	0.912219451507829
0.922628303127792	1.03655036040059	0.572459536878679	1.03655036040059	0.350168766249113	1.03655036040059	0.505661614937831	0.922628303127760
0.932071805029898	1.05249728902213	0.602756079021749	1.05249728902213	0.329315726008149	1.05249728902213	0.521496257685625	0.932071805029892
0.940599481908409	1.06844421764368	0.632444746858422	1.06844421764368	0.308154735049987	1.06844421764368	0.537826757551189	0.940599481908408
0.948264193357143	1.08439114626523	0.661362046246854	1.08439114626523	0.286902147110289	1.08439114626523	0.554668642152366	0.948264193357143
0.955121100887236	1.10033807488678	0.689357926389685	1.10033807488678	0.265763174497550	1.10033807488678	0.572037925349720	0.955121100887236
0.961226698588939	1.11628500350832	0.716298069173491	1.11628500350832	0.244928629415449	1.11628500350832	0.589951122473088	0.961226698588939
0.966637922276542	1.13223193212987	0.742065732310957	1.13223193212987	0.224572189965585	1.13223193212987	0.608425266024938	0.966637922276542
0.971411348319529	1.14817886075142	0.766563107227099	1.14817886075142	0.204848241092430	1.14817886075142	0.627477921875473	0.971411348319529
0.975602489772130	1.16412578937297	0.789712171288623	1.16412578937297	0.185890318483507	1.16412578937297	0.647127205964889	0.975602489772130
0.979265193988461	1.18007271799451	0.811455032719067	1.18007271799451	0.167810161269394	1.18007271799451	0.667391801528648	0.979265193988461
0.982451142752774	1.19601964661606	0.831753784487044	1.19601964661606	0.150697358265729	1.19601964661606	0.688290976862160	0.982451142752774
0.985209453142888	1.21196657523761	0.850589899799148	1.21196657523761	0.134619553343740	1.21196657523761	0.709844603641766	0.985209453142888
0.987586374935632	1.22791350385916	0.867963215897140	1.22791350385916	0.119623159038493	1.22791350385916	0.732073175819424	0.987586374935632
0.989625078389374	1.24386043248070	0.883890564133348	1.24386043248070	0.105734514256025	1.24386043248070	0.754997829109092	0.989625078389374
0.991365524712463	1.25980736110225	0.898404112440089	1.25980736110225	0.0929614122723740	1.25980736110225	0.778640361083311	0.991365524712463
0.992844410440687	1.27575428972380	0.911549491166227	1.27575428972380	0.0812949192744605	1.27575428972380	0.803023251899105	0.992844410440687
0.994095176278349	1.29170121834535	0.923383774856598	1.29170121834535	0.0707114014217518	1.29170121834535	0.828169685672920	0.994095176278349
0.995148070670314	1.30764814696689	0.933973391094019	1.30764814696689	0.0611746795762944	1.30764814696689	0.854103572524893	0.995148070670314
0.996030258420435	1.32359507558844	0.943392023346216	1.32359507558844	0.0526382350742187	1.32359507558844	0.880849571313448	0.996030258420435
0.996765965003455	1.33954200420999	0.951718568307983	1.33954200420999	0.0450473966954713	1.33954200420999	0.908433113081812	0.996765965003455
0.997376647777604	1.35548893283154	0.959035200022502	1.35548893283154	0.0383414477551024	1.35548893283154	0.936880425238748	0.997376647777604
0.997881186038682	1.37143586145308	0.965425583662034	1.37143586145308	0.0324556023766478	1.37143586145308	0.966218556496508	0.997881186038682
0.998296082710207	1.38738279007463	0.970973271805561	1.38738279007463	0.0273228109046462	1.38738279007463	0.996475402589705	0.998296082710207
0.998635671389423	1.40332971869618	0.975760305897143	1.40332971869618	0.0228753654922802	1.40332971869618	1.02767973279957	0.998635671389423
0.998912323421941	1.41927664731773	0.979866035775069	1.41927664731773	0.0190462876468712	1.41927664731773	1.05986121730879	0.998912323421941
0.999136650621570	1.43522357593927	0.983366161122975	1.43522357593927	0.0157704894985953	1.43522357593927	1.09305045541299	0.999136650621570
0.999317700156141	1.45117050456082	0.986331990715975	1.45117050456082	0.0129857094401666	1.45117050456082	1.12727900461561	0.999317700156141
0.999463138961503	1.46711743318237	0.988829908629090	1.46711743318237	0.0106332303324130	1.46711743318237	1.16257941063392	0.999463138961503
0.999579425807766	1.48306436180392	0.990921031260138	1.48306436180392	0.00865839454762729	1.48306436180392	1.19898523834460	0.999579425807766
0.999671969813589	1.49901129042546	0.992661035126598	1.49901129042546	0.00701093468699077	1.49901129042546	1.23653110369844	0.999671969813589
0.999745274780813	1.51495821904701	0.994100132881809	1.51495821904701	0.00564514189900378	1.51495821904701	1.27525270663445	0.999745274780813
0.999803069202354	1.53090514766856	0.995283173754561	1.53090514766856	0.00451989544779352	1.53090514766856	1.31518686502453	0.999803069202354
0.999848422184221	1.54685207629011	0.996249844495820	1.54685207629011	0.00359857768840044	1.54685207629011	1.35637154968132	0.999848422184221
0.999883845823675	1.56279900491165	0.997034947734772	1.56279900491165	0.00284889808890254	1.56279900491165	1.39884592046209	0.999883845823675
0.999911384808009	1.57874593353320	0.997668736205938	1.57874593353320	0.00224264860207102	1.57874593353320	1.44265036350341	0.999911384808009
0.999932694151329	1.59469286215475	0.998177283410433	1.59469286215475	0.00175541074089558	1.59469286215475	1.48782652962166	0.999932694151329
0.999949106080016	1.61063979077630	0.998582873727237	1.61063979077630	0.00136623235277866	1.61063979077630	1.53441737391611	0.999949106080016
0.999961687121175	1.62658671939784	0.998904397622503	1.62658671939784	0.00105728949867279	1.62658671939784	1.58246719661218	0.999961687121175
0.999971286451846	1.64253364801939	0.999157740268026	1.64253364801939	0.000813546183820013	1.64253364801939	1.63202168518363	0.999971286451846
0.999978576539132	1.65848057664094	0.999356154453146	1.65848057664094	0.000622422085986174	1.65848057664094	1.68312795779384	0.999978576539132
0.999984087050608	1.67442750526249	0.999510611064871	1.67442750526249	0.000473475985736771	1.67442750526249	1.73583460809746	0.999984087050608
0.999988232947537	1.69037443388403	0.999630122554279	1.69037443388403	0.000358110393257283	1.69037443388403	1.79019175144491	0.999988232947537
0.999991337596463	1.70632136250558	0.999722036663611	1.70632136250558	0.000269300932852268	1.70632136250558	1.84625107253390	0.999991337596463
0.999993651652653	1.72226829112713	0.999792299240957	1.72226829112713	0.000201352411696276	1.72226829112713	1.90406587455304	0.999993651652653
0.999995368385472	1.73821521974868	0.999845686218986	1.73821521974868	0.000149682166486365	1.73821521974868	1.96369112986438	0.999995368385472
0.999996636034201	1.75416214837022	0.999886005796125	1.75416214837022	0.000110630238076226	1.75416214837022	2.02518353227313	0.999996636034201
0.999997567705137	1.77010907699177	0.999916272558517	1.77010907699177	8.12951466206968e-05	1.77010907699177	2.08860155093409	0.999997567705137
0.999998249248630	1.78605600561332	0.999938855750888	1.78605600561332	5.93934977423105e-05	1.78605600561332	2.15400548594623	0.999998249248630
0.999998745488907	1.80200293423487	0.999955604179577	1.80200293423487	4.31413093299193e-05	1.80200293423487	2.22145752568815	0.999998745488907
0.999999105120511	1.81794986285641	0.999967950347239	1.81794986285641	3.11547732713136e-05	1.81794986285641	2.29102180594896	0.999999105120511
0.999999364533128	1.83389679147796	0.999976996410900	1.83389679147796	2.23681222278893e-05	1.83389679147796	2.36276447091092	0.999999364533128
0.999999550781184	1.84984372009951	0.999983584454673	1.84984372009951	1.59663265106857e-05	1.84984372009951	2.43675373604162	0.999999550781184
0.999999683875591	1.86579064872106	0.999988353403686	1.86579064872106	1.13304719043805e-05	1.86579064872106	2.51305995295570	0.999999683875591
0.999999778541861	1.88173757734260	0.999991784700199	1.88173757734260	7.99384166194628e-06	1.88173757734260	2.59175567630764	0.999999778541861
0.999999845560891	1.89768450596415	0.999994238635933	1.89768450596415	5.60692495721327e-06	1.89768450596415	2.67291573277929	0.999999845560891
0.999999892785504	1.91363143458570	0.999995983001426	1.91363143458570	3.90978407749376e-06	1.91363143458570	2.75661729222774	0.999999892785504
0.999999925906705	1.92957836320725	0.999997215485042	1.92957836320725	2.71042166377633e-06	1.92957836320725	2.84293994106116	0.999999925906705
0.999999949027948	1.94552529182879	0.999998081039081	1.94552529182879	1.86798886647743e-06	1.94552529182879	2.93196575791236	0.999999949027948
0.999999965093056	1.96147222045034	0.999998685233329	1.96147222045034	1.27985972719902e-06	1.96147222045034	3.02377939168208	0.999999965093056
0.999999976203299	1.97741914907189	0.999999104440158	1.97741914907189	8.71763140564141e-07	1.97741914907189	3.11846814202615	0.999999976203299
0.999999983850999	1.99336607769344	0.999999393541081	1.99336607769344	5.90309918391263e-07	1.99336607769344	3.21612204236310	0.999999983850999
0.999999989090688	2.00931300631498	0.999999591712021	2.00931300631498	3.97378667749138e-07	2.00931300631498	3.31683394548113	0.999999989090688
0.999999992663809	2.02525993493653	0.999999726732556	2.02525993493653	2.65931253018437e-07	2.02525993493653	3.42069961182583	0.999999992663809
0.999999995089062	2.04120686355808	0.999999818171046	2.04120686355808	1.76918016681071e-07	2.04120686355808	3.52781780055257	0.999999995089062
0.999999996727516	2.05715379217963	0.999999879720985	2.05715379217963	1.17006530797603e-07	2.05715379217963	3.63829036343027	0.999999996727516
0.999999997829254	2.07310072080117	0.999999920901860	2.07310072080117	7.69273935663506e-08	2.07310072080117	3.75222234168561	0.999999997829254
0.999999998566632	2.08904764942272	0.999999948288130	2.08904764942272	5.02785014733220e-08	2.08904764942272	3.86972206588000	0.999999998566632
0.999999999057844	2.10499457804427	0.999999966390672	2.10499457804427	3.26671721895266e-08	2.10499457804427	3.99090125891406	0.999999999057844
0.999999999383543	2.12094150666582	0.999999978284334	2.12094150666582	2.10992088245376e-08	2.12094150666582	4.11587514225775	0.999999999383543
0.999999999598489	2.13688843528736	0.999999986051471	2.13688843528736	1.35470175077757e-08	2.13688843528736	4.24476254550704	0.999999999598489
0.999999999739681	2.15283536390891	0.999999991093157	2.15283536390891	8.64652460741411e-09	2.15283536390891	4.37768601937125	0.999999999739681
}\datatwo

\begin{tikzpicture}
\definecolor{color1}{rgb}{1,0.7,0.1}
\definecolor{color2}{rgb}{0.15,0.45,0.55}

 \begin{groupplot}[
     group style = {group size = 2 by 2,horizontal sep=1.4cm,vertical sep=1.4cm},
     width = 1\textwidth]

\nextgroupplot[height = 5.5cm,
     width = 5.5cm,
          xlabel={$\overline{c}$},
         ylabel={$y/y_{90}$},
         xtick={0, 0.25, 0.5 , 0.75, 1},
         ytick={0, 0.5, 1,  1.5,2 },
       %yticklabels={0, 0.5, \phantom{0 , 1,  1.5,2 },
     grid=both,
     xmin=0,
     xmax=1,
     ymin=0,
     ymax=2,
    legend columns=1,
   legend style={cells={align=left},anchor = north east,at={(-0.4,1)},font=\small},
   clip mode = individual,
   ]

\addplot [fill=black,mark=*,only marks,fill opacity=0,draw opacity=1, mark size=2.5,clip mode=individual]
table[row sep=crcr]{%
0 -1	\\
}; \addlegendentry{$\overline{c}_\text{meas} \, \,$};

\addplot [fill=black,mark=diamond,only marks,fill opacity=0,draw opacity=1, mark size=3,clip mode=individual]
table[row sep=crcr]{%
0 -1	\\
}; \addlegendentry{$\overline{c}_\text{trap,meas} \, \,$};

\addplot [fill=blue,draw=blue,mark=*,only marks,fill opacity=1,draw opacity=0.5, mark size=3,clip mode=individual]
table[row sep=crcr]{%
0 -1	\\
}; \addlegendentry{$y_{50}$};

\addplot [fill=blue,draw=blue,mark=pentagon*
,only marks,fill opacity=1,draw opacity=1, mark size=2.5,clip mode=individual,line width =1.3]
table[row sep=crcr]{%
0 -1	\\
}; \addlegendentry{$y_{90}$};

\addplot [fill=teal,draw=teal,mark=diamond*,only marks,fill opacity=1,draw opacity=0.5, mark size=3.5,clip mode=individual]
table[row sep=crcr]{%
0 -1	\\
}; \addlegendentry{$y_{50_\text{trap}}$};

%\addplot [fill=black,draw=black,mark=square*,only marks,fill opacity=1,draw opacity=0.5, mark size=2.5,clip mode=individual]
%table[row sep=crcr]{%
%0 -1	\\
%}; \addlegendentry{$\overline{c}_\text{max}, y_{\overline{c}_\text{max}}$};

\addplot [fill=orange,draw=orange,mark=*,only marks,fill opacity=1,draw opacity=0.5, mark size=2.5,clip mode=individual]
table[row sep=crcr]{%
0 -1	\\
}; \addlegendentry{$y_\star$};

\addplot[color=violet,mark=none,line width = 2,draw opacity=0.4] table[row sep=crcr]{%
0 -1	\\
}; \addlegendentry{Eq. (\ref{eq:voidfraction1}) \,};

\addplot[color=blue,mark=none,line width = 2,draw opacity=1] table[row sep=crcr]{%
0 -1	\\
}; \addlegendentry{Eq. (\ref{eq:superposition}) \,};

\addplot[color=teal,mark=none,line width = 2,draw opacity=1] 
table[row sep=crcr]{%
0 -1	\\
}; \addlegendentry{Eq. (\ref{entrappedair}) \,};

\addplot[color=black,mark=none,line width = 2,draw opacity=1] 
table[row sep=crcr]{%
0 -1	\\
}; \addlegendentry{Eq. (\ref{entrainedair}) \,};

\addplot[color=orange,mark=none,line width = 2,draw opacity=1] table[row sep=crcr]{%
0 -1	\\
}; \addlegendentry{Eq. (\ref{eq:voidfractionfinal1}) \,};

\addplot [fill=black,mark=*,only marks,fill opacity=0,draw opacity=1, mark size=2.5,clip mode=individual] table [y=series2,  x =series1] {\dataone};

\addplot [fill=black,mark=diamond,only marks,fill opacity=0,draw opacity=1, mark size=3,clip mode=individual] table [y=series4,  x =series3] {\dataone};

\addplot[color=black,mark=none,line width = 2,draw opacity=1] 
table [y=series6,  x =series5] {\datatwo};

\addplot[color=blue,mark=none,line width = 2,draw opacity=1] 
table [y=series2,  x =series1] {\datatwo};

\addplot[color=teal,mark=none,line width = 2,draw opacity=1] 
table [y=series4,  x =series3] {\datatwo};

\node[align=center,color=black] at (rel axis cs:0.1,0.9) {(\textit{a})};

\addplot [fill=blue,draw=blue,mark=*,only marks,fill opacity=1,draw opacity=0.5, mark size=3,clip mode=individual]
table[row sep=crcr]{%
0.5 0.7	\\
};

%\addplot [fill=black,draw=black,mark=square*,only marks,fill opacity=1,draw opacity=0.5, mark size=2.5,clip mode=individual]
%table[row sep=crcr]{%
%0.5 0.85\\
%};

\addplot [fill=teal,draw=teal,mark=diamond*,only marks,fill opacity=1,draw opacity=0.5, mark size=3.5,clip mode=individual]
table[row sep=crcr]{%
0.5 1	\\
};

\addplot [fill=blue,draw=blue,mark=pentagon*
,only marks,fill opacity=1,draw opacity=1, mark size=2.5,clip mode=individual,line width =1.3]
table[row sep=crcr]{%
0.9 1	\\
}; 

\nextgroupplot[height = 5.5cm,
     width = 5.5cm,
       xlabel={$\overline{c}$},
         ylabel={$y/y_{90}$},
         xtick={0, 0.25, 0.5 , 0.75, 1},
         ytick={0, 0.5, 1,  1.5,2 },
       %yticklabels={0, 0.5, \phantom{0 , 1,  1.5,2 },
     grid=both,
     xmin=0,
     xmax=1,
     ymin=0,
     ymax=2,
    legend columns=5,
   legend style={cells={align=left},anchor = south west,at={(0,1.2)},font=\normalsize},
   clip mode = individual,
   ]

\node[align=center,color=black] at (rel axis cs:0.1,0.9) {(\textit{b})};   

\addplot [fill=black,mark=*,only marks,fill opacity=0,draw opacity=1, mark size=2.5,clip mode=individual] table [y=series2,  x =series1] {\dataone};

\addplot[color=blue,mark=none,line width = 2,draw opacity=1] 
table [y=series2,  x =series1] {\datatwo};

\addplot[color=violet,mark=none,line width = 2,draw opacity=0.4] 
table [y=series2,  x =series7] {\datatwo};

\addplot[color=orange,mark=none,line width = 2,draw opacity=1] 
table [y=series2,  x =series8] {\datatwo};

\addplot [fill=orange,draw=orange,mark=*,only marks,fill opacity=1,draw opacity=0.5, mark size=2.5,clip mode=individual]
table[row sep=crcr]{%
0.19 0.4918\\
};

\end{groupplot}

\end{tikzpicture}

%% file: Figures/fig4.tex
\pgfplotstableread{
ycmax ycmaxest
0.0511000000000000	0.0535437328385900
0.0576600000000000	0.0601448533333333
0.0561400000000000	0.0579063054545455
0.0578400000000000	0.0606234444444444
0.0563100000000000	0.0582713684210526
0.0853200000000000	0.0855894109090909
0.0892300000000000	0.0899723478260870
0.0855500000000000	0.0859828545454545
0.0838700000000000	0.0860738352941177
0.0401700000000000	0.0415216275303644
0.0486200000000000	0.0480222857142857
0.0527200000000000	0.0534289473684211
0.0542200000000000	0.0533748421052632
0.0485900000000000	0.0493733714285714
0.0553200000000000	0.0570882666666667
0.0553200000000000	0.0570882666666667
0.0528400000000000	0.0554171076923077
0.0541300000000000	0.0551578151260504
0.0529400000000000	0.0535845000000000
0.0569000000000000	0.0584583214285714
}\ycmax

\pgfplotstableread{
H Hest
0.00592724115310302	0.00712123636363636
0.0112627700936106	0.0124339047619048
0.0116861208092958	0.0123135709090909
0.0113859187009452	0.0104856000000000
0.0150456299717998	0.0148609122807018
0.0103034956526955	0.0104968145454545
0.0122092850012479	0.0126488695652174
0.0150275629211449	0.0144948000000000
0.0169350213299083	0.0162938000000000
0.00647183663450477	0.00722235897435897
0.00961562020150922	0.00952312087912087
0.0146144518019717	0.0146723636363636
0.0125289496004090	0.0121556491228070
0.0173949912576180	0.0169375238095238
0.0214821337433925	0.0213432380952381
0.0214821337433925	0.0213432380952381
0.0289101145802956	0.0306344000000000
0.0136327594835517	0.0131912268907563
0.0168366096560235	0.0160625000000000
0.0167109746729279	0.0169375238095238
}\Hest

\pgfplotstableread{
cmax cmaxest
0.140189421665960	0.123352420007266
0.258160484851063	0.250397215149194
0.341064144442504	0.336490613220157
0.292702875575399	0.280776480413255
0.378623142143641	0.374899535432743
0.188869375775476	0.188804042519248
0.326048992493397	0.325434732426922
0.351056694923316	0.350918116079676
0.380929625434346	0.377210140539539
0.297504806751265	0.289259933929112
0.343046252716164	0.342460697459400
0.500293419244245	0.499717425095800
0.419958186111252	0.419170987699780
0.521944271708461	0.521427902173139
0.533256678270059	0.531097386295309
0.533256678270059	0.531097386295309
0.556174409748348	0.552988640289022
0.482887559650139	0.481425540343282
0.465155032802074	0.464829586403983
0.407894341214303	0.405986162521839
}\cmax

\pgfplotstableread{
Cmean H Htrap y50 y50trap
0.195912408759124	0.0911865988221989	0.0758048370049632	0.810882548108825	0.836585729182184
0.241914893617021	0.155404284206896	0.128987075286551	0.784869976359338	0.874893617021277
0.250294117647059	0.167173849984204	0.138698852930871	0.762352941176471	0.894385026737968
0.216935483870968	0.158792779952655	0.122531417906815	0.795698924731183	0.895260852250100
0.330142857142857	0.209083240297384	0.172531641287209	0.717293233082707	0.902255639097745
0.150404624277457	0.106215233743853	0.104794956929608	0.857109826589596	0.907514450867052
0.178189655172414	0.119449934881836	0.111965384833837	0.831709145427287	0.928785607196402
0.242121212121212	0.147659109786041	0.142805970503322	0.778787878787879	0.910927456382002
0.253461538461538	0.168961601615368	0.139210507265230	0.783589743589744	0.933936651583710
0.192535211267606	0.133004808348731	0.106953263893246	0.809317443120260	0.897331356560415
0.257363636363636	0.170067566351419	0.181831316632724	0.771428571428572	0.927272727272727
0.349590163934426	0.233055619729089	0.204847159959302	0.704918032786885	0.999137187230371
0.346639344262295	0.199798264980689	0.221480070433763	0.735116479723900	0.967213114754098
0.459576271186441	0.286799961379971	0.249560481204751	0.624697336561743	1.00338983050847
0.494913793103448	0.300244500863641	0.219526133595722	0.613026819923372	0.982758620689655
0.494913793103448	0.300244500863641	0.219526133595722	0.613026819923372	0.982758620689655
0.553775510204082	0.382621490514514	0.239014695652167	0.506122448979592	0.960753532182104
0.350410958904110	0.211940808914019	0.175632577113397	0.733279613215149	0.981735159817352
0.421792452830189	0.257516115679581	0.237550756113762	0.666273584905661	0.972877358490566
0.389838709677419	0.233058235679510	0.199948949118453	0.700460829493088	0.930107526881721
}\data

\begin{tikzpicture}
\definecolor{color1}{rgb}{1,0.7,0.1}
\definecolor{color2}{rgb}{0.15,0.45,0.55}

 \begin{groupplot}[
     group style = {group size = 3 by 2,horizontal sep=1.4cm,vertical sep=1.4cm},
     width = 1\textwidth]

\nextgroupplot[height = 4.6cm,
     width = 4.6cm,
     ylabel={$\langle \overline{c} \rangle_\text{ent} / \langle \overline{c} \rangle$},
    ylabel={$\mathcal{H}/y_{90} ,\mathcal{H}_{\text{trap}}/y_{90}$},
   xlabel={$\langle \overline{c} \rangle$ },
     grid=both,
     xmin=0,
     xmax=0.8,
     ymin=0,
     ymax=0.6,
     %ymode=log, 
    domain=0:0.8,samples=400,
    legend columns=4,
   legend style={cells={align=left},anchor = south west,at={(0,1.2)},font=\small},
   clip mode=individual
   ]

\addplot [fill=blue,mark=*,only marks,fill opacity=0.6,draw opacity=1, mark size=3,clip mode=individual]
table[row sep=crcr]{%
0 -0.001	\\
}; \addlegendentry{$\mathcal{H}$ \, \,};

\addplot [fill=teal,mark=*,only marks,fill opacity=0.7,draw opacity=1, mark size=3,clip mode=individual]
table[row sep=crcr]{%
0 -0.001	\\
}; \addlegendentry{$\mathcal{H}_\text{trap}$ \, \,};

\addplot [fill=black,mark=*,only marks,fill opacity=0.1,draw opacity=0, mark size=3,clip mode=individual]
table[row sep=crcr]{%
0 -0.001	\\
}; \addlegendentry{$\mathcal{H}$ \citep{Kramer2023JFM} \, };

\addplot+[densely dotted,mark=none,color=red,line width=1.5,draw opacity=0.6] 
table[row sep=crcr]{%
0 -0.001	\\
};\addlegendentry{$\mathcal{H}/y_{90} = 0.7 \langle \overline{c} \rangle$};

\addplot [fill=blue,mark=square*,only marks,fill opacity=0.6,draw opacity=1, mark size=3,clip mode=individual]
table[row sep=crcr]{%
0 -0.001	\\
}; \addlegendentry{$y_{50}$ \, \,};

\addplot [fill=teal,mark=square*,only marks,fill opacity=0.7,draw opacity=1, mark size=3,clip mode=individual]
table[row sep=crcr]{%
0 -0.001	\\
}; \addlegendentry{$y_{50_\text{trap}}$};

\addplot [fill=black,mark=square*,only marks,fill opacity=0.1,draw opacity=0, mark size=3,clip mode=individual]
table[row sep=crcr]{%
0 -0.001	\\
}; \addlegendentry{$y_{50}$ \citep{Kramer2023JFM}};

\addplot+[densely dashed,mark=none,color=red,line width=1.5,draw opacity=0.6] 
table[row sep=crcr]{%
0 -0.001	\\
};\addlegendentry{$y_{50}/y_{90} = 1- 0.9 \langle \overline{c} \rangle$};

\addplot+[densely dashdotted,mark=none,color=gray,line width=1] table[row sep=crcr]{%
0 -0.001	\\
};\addlegendentry{1:1}

 \node[align=center,color=black] at (rel axis cs:0.1,0.9) {(\textit{a})};

\addplot [fill=black,mark=*,only marks,fill opacity=0.1,draw opacity=0, mark size=3,clip mode=individual]
table[row sep=crcr]{%
0.210304241	0.151259543	\\
0.248783264	0.180498742	\\
0.206732801	0.147181215	\\
0.279105263	0.193159365	\\
0.23847191	0.174122592	\\
0.20681844	0.145186083	\\
0.283871391	0.195949081	\\
0.256571514	0.178742795	\\
0.239573671	0.171211472	\\
0.296059343	0.203810188	\\
0.265983683	0.182126636	\\
0.271521283	0.191657461	\\
0.330622449	0.227594002	\\
0.291258065	0.203992473	\\
0.264347079	0.183417799	\\
0.355041237	0.242815906	\\
0.314695496	0.218238864	\\
0.284164688	0.194807688	\\
0.324873936	0.227247438	\\
0.296886882	0.208388721	\\
0.291487752	0.202118816	\\
0.353172437	0.238021501	\\
0.328058226	0.226175675	\\
0.310328678	0.21349486	\\
0.329708621	0.225579261	\\
0.32638613	0.224485838	\\
0.331049133	0.222496314	\\
0.312257143	0.208189201	\\
0.065649346	0.04779477	\\
0.095667944	0.070651501	\\
0.108748755	0.079496643	\\
0.122272506	0.090724429	\\
0.135834774	0.101290297	\\
0.157674889	0.116484604	\\
0.171409974	0.127952848	\\
0.070865962	0.051483813	\\
0.092232189	0.072436954	\\
0.11042946	0.082348192	\\
0.119224335	0.088027197	\\
0.134156666	0.099998926	\\
0.140477574	0.104611924	\\
0.017426543	0.011130023	\\
0.045031428	0.036027673	\\
0.073772244	0.054700832	\\
0.093441974	0.06818538	\\
0.108060034	0.077646595	\\
0.119799689	0.088478872	\\
0.122641331	0.091572198	\\
0.018916814	0.017361945	\\
0.034551307	0.023144904	\\
0.055588428	0.036667353	\\
0.069244114	0.052970742	\\
0.0918108	0.066525148	\\
0.101487076	0.074978781	\\
0.111393785	0.082961194	\\
0.018905459	0.014137841	\\
0.023320165	0.017725335	\\
0.039796359	0.029510569	\\
0.057974646	0.042007748	\\
0.080194264	0.05783353	\\
0.090094536	0.065679737	\\
0.098966069	0.072765603	\\
0.063718126	0.012724782	\\
0.013097907	0.009770533	\\
0.01917624	0.013534465	\\
0.017546715	0.012530979	\\
0.024851114	0.016736204	\\
0.042646458	0.030318235	\\
0.051559631	0.037622952	\\
0.064179413	0.04779185	\\
0.073417071	0.053915551	\\
0.02020456	0.013201097	\\
0.017733803	0.014798363	\\
0.02026311	0.012293463	\\
0.021603408	0.013685719	\\
0.021898726	0.017472712	\\
0.036833829	0.02205923	\\
0.031747988	0.025720129	\\
0.049465667	0.035658672	\\
0.054525746	0.042768721	\\
0.01928167	0.018094713	\\
0.027292583	0.022177064	\\
0.022357173	0.015315636	\\
0.021557539	0.016217388	\\
0.024718685	0.018911379	\\
0.027741851	0.020969883	\\
0.025488662	0.019927723	\\
0.03769575	0.028029038	\\
0.053810098	0.038044104	\\
0.030483786	0.027008297	\\
0.031653926	0.015313973	\\
0.024050448	0.01825619	\\
0.036601582	0.028776756	\\
0.031667916	0.021318189	\\
0.022403435	0.017420887	\\
0.041460462	0.025932705	\\
0.041059301	0.029696196	\\
0.050252268	0.038867822	\\
0.121474531	0.087446946	\\
0.160595649	0.123296793	\\
0.183001124	0.135203995	\\
0.180082565	0.131923612	\\
0.188725198	0.138579949	\\
0.189353903	0.134247752	\\
0.208807975	0.156293245	\\
0.019091864	0.014152788	\\
0.096533695	0.070057903	\\
0.134112459	0.096671804	\\
0.165133618	0.120434932	\\
0.155809306	0.114152235	\\
0.162094341	0.120754883	\\
0.167096062	0.119561699	\\
0.18257986	0.133317611	\\
0.011755388	0.008763902	\\
0.065565457	0.044404983	\\
0.113364429	0.083038032	\\
0.152130619	0.110764563	\\
0.14387939	0.107576989	\\
0.160941962	0.112100894	\\
0.153713147	0.110056012	\\
0.174125809	0.127566673	\\
0.016251599	0.011439078	\\
0.040914672	0.031202614	\\
0.098193134	0.069818301	\\
0.133661053	0.097941067	\\
0.135406061	0.100514534	\\
0.14372617	0.102187768	\\
0.145782882	0.104514758	\\
0.16354797	0.119217823	\\
0.01205209	0.00890273	\\
0.041207114	0.032950125	\\
0.079273341	0.053620222	\\
0.116208375	0.082504078	\\
0.129746394	0.093726888	\\
0.134340982	0.097540526	\\
0.142110403	0.102421424	\\
0.152651898	0.112344082	\\
0.01332882	0.009438205	\\
0.015794479	0.012115635	\\
0.040729425	0.032946175	\\
0.082688526	0.057953611	\\
0.103385424	0.073217572	\\
0.111528816	0.082833715	\\
0.128097057	0.089748752	\\
0.132158463	0.094664439	\\
0.013415536	0.01149926	\\
0.017103841	0.012522536	\\
0.029193811	0.021420643	\\
0.062426818	0.039138532	\\
0.080545833	0.057482911	\\
0.093076896	0.067596159	\\
0.103721362	0.076469459	\\
0.118694494	0.083376843	\\
0.027329081	0.0186198	\\
0.013080055	0.013448553	\\
0.025852237	0.019004796	\\
0.041529981	0.030212024	\\
0.096606857	0.0447676	\\
0.079960702	0.05726886	\\
0.092865045	0.066206096	\\
0.100496067	0.075425907	\\
0.030566641	0.017180917	\\
0.021440589	0.01207438	\\
0.022584105	0.018108641	\\
0.038108744	0.023953256	\\
0.051308911	0.034874938	\\
0.056112547	0.060634897	\\
0.077607283	0.057882533	\\
0.089633251	0.069541045	\\
0.050252268	0.038867822	\\
0.121474531	0.087446946	\\
0.160595649	0.123296793	\\
0.169874043	0.123773047	\\
0.191715974	0.139356766	\\
0.215947998	0.158536653	\\
0.23270139	0.167191651	\\
0.234023058	0.170936173	\\
0.019091864	0.014152788	\\
0.096533695	0.070057903	\\
0.134112459	0.096671804	\\
0.149246336	0.109893662	\\
0.165311407	0.117463892	\\
0.18117404	0.132213718	\\
0.197402327	0.144663959	\\
0.192437806	0.13983058	\\
0.011755388	0.008763902	\\
0.065565457	0.044404983	\\
0.113364429	0.083038032	\\
0.137388212	0.100872771	\\
0.157331822	0.114424255	\\
0.165776391	0.11951628	\\
0.174597808	0.125298427	\\
0.175913228	0.127483469	\\
0.016251599	0.011439078	\\
0.040914672	0.031202614	\\
0.098193134	0.069818301	\\
0.129136844	0.093099455	\\
0.146485428	0.106264987	\\
0.161282554	0.110742056	\\
0.165342579	0.118283486	\\
0.162922679	0.11669329	\\
0.01205209	0.008902737	\\
0.041207114	0.032950125	\\
0.079273341	0.053620222	\\
0.115670714	0.086032037	\\
0.139852978	0.09925834	\\
0.144443596	0.102693775	\\
0.157761614	0.111640271	\\
0.160808544	0.113769748	\\
0.01332882	0.009438205	\\
0.015794479	0.012115635	\\
0.040729425	0.032946175	\\
0.091743431	0.065120189	\\
0.113912272	0.082013623	\\
0.125013774	0.09117702	\\
0.139419259	0.099671594	\\
0.147817803	0.102934666	\\
0.107584869	0.075849004	\\
0.158060233	0.119892734	\\
0.248056292	0.181194154	\\
0.232724866	0.172524673	\\
0.209531297	0.151792978	\\
0.237137182	0.174168154	\\
0.24588071	0.186524093	\\
0.238421059	0.172850184	\\
0.041747219	0.030885872	\\
0.129249748	0.097060293	\\
0.213729089	0.152892701	\\
0.19005708	0.139998004	\\
0.177845779	0.127937183	\\
0.195903376	0.143927293	\\
0.214973049	0.156759788	\\
0.209102525	0.151469019	\\
0.019942857	0.014444283	\\
0.101937861	0.077537145	\\
0.19222486	0.138071621	\\
0.162438661	0.118281235	\\
0.166140847	0.114819633	\\
0.172543727	0.124522703	\\
0.177655376	0.132237816	\\
0.180049103	0.12919643	\\
0.017301045	0.013443152	\\
0.080263005	0.057062277	\\
0.163425857	0.116899343	\\
0.148309252	0.105631868	\\
0.168681693	0.119086495	\\
0.177405891	0.125389564	\\
0.170420591	0.125067141	\\
0.172729484	0.121433517	\\
0.018143071	0.010024094	\\
0.056247521	0.042740628	\\
0.134542542	0.098245522	\\
0.129169256	0.096825381	\\
0.161150393	0.114821757	\\
0.166383071	0.115963313	\\
0.160809382	0.115888753	\\
0.163639579	0.115038808	\\
0.017296077	0.012362163	\\
0.028689176	0.021926023	\\
0.080278553	0.062119306	\\
0.107966894	0.076724635	\\
0.142569685	0.09646022	\\
0.15064514	0.108276107	\\
0.151817466	0.107300768	\\
0.156854293	0.108934067	\\
0.020558209	0.014347373	\\
0.024754779	0.019042711	\\
0.052258584	0.039170624	\\
0.077749635	0.058400802	\\
0.119831382	0.083338735	\\
0.136578185	0.092424873	\\
0.138768937	0.09671609	\\
0.146403293	0.100776338	\\
0.030970027	0.020502301	\\
0.028363423	0.02164774	\\
0.045310126	0.030067135	\\
0.059620378	0.043303558	\\
0.095794242	0.066304344	\\
0.109437629	0.080488523	\\
0.120897422	0.086139393	\\
0.129725759	0.090167323	\\
0.023794583	0.018382668	\\
0.021975598	0.01511282	\\
0.036465252	0.027126421	\\
0.048360074	0.036011007	\\
0.074627385	0.0547381	\\
0.097332954	0.065867206	\\
0.110574482	0.07833094	\\
0.121469626	0.083712175	\\
0.198709677	0.141451363	\\
0.151734266	0.102430366	\\
0.164856115	0.113603622	\\
0.162418478	0.10756734	\\
0.149237288	0.101319409	\\
0.151942675	0.102274087	\\
0.146686047	0.099785513	\\
0.153562092	0.099267205	\\
0.24615	0.172047658	\\
0.238217822	0.170220057	\\
0.240756303	0.173482178	\\
0.244017857	0.173363791	\\
0.25441989	0.168760817	\\
0.246363636	0.17523271	\\
0.245333333	0.172676159	\\
0.230075188	0.164091062	\\
0.248318584	0.181574421	\\
0.312333333	0.224858218	\\
0.303024691	0.213679441	\\
0.327916667	0.220921202	\\
0.296348315	0.199611344	\\
0.31390625	0.204100177	\\
0.306090909	0.196343589	\\
0.293691275	0.185954418	\\
0.230075188	0.164091062	\\
0.345597015	0.226564687	\\
0.323920455	0.206298246	\\
0.348243243	0.226090715	\\
0.395625	0.286542932	\\
0.402910448	0.281557595	\\
0.398157895	0.270310266	\\
0.395228758	0.258320589	\\
0.369178767	0.248978802	\\
0.373228481	0.242197518	\\
0.361706471	0.237234056	\\
0.349253247	0.228418663	\\
0.445357143	0.280169898	\\
0.43175	0.273993212	\\
0.390645161	0.235740032	\\
0.532142857	0.423810559	\\
0.532142857	0.423810559	\\
0.542688679	0.372860389	\\
0.545816327	0.375666077	\\
0.503518519	0.334648864	\\
0.496192053	0.31827612	\\
0.476746032	0.309000814	\\
0.449411765	0.287350914	\\
0.5	0.316698749	\\
0.452702703	0.280291479	\\
0.370298013	0.239782921	\\
0.5852	0.493277422	\\
0.5852	0.493277422	\\
0.588658537	0.459755011	\\
0.602959184	0.440250705	\\
0.596626506	0.400493212	\\
0.577931034	0.378537224	\\
0.556269841	0.354294742	\\
0.535571429	0.340033643	\\
0.608196721	0.612191504	\\
0.62469697	0.580530682	\\
0.6405	0.543734663	\\
0.647357143	0.515475515	\\
0.6405	0.543734663	\\
0.653852459	0.498284363	\\
0.635454545	0.446746152	\\
0.622183099	0.405192349	\\
0.6478875	0.736865873	\\
0.677978723	0.783769899	\\
0.698043478	0.600255056	\\
0.694259259	0.608824628	\\
0.709375	0.577533336	\\
0.69382716	0.514388657	\\
0.635454545	0.446746152	\\
0.678454545	0.467278485	\\
0.195912409	0.100099095	\\
0.241914894	0.170732119	\\
0.250294118	0.169041403	\\
0.216935484	0.157405683	\\
0.330142857	0.217555993	\\
0.150404624	0.106063161	\\
0.250294118	0.169041403	\\
0.242121212	0.156487403	\\
0.253461538	0.174396076	\\
0.192535211	0.125865521	\\
0.257363636	0.173560481	\\
0.349590164	0.231972568	\\
0.346639344	0.229313181	\\
0.459576271	0.300410262	\\
0.494913793	0.323749416	\\
0.494913793	0.323749416	\\
0.55377551	0.38261538	\\
};

\addplot [fill=blue,mark=*,only marks,fill opacity=1,draw opacity=1, mark size=3,clip mode=individual]
table [y=H,  x =Cmean] {\data};

\addplot [fill=teal,mark=*,only marks,fill opacity=1,draw opacity=1, mark size=3,clip mode=individual]
table [y=Htrap,  x =Cmean] {\data};

% \addplot+[solid,mark=none,color=black,line width=2] {0.6*x}; 
%$\mathcal{H}_{\text{trap}}= y_{84_\text{trap}} - y_{50_\text{trap}}$ (m),

 \addplot+[densely dotted,mark=none,color=red,line width=1.5,draw opacity=0.6] {0.7*x};

\nextgroupplot[height = 4.6cm,
     width = 4.6cm,
     xlabel={$\mathcal{H}$ (m)},
     ylabel style={align=center},
    ylabel={$y_{84} - y_{50}$ (m)},
      grid=both,
     xmin=0,
     xmax=0.04,
     ymin=0,
     ymax=0.04,
     %ymode=log, 
    domain=0:0.8,samples=400,
    legend columns=1,
   legend style={cells={align=left},anchor = south east,at={(1,0)},font=\small},
   clip mode=individual
   ]

 \node[align=center,color=black] at (rel axis cs:0.1,0.9) {(\textit{b})}; 
 
 \addplot+[densely dashdotted,mark=none,color=gray,line width=1] {x};

\addplot [fill=blue,mark=*,only marks,fill opacity=1,draw opacity=1, mark size=3,clip mode=individual]
table [y=Hest,  x =H] {\Hest};

\nextgroupplot[height = 4.6cm,
     width = 4.6cm,
        ylabel={$y_{50}/y_{90} ,y_{50_\text{trap}}/y_{90}$},
   xlabel={$\langle \overline{c} \rangle$ },
     grid=both,
     xmin=0,
     xmax=0.8,
     ymin=0,
     ymax=1.5,
     %xmode=log, 
     %ymode=log, 
    legend columns=1,
      legend style={cells={align=left},anchor = north east,at={(1,1)},font=\small},
   clip mode=individual,
    yticklabel style={
        /pgf/number format/fixed,
        /pgf/number format/precision=2},
    xticklabel style={
        /pgf/number format/fixed,
        /pgf/number format/precision=2},
          domain=0:0.8,samples=400,
   ]

 \node[align=center,color=black] at (rel axis cs:0.1,0.9) {(\textit{c})};

\addplot [fill=black,mark=square*,only marks,fill opacity=0.1,draw opacity=0, mark size=3,clip mode=individual]
table[row sep=crcr]{%
0.210304241	0.796918615	\\
0.248783264	0.760981396	\\
0.206732801	0.800179061	\\
0.279105263	0.741193119	\\
0.23847191	0.771669342	\\
0.20681844	0.800981734	\\
0.283871391	0.738491492	\\
0.256571514	0.760992734	\\
0.239573671	0.7650187	\\
0.296059343	0.728685077	\\
0.265983683	0.754677499	\\
0.271521283	0.747407044	\\
0.330622449	0.70014634	\\
0.291258065	0.737234527	\\
0.264347079	0.759099573	\\
0.355041237	0.677540236	\\
0.314695496	0.717828422	\\
0.284164688	0.743620043	\\
0.324873936	0.714855453	\\
0.296886882	0.740243386	\\
0.291487752	0.741595231	\\
0.353172437	0.690621403	\\
0.328058226	0.711544352	\\
0.310328678	0.725417043	\\
0.329708621	0.716348296	\\
0.32638613	0.711024546	\\
0.331049133	0.716930128	\\
0.312257143	0.728182553	\\
0.065649346	0.94729808	\\
0.095667944	0.907720286	\\
0.108748755	0.89181895	\\
0.122272506	0.883236816	\\
0.135834774	0.866353587	\\
0.157674889	0.84290686	\\
0.171409974	0.831693166	\\
0.070865962	0.931731401	\\
0.092232189	0.906795394	\\
0.11042946	0.893878461	\\
0.119224335	0.883319417	\\
0.134156666	0.869613698	\\
0.140477574	0.861691112	\\
0.017426543	0.984416429	\\
0.045031428	0.960470694	\\
0.073772244	0.92696475	\\
0.093441974	0.908906114	\\
0.108060034	0.896144899	\\
0.119799689	0.88129123	\\
0.122641331	0.880302399	\\
0.018916814	0.983648612	\\
0.034551307	0.965931644	\\
0.055588428	0.947224372	\\
0.069244114	0.932856729	\\
0.0918108	0.909153798	\\
0.101487076	0.899761658	\\
0.111393785	0.8890022	\\
0.018905459	0.985549065	\\
0.023320165	0.979537836	\\
0.039796359	0.960668877	\\
0.057974646	0.945086133	\\
0.080194264	0.921822157	\\
0.090094536	0.909955198	\\
0.098966069	0.904729719	\\
0.063718126	0.991309014	\\
0.013097907	0.98787031	\\
0.01917624	0.984768035	\\
0.017546715	0.984287634	\\
0.024851114	0.977572706	\\
0.042646458	0.961769532	\\
0.051559631	0.948288913	\\
0.064179413	0.941091043	\\
0.073417071	0.92839268	\\
0.02020456	0.980473737	\\
0.017733803	0.984390475	\\
0.02026311	0.983309208	\\
0.021603408	0.982666174	\\
0.021898726	0.980386564	\\
0.036833829	0.973222586	\\
0.031747988	0.96963991	\\
0.049465667	0.949996332	\\
0.054525746	0.949302275	\\
0.01928167	0.990991144	\\
0.027292583	0.981699281	\\
0.022357173	0.979077652	\\
0.021557539	0.984675561	\\
0.024718685	0.981739699	\\
0.027741851	0.974617389	\\
0.025488662	0.976182623	\\
0.03769575	0.965503606	\\
0.053810098	0.95259709	\\
0.030483786	0.965853293	\\
0.031653926	0.975866894	\\
0.024050448	0.97878263	\\
0.036601582	0.973080337	\\
0.031667916	0.970263828	\\
0.022403435	0.979506366	\\
0.041460462	0.958281264	\\
0.041059301	0.958964904	\\
0.050252268	0.948474076	\\
0.121474531	0.882956213	\\
0.160595649	0.846139317	\\
0.183001124	0.822124827	\\
0.180082565	0.822147006	\\
0.188725198	0.813025547	\\
0.189353903	0.811159547	\\
0.208807975	0.794738016	\\
0.019091864	0.983269896	\\
0.096533695	0.906278531	\\
0.134112459	0.869370657	\\
0.165133618	0.838392028	\\
0.155809306	0.847277653	\\
0.162094341	0.84063204	\\
0.167096062	0.833623621	\\
0.18257986	0.816559021	\\
0.011755388	0.99090199	\\
0.065565457	0.929044575	\\
0.113364429	0.889998047	\\
0.152130619	0.854721896	\\
0.14387939	0.859297519	\\
0.160941962	0.840049815	\\
0.153713147	0.84741885	\\
0.174125809	0.827114535	\\
0.016251599	0.985992741	\\
0.040914672	0.961356431	\\
0.098193134	0.903236437	\\
0.133661053	0.87021502	\\
0.135406061	0.86885459	\\
0.14372617	0.857472477	\\
0.145782882	0.856495456	\\
0.16354797	0.839009264	\\
0.01205209	0.990967025	\\
0.041207114	0.972339268	\\
0.079273341	0.922737368	\\
0.116208375	0.889603576	\\
0.129746394	0.87597793	\\
0.134340982	0.870781748	\\
0.142110403	0.86022971	\\
0.152651898	0.853136301	\\
0.01332882	0.990506662	\\
0.015794479	0.98769191	\\
0.040729425	0.96086237	\\
0.082688526	0.920868633	\\
0.103385424	0.895646215	\\
0.111528816	0.892117161	\\
0.128097057	0.87060614	\\
0.132158463	0.870521614	\\
0.013415536	0.990284211	\\
0.017103841	0.98453091	\\
0.029193811	0.972659753	\\
0.062426818	0.945814288	\\
0.080545833	0.922116759	\\
0.093076896	0.909397982	\\
0.103721362	0.900675836	\\
0.118694494	0.882205786	\\
0.027329081	0.970418598	\\
0.013080055	0.993139508	\\
0.025852237	0.976084765	\\
0.041529981	0.961418132	\\
0.096606857	0.938481003	\\
0.079960702	0.922485377	\\
0.092865045	0.910635474	\\
0.100496067	0.904305167	\\
0.030566641	0.968518195	\\
0.021440589	0.983045767	\\
0.022584105	0.982116338	\\
0.038108744	0.967030649	\\
0.051308911	0.951998475	\\
0.056112547	0.954862479	\\
0.077607283	0.928953128	\\
0.089633251	0.912348994	\\
0.050252268	0.948474076	\\
0.121474531	0.882956213	\\
0.160595649	0.846139317	\\
0.169874043	0.834512299	\\
0.191715974	0.813899684	\\
0.215947998	0.785512256	\\
0.23270139	0.770194332	\\
0.234023058	0.767679963	\\
0.019091864	0.983269896	\\
0.096533695	0.906278531	\\
0.134112459	0.869370657	\\
0.149246336	0.853988553	\\
0.165311407	0.835653956	\\
0.18117404	0.820921005	\\
0.197402327	0.80265135	\\
0.192437806	0.806476937	\\
0.011755388	0.99090199	\\
0.065565457	0.929044575	\\
0.113364429	0.889998047	\\
0.137388212	0.86405291	\\
0.157331822	0.845060813	\\
0.165776391	0.837836552	\\
0.174597808	0.83072157	\\
0.175913228	0.826842336	\\
0.016251599	0.985992741	\\
0.040914672	0.961356431	\\
0.098193134	0.903236437	\\
0.129136844	0.87429617	\\
0.146485428	0.855901166	\\
0.161282554	0.841886483	\\
0.165342579	0.834656387	\\
0.162922679	0.838895388	\\
0.01205209	0.990967025	\\
0.041207114	0.972339268	\\
0.079273341	0.922737368	\\
0.115670714	0.887886275	\\
0.139852978	0.862993982	\\
0.144443596	0.860419122	\\
0.157761614	0.845237863	\\
0.160808544	0.842512327	\\
0.01332882	0.990506662	\\
0.015794479	0.98769191	\\
0.040729425	0.96086237	\\
0.091743431	0.910953319	\\
0.113912272	0.889644563	\\
0.125013774	0.875639435	\\
0.139419259	0.862839455	\\
0.147817803	0.854094657	\\
0.107584869	0.895526687	\\
0.158060233	0.848920161	\\
0.248056292	0.756399709	\\
0.232724866	0.767969781	\\
0.209531297	0.792574917	\\
0.237137182	0.769558824	\\
0.24588071	0.761711221	\\
0.238421059	0.762345196	\\
0.041747219	0.958055661	\\
0.129249748	0.874324288	\\
0.213729089	0.790154029	\\
0.19005708	0.815173511	\\
0.177845779	0.824562281	\\
0.195903376	0.808157813	\\
0.214973049	0.786988412	\\
0.209102525	0.795615001	\\
0.019942857	0.981214343	\\
0.101937861	0.89586243	\\
0.19222486	0.811134097	\\
0.162438661	0.839648041	\\
0.166140847	0.83171017	\\
0.172543727	0.830320768	\\
0.177655376	0.824443073	\\
0.180049103	0.823589305	\\
0.017301045	0.987879065	\\
0.080263005	0.916529372	\\
0.163425857	0.844915584	\\
0.148309252	0.855606865	\\
0.168681693	0.834487969	\\
0.177405891	0.827170094	\\
0.170420591	0.836966886	\\
0.172729484	0.828644047	\\
0.018143071	0.987089196	\\
0.056247521	0.946633546	\\
0.134542542	0.870414477	\\
0.129169256	0.878867728	\\
0.161150393	0.844971161	\\
0.166383071	0.836676874	\\
0.160809382	0.839194055	\\
0.163639579	0.837679616	\\
0.017296077	0.984501948	\\
0.028689176	0.977146492	\\
0.080278553	0.923871068	\\
0.107966894	0.895873784	\\
0.142569685	0.860824823	\\
0.15064514	0.85173519	\\
0.151817466	0.85486981	\\
0.156854293	0.845829961	\\
0.020558209	0.982223987	\\
0.024754779	0.975717338	\\
0.052258584	0.949148912	\\
0.077749635	0.924338969	\\
0.119831382	0.880442658	\\
0.136578185	0.869402895	\\
0.138768937	0.866220848	\\
0.146403293	0.857579478	\\
0.030970027	0.975586069	\\
0.028363423	0.97431619	\\
0.045310126	0.953542377	\\
0.059620378	0.942477433	\\
0.095794242	0.906047365	\\
0.109437629	0.897149555	\\
0.120897422	0.882504042	\\
0.129725759	0.872129154	\\
0.023794583	0.981691577	\\
0.021975598	0.980114788	\\
0.036465252	0.96519178	\\
0.048360074	0.952754008	\\
0.074627385	0.929109643	\\
0.097332954	0.903884661	\\
0.110574482	0.8902358	\\
0.121469626	0.883562666	\\
0.198709677	0.814882698	\\
0.151734266	0.861964123	\\
0.164856115	0.853717026	\\
0.162418478	0.853864734	\\
0.149237288	0.86779661	\\
0.151942675	0.862322391	\\
0.146686047	0.870016611	\\
0.153562092	0.861500156	\\
0.24615	0.769736842	\\
0.238217822	0.789250354	\\
0.240756303	0.792316927	\\
0.244017857	0.788690476	\\
0.25441989	0.785548247	\\
0.246363636	0.787878788	\\
0.245333333	0.787878788	\\
0.230075188	0.80075188	\\
0.248318584	0.786504425	\\
0.312333333	0.71114582	\\
0.303024691	0.730070276	\\
0.327916667	0.707999094	\\
0.296348315	0.743998496	\\
0.31390625	0.735212853	\\
0.306090909	0.73874743	\\
0.293691275	0.746902013	\\
0.230075188	0.80075188	\\
0.345597015	0.713932661	\\
0.323920455	0.733114636	\\
0.348243243	0.715442486	\\
0.395625	0.633704434	\\
0.402910448	0.643021614	\\
0.398157895	0.652529791	\\
0.395228758	0.666927847	\\
0.369178767	0.692261906	\\
0.373228481	0.696032767	\\
0.361706471	0.701796137	\\
0.349253247	0.708222072	\\
0.445357143	0.636567843	\\
0.43175	0.646978112	\\
0.390645161	0.68969896	\\
0.532142857	0.447965918	\\
0.532142857	0.447965918	\\
0.542688679	0.518399247	\\
0.545816327	0.520889508	\\
0.503518519	0.567784203	\\
0.496192053	0.594139807	\\
0.476746032	0.611615115	\\
0.449411765	0.630672915	\\
0.5	0.594996203	\\
0.452702703	0.641531596	\\
0.370298013	0.696483224	\\
0.5852	0.375625897	\\
0.5852	0.375625897	\\
0.588658537	0.415611086	\\
0.602959184	0.433094184	\\
0.596626506	0.486636968	\\
0.577931034	0.514946693	\\
0.556269841	0.548240128	\\
0.535571429	0.560470014	\\
0.608196721	0.201443157	\\
0.62469697	0.262957727	\\
0.6405	0.30029199	\\
0.647357143	0.322682718	\\
0.6405	0.30029199	\\
0.653852459	0.349925645	\\
0.635454545	0.432359788	\\
0.622183099	0.4693114	\\
0.6478875	0.033730738	\\
0.677978723	-0.007029644	\\
0.698043478	0.207276843	\\
0.694259259	0.194805037	\\
0.709375	0.239227041	\\
0.69382716	0.323515309	\\
0.635454545	0.432359788	\\
0.678454545	0.385811288	\\
0.195912409	0.810882548	\\
0.241914894	0.784869976	\\
0.250294118	0.764664313	\\
0.216935484	0.795698925	\\
0.330142857	0.710363498	\\
0.150404624	0.857109827	\\
0.250294118	0.764664313	\\
0.242121212	0.778787879	\\
0.253461538	0.780725671	\\
0.192535211	0.809317443	\\
0.257363636	0.767193571	\\
0.349590164	0.703877155	\\
0.346639344	0.718246822	\\
0.459576271	0.612388514	\\
0.494913793	0.600079415	\\
0.494913793	0.600079415	\\
0.55377551	0.524578363	\\
};

\addplot [fill=blue,mark=square*,only marks,fill opacity=1,draw opacity=1, mark size=3,clip mode=individual]
table [y=y50,  x =Cmean] {\data};
 
\addplot [fill=teal,mark=square*,only marks,fill opacity=1,draw opacity=1, mark size=3,clip mode=individual]
table [y=y50trap,  x =Cmean] {\data};

\addplot+[densely dashed,mark=none,color=red,line width=1.5,draw opacity=0.6] {1-0.9*x};

\begin{comment}
\nextgroupplot[height = 4.6cm,
     width = 4.6cm,
     ylabel={$(y_{50}+y_{50_\text{trap}})/2$ (m)},    
    xlabel={$y_{\overline{c}_\text{max}}$ (m)},
      grid=both,
     xmin=0,
     xmax=0.1,
     ymin=0,
     ymax=0.1,
       xtick={0, 0.05, 0.1},
    ytick={0, 0.05, 0.1},
    xticklabels={0, 0.05, 0.1},
    yticklabels={0, 0.05, 0.1},
    domain=0:0.8,samples=400,
    legend columns=1,
   legend style={cells={align=left},anchor = south east,at={(1,0)},font=\small},
   clip mode=individual
   ]

 \node[align=center,color=black] at (rel axis cs:0.1,0.9) {(\textit{d})}; 
 
 \addplot+[densely dashdotted,mark=none,color=gray,line width=1] {x};

\addplot [fill=black,mark=*,only marks,fill opacity=1,draw opacity=1, mark size=3,clip mode=individual]
table [y=ycmaxest,  x =ycmax] {\ycmax};

\nextgroupplot[height = 4.6cm,
     width = 4.6cm,
     ylabel={$\overline{c}_\text{ent}(y_{\overline{c}_\text{max}})$ [Eq. (\ref{entrainedair})]},    
    xlabel={$\overline{c}_\text{max}$},
      grid=both,
     xmin=0,
     xmax=0.8,
       ymin=0,
     ymax=0.8,
     %ymode=log, 
    domain=0:0.8,samples=400,
    legend columns=1,
   legend style={cells={align=left},anchor = south east,at={(1,0)},font=\small},
   clip mode=individual
   ]

 \node[align=center,color=black] at (rel axis cs:0.1,0.9) {(\textit{e})}; 
 
 \addplot+[densely dashdotted,mark=none,color=gray,line width=1] {x};

\addplot [fill=black,mark=*,only marks,fill opacity=1,draw opacity=1, mark size=3,clip mode=individual]
table [y=cmaxest,  x =cmax] {\cmax};

\end{comment}

\end{groupplot}
\end{tikzpicture}

%% file: Figures/fig5.tex
\pgfplotstableread{
Label series1 series2 series3 
1	0.103326134979525	0.0438761063753310	2.69899462449981e-05
\phantom{2}	0.0889221881200783	0.0827945457616167	1.13209445860404e-05
\phantom{3}	0.107007509202619	0.0689182292366671	8.01501616703206e-05
\phantom{4}	0.161696297264585	0.0238331718015040	0.00877210872488240
5	0.104243255680377	0.0637889258789011	0.0495369668244398
\phantom{6}	0.107076501579141	0.0968296726947329	0.0288974800414300
\phantom{7}	0.0904777329241351	0.106418501832028	0.0377372135226777
\phantom{8}	0.119725112745379	0.0575603757571307	0.0589926957412826
\phantom{9}	0.111021461278456	0.109633722657683	0.0262306050778004
10	0.103759303231469	0.0751540817684537	0.0746497356121365
\phantom{11}	0.120121858223439	0.119744719840463	0.0740960771842494
\phantom{12}	0.0974787834213479	0.135918544129394	0.0864516569703318
\phantom{13}	0.0583514060858485	0.0987551374792237	0.162814710451790
\phantom{14}	0.0809722722427064	0.195951251051168	0.0556449578393385
15	0.109878053329553	0.135223426009096	0.118577684490323
\phantom{16}	0.0741177132048594	0.104414041651178	0.198757526138715
\phantom{17}	0.0939891143950453	0.220452114105196	0.122112642191818
\phantom{18}	0.0884260209905622	0.185937235277409	0.198857915319036
\phantom{19}	0.0884260209905622	0.185937235277409	0.198857915319036
20	0.107066054090975	0.208849378425737	0.229817498035009
}\testdata

\pgfplotstableread{
Cmeantot Cmeantotfit
0.195912408759124	0.194284536693525
0.241914893617021	0.236228511257618
0.250294117647059	0.253511621413158
0.216935483870968	0.217531493055489
0.330142857142857	0.313837585765105
0.150404624277457	0.147175251408611
0.178189655172414	0.171705412937109
0.242121212121212	0.232785967721742
0.253461538461538	0.234543654803834
0.192535211267606	0.175845588277615
0.257363636363636	0.246726609671881
0.349590163934426	0.332453663247137
0.346639344262295	0.319734166634999
0.459576271186441	0.436524193187607
0.494913793103448	0.473110477906610
0.494913793103448	0.473110477906610
0.553775510204082	0.545637639517020
0.350410958904110	0.319873282160351
0.421792452830189	0.377278268587441
0.389838709677419	0.363641508500744
}\cmeandata

\begin{tikzpicture}
 \begin{groupplot}[
     group style = {group size = 2 by 1,horizontal sep=1.7cm,vertical sep=1.4cm},
     width = 1\textwidth]

\nextgroupplot[height = 4.6cm,
     width = 4.6cm,
        ylabel={$\langle \overline{c} \rangle$ [Eq. (\ref{eq:meanair})]},
   xlabel={$\langle \overline{c} \rangle$ [Eq. (\ref{eq:meanair0})]},
     grid=both,
     xmin=0,
     xmax=0.8,
     ymin=0,
     ymax=0.8,
     %xmode=log, 
     %ymode=log, 
    legend columns=1,
    legend style={cells={align=left},anchor = south west,at={(0,1.1)},font=\normalsize},
   clip mode=individual,
             domain=0:0.8,samples=400,
   ]

 \addplot+[densely dashdotted,mark=none,color=gray,line width=1,draw opacity=1] {x}; \addlegendentry{1:1}

\addplot [fill=black,mark=*,only marks,fill opacity=1,draw opacity=1, mark size=3,clip mode=individual] table [x=Cmeantot, y=Cmeantotfit] {\cmeandata};

 \node[align=center,color=black] at (rel axis cs:0.1,0.9) {(\textit{a})};

\nextgroupplot[
        ybar stacked,
        width=8.7cm,
        height=5.5cm,
        ymin=0,
        ymax=0.6,
        xmin=-1,
        xmax=20,
        bar width=0.26cm,
         layers/my layer set/.define layer set={
            background,
            main,
            foreground
        }
        axis on top,
        xtick=data,
        ytick={0, 0.2, 0.4, 0.6},
        extra y ticks = 0.1,
        extra y tick labels={0.1},
        extra y tick style={grid=major,major grid style={thick,dashed,draw=black},axis on top},
        xlabel={profile (ordered by $\langle \overline{c} \rangle)$},
        ylabel={$\langle \overline{c} \rangle$},
        legend style={cells={align=left},anchor = south west,at={(0,1.1)},font=\normalsize},
        xmajorgrids,
        ymajorgrids,
        %reverse legend=true, 
          legend columns=3,
        xticklabels from table={\testdata}{Label},
        xticklabel style={text width=2.5cm,align=center},
    ]

\addplot [fill=teal,draw=teal,fill opacity=0.5,draw opacity=1,on layer = background] table [y=series1, meta=Label, x expr=\coordindex] {\testdata}; \addlegendentry{$\langle \overline{c} \rangle_{\text{TWL}_\text{trap}}$}

\addplot [fill=black,draw=black,fill opacity=0.5,draw opacity=1,on layer = background] table [y=series2, meta=Label, x expr=\coordindex] {\testdata}; \addlegendentry{$\langle \overline{c} \rangle_{\text{TWL}_\text{ent}}$}

\addplot [fill=violet,draw=violet,fill opacity=0.5,draw opacity=1,on layer = background] table [y=series3, meta=Label, x expr=\coordindex] {\testdata};  \addlegendentry{$\langle \overline{c} \rangle_\text{TBL}$}

 \node[align=center,color=black] at (rel axis cs:0.05,0.9) {(\textit{b})};

\end{groupplot}

\end{tikzpicture}

%% file: Figures/fig6.tex
\pgfplotstableread{
Fr Cmeantot CmeanTBL
6.84387903488895	0.198709677419355	2
5.06207237515353	0.151734265734266	2
5.63177455752933	0.164856115107914	2
5.74402460802221	0.162418478260870	2
5.80854734632257	0.149237288135593	2
5.97906760525525	0.151942675159236	2
7.87808062736095	0.246150000000000	0.0512028909088554
7.95210004051439	0.238217821782178	0.0312593296214322
8.20195127013827	0.240756302521008	0.0333623090801983
8.51524728603490	0.244017857142857	0.0405783469865276
8.62781759208994	0.254419889502763	0.0474291852384124
8.70811070793598	0.246363636363636	0.0405217124930359
9.54931612978074	0.312333333333333	0.0519078477379939
9.62464281802124	0.303024691358025	0.0431925299022636
10.6226780501364	0.327916666666667	0.0628546910809664
9.57866481673108	0.296348314606742	0.0559757112614108
9.92151710403022	0.313906250000000	0.0716644585631347
9.85627571911217	0.306090909090909	0.0698314613861278
11.6675922489532	0.395625000000000	0.105304824124605
11.5452092133098	0.402910447761194	0.108185615876588
12.3944794302434	0.398157894736842	0.122771514558481
12.3469535550506	0.395228758169935	0.130631062732169
12.1437093913529	0.369178767123288	0.111585413870620
12.2530192366300	0.373228481012658	0.124553457631941
14.9759363269678	0.532142857142857	0.198536896895234
14.9759363269678	0.532142857142857	0.198536896895234
15.8767214136979	0.542688679245283	0.276639901923356
15.9860141585336	0.545816326530612	0.279272828888668
14.8803450055785	0.503518518518519	0.220124121368344
13.8438796599844	0.496192052980133	0.223026178255519
13.8109861231251	0.585200000000000	0.277833876334183
20.0887070881820	0.585200000000000	0.277833876334183
19.5725774234842	0.588658536585366	0.302689798737681
19.5584637069722	0.602959183673469	0.333476028211824
19.5892106396083	0.596626506024097	0.350909909387011
19.1864057145378	0.577931034482759	0.334996657271473
24.0222725464647	0.608196721311476	0.279575897490009
23.6965002484406	0.624696969696970	0.318248112536774
24.8593179446417	0.640500000000000	0.367060146364456
25.1402031325835	0.647357142857143	0.396817737564261
24.8593179446417	0.640500000000000	0.367060146364456
23.9510256229262	0.653852459016393	0.402943633409575
24.6788924747903	0.647887500000000	0.282284762119253
32.2248924470911	0.677978723404255	0.397832957879741
26.1863018303605	0.698043478260870	0.488236731108928
25.0184834999851	0.694259259259259	0.448267799358133
25.7506861224808	0.709375000000000	0.521322865954060
26.2965177700794	0.693827160493827	0.503951697432207
}\Froude

\pgfplotstableread{
Fr Cmeantot CmeanTBL
10.4676061131175	0.195912408759124	0.00877985768753531
9.71257729477519	0.241914893617021	0.0617924744423729
10.4254632912701	0.250294117647059	0.0800269015212798
9.40118231857587	0.216935483870968	0.0536913811347403
11.8188374880546	0.330142857142857	0.0800318107935911
10.5730404422561	0.150404624277457	2.69899462449981e-05
10.2754511922383	0.178189655172414	0.000935864752448654
11.6781133459465	0.242121212121212	0.0323052317216213
12.2219330709387	0.253461538461538	0.0387307757121574
16.0605008998134	0.192535211267606	0.00609141228695123
14.5374006437004	0.257363636363636	0.0304470198017298
15.1851717040304	0.349590163934426	0.0606097057756580
15.0824151216053	0.346639344262295	0.0914409167934811
21.0773243211065	0.459576271186441	0.126593711237203
18.2067734211409	0.494913793103448	0.201604645976794
18.2067734211409	0.494913793103448	0.201604645976794
20.2038436512154	0.553775510204082	0.232025996295962
}\FroudeKillen

\pgfplotstableread{
Xstar Cmeantot CmeanTBL CmeanTWLent CmeanTWLtrap 
0.200000000000000	0.502339509638780	0.0225124556090649	0.0611104105958260	0.414632432481023
0.800000000000000	0.620294599018003	0.158442242159931	0.154359181231100	0.307756235034165
1.40000000000000	0.641779788838613	0.205197183387897	0.205180175320805	0.268601293709520
2.00000000000000	0.556244830438379	0.137670208037796	0.173588939621335	0.268824189540098
2.40000000000000	0.846520146520146	0.205209771265618	0.318886077831747	0.311245632367258
0.234567901234568	0.385652882762709	6.92049903717899e-05	0.112478775333622	0.264997393056329
0.481481481481482	0.456896551724138	0.00239965321140681	0.214148180137436	0.228493507326586
0.851851851851852	0.620823620823621	0.0828339274913368	0.255052097807804	0.276231653829015
1.34567901234568	0.649901380670611	0.0993096813132240	0.274758433041137	0.232667255823213
0.166666666666667	0.310540663334848	0.00982485852734070	0.119785009131274	0.173524281769821
0.500000000000000	0.415102639296188	0.0491080964544029	0.181659686741478	0.181002849291087
1	0.563855103120043	0.0977575899607386	0.320551807114783	0.134252767458599
1.33333333333333	0.559095716552089	0.147485349666905	0.224158351301348	0.160361706013207
2.33333333333333	0.741252050300711	0.204183405221296	0.359766003768329	0.155157629310335
3.16666666666667	0.798248053392658	0.325168783833539	0.303097143546858	0.144920540116113
4	0.798248053392658	0.325168783833539	0.303097143546858	0.144920540116113
4.83333333333333	0.893186306780777	0.374235477896713	0.340135301227999	0.174692956951196
5	0.898489638215666	0.422567904042773	0.256850933879876	0.152344646067315
8	1.03	0.515592406870376	0.271941928921899	0.193089513448299
16.5000000000000	0.999586435070306	0.314999624714424	0.354646525230017	0.285127401583613
}\Cdevelopment

\pgfplotstableread{
xLi Cmean
0	0
0.0540423692174665	0.00817210227015499
0.108084738434933	0.0162181983978310
0.162127107652400	0.0241402312794109
0.216169476869866	0.0319401138536404
0.270211846087332	0.0396197295635457
0.324254215304799	0.0471809328112305
0.378296584522266	0.0546255494056597
0.432338953739732	0.0619553770035383
0.486381322957198	0.0691721855433934
0.540423692174665	0.0762777176729621
0.594466061392131	0.0832736891699898
0.648508430609598	0.0901617893565405
0.702550799827065	0.0969436815069185
0.756593169044531	0.103621003249300
0.810635538261997	0.110195366961173
0.864677907479464	0.116668360158679
0.918720276696931	0.123041545879951
0.972762645914397	0.129316463062542
1.02680501513186	0.135494626915033
1.08084738434933	0.141577529282913
1.13488975356680	0.147566639008811
1.18893212278426	0.153463402287187
1.24297449200173	0.159269243013537
1.29701686121920	0.164985563128226
1.35105923043666	0.170613742955017
1.40510159965413	0.176155141534375
1.45914396887160	0.181611096951638
1.51318633808906	0.186982926660126
1.56722870730653	0.192271927799263
1.62127107652399	0.197479377507804
1.67531344574146	0.202606533232222
1.72935581495893	0.207654633030349
1.78339818417639	0.212624895870323
1.83744055339386	0.217518521924944
1.89148292261133	0.222336692861469
1.94552529182879	0.227080572126960
1.99956766104626	0.231751305229216
2.05361003026373	0.236350020013380
2.10765239948119	0.240877826934285
2.16169476869866	0.245335819324591
2.21573713791613	0.249725073658798
2.26977950713359	0.254046649813176
2.32382187635106	0.258301591321702
2.37786424556853	0.262490925628036
2.43190661478599	0.266615664333625
2.48594898400346	0.270676803441970
2.53999135322093	0.274675323599131
2.59403372243839	0.278612190330532
2.64807609165586	0.282488354274096
2.70211846087332	0.286304751409806
2.75616083009079	0.290062303285712
2.81020319930826	0.293761917240456
2.86424556852572	0.297404486622374
2.91828793774319	0.300990891005206
2.97233030696066	0.304521996400494
3.02637267617812	0.307998655466693
3.08041504539559	0.311421707715067
3.13445741461306	0.314791979712403
3.18849978383052	0.318110285280607
3.24254215304799	0.321377425693212
3.29658452226546	0.324594189868869
3.35062689148292	0.327761354561842
3.40466926070039	0.330879684549574
3.45871162991786	0.333949932817360
3.51275399913532	0.336972840740166
3.56679636835279	0.339949138261655
3.62083873757026	0.342879544070441
3.67488110678772	0.345764765773638
3.72892347600519	0.348605500067721
3.78296584522265	0.351402432906759
3.83700821444012	0.354156239668055
3.89105058365759	0.356867585315227
3.94509295287505	0.359537124558779
3.99913532209252	0.362165502014195
4.05317769130999	0.364753352357592
4.10722006052745	0.367301300478977
4.16126242974492	0.369809961633141
4.21530479896239	0.372279941588224
4.26934716817985	0.374711836771990
4.32338953739732	0.377106234415846
4.37743190661479	0.379463712696644
4.43147427583225	0.381784840876290
4.48551664504972	0.384070179439207
4.53955901426719	0.386320280227675
4.59360138348465	0.388535686575084
4.64764375270212	0.390716933437133
4.70168612191959	0.392864547521005
4.75572849113705	0.394979047412557
4.80977086035452	0.397060943701535
4.86381322957198	0.399110739104873
4.91785559878945	0.401128928588080
4.97189796800692	0.403115999484764
5.02594033722438	0.405072431614306
5.07998270644185	0.406998697397720
5.13402507565932	0.408895261971736
5.18806744487678	0.410762583301108
5.24210981409425	0.412601112289207
5.29615218331172	0.414411292886895
5.35019455252918	0.416193562199732
5.40423692174665	0.417948350593516
5.45827929096412	0.419676081798213
5.51232166018158	0.421377173010268
5.56636402939905	0.423052034993350
5.62040639861651	0.424701072177538
5.67444876783398	0.426324682756977
5.72849113705145	0.427923258786035
5.78253350626892	0.429497186273968
5.83657587548638	0.431046845278133
5.89061824470385	0.432572609995759
5.94466061392131	0.434074848854304
5.99870298313878	0.435553924600422
6.05274535235625	0.437010194387555
6.10678772157371	0.438444009862171
6.16083009079118	0.439855717248684
6.21487246000865	0.441245657433049
6.26891482922611	0.442614166045082
6.32295719844358	0.443961573539501
6.37699956766105	0.445288205275725
6.43104193687851	0.446594381596433
6.48508430609598	0.447880417904922
6.53912667531345	0.449146624741268
6.59316904453091	0.450393307857309
6.64721141374838	0.451620768290476
6.70125378296584	0.452829302436489
6.75529615218331	0.454019202120919
6.80933852140078	0.455190754669666
6.86338089061824	0.456344242978331
6.91742325983571	0.457479945580533
6.97146562905318	0.458598136715161
7.02550799827064	0.459699086392601
7.07955036748811	0.460783060459932
7.13359273670558	0.461850320665118
7.18763510592304	0.462901124720217
7.24167747514051	0.463935726363611
7.29571984435798	0.464954375421272
7.34976221357544	0.465957317867090
7.40380458279291	0.466944795882269
7.45784695201038	0.467917047913808
7.51188932122784	0.468874308732075
7.56593169044531	0.469816809487500
7.61997405966278	0.470744777766392
7.67401642888024	0.471658437645889
7.72805879809771	0.472558009748075
7.78210116731517	0.473443711293245
7.83614353653264	0.474315756152362
7.89018590575011	0.475174354898700
7.94422827496757	0.476019714858691
7.99827064418504	0.476852040161987
8.05231301340251	0.477671531790753
8.10635538261997	0.478478387628199
8.16039775183744	0.479272802506361
8.21444012105491	0.480054968253147
8.26848249027237	0.480825073738659
8.32252485948984	0.481583304920800
8.37656722870731	0.482329844890178
8.43060959792477	0.483064873914313
8.48465196714224	0.483788569481171
8.53869433635971	0.484501106342020
8.59273670557717	0.485202656553628
8.64677907479464	0.485893389519807
8.70082144401211	0.486573472032324
8.75486381322957	0.487243068311172
8.80890618244704	0.487902340044227
8.86294855166450	0.488551446426289
8.91699092088197	0.489190544197524
8.97103329009944	0.489819787681311
9.02507565931690	0.490439328821509
9.07911802853437	0.491049317219144
9.13316039775184	0.491649900168536
9.18720276696930	0.492241222692863
9.24124513618677	0.492823427579183
9.29528750540424	0.493396655412913
9.34932987462170	0.493961044611772
9.40337224383917	0.494516731459210
9.45741461305664	0.495063850137313
9.51145698227410	0.495602532759207
9.56549935149157	0.496132909400954
9.61954172070904	0.496655108132969
9.67358408992650	0.497169255050940
9.72762645914397	0.497675474306278
9.78166882836144	0.498173888136096
9.83571119757890	0.498664616892724
9.88975356679637	0.499147779072775
9.94379593601383	0.499623491345754
9.99783830523130	0.500091868582231
10.0518806744488	0.500553023881581
10.1059230436662	0.501007068599291
10.1599654128837	0.501454112373854
10.2140077821012	0.501894263153237
10.2680501513186	0.502327627220954
10.3220925205361	0.502754309221724
10.3761348897536	0.503174412186745
10.4301772589710	0.503588037558569
10.4842196281885	0.503995285215601
10.5382619974060	0.504396253496213
10.5923043666234	0.504791039222492
10.6463467358409	0.505179737723621
10.7003891050584	0.505562442858894
10.7544314742758	0.505939247040384
10.8084738434933	0.506310241255258
}\Cdevelopmentcalc

\pgfplotstableread{
xLi Cmeantot
7.49056603773585	0.948339160839161
5.42857142857143	1.03035071942446
4.62500000000000	1.01511548913044
3.36893203883495	0.932733050847458
3.01785714285714	0.949641719745223
2.43511450381679	0.916787790697674
2.04054054054054	0.959763071895425
14	1.02562500000000
9.22727272727273	0.992574257425743
7.03571428571428	1.00315126050420
5.61764705882353	1.01674107142857
4.69620253164557	1.06008287292818
3.89130434782609	1.06008287292818
3.28571428571429	1.02222222222222
2.62903225806452	0.958646616541353
1.48618784530387	0.801027690550956
14	1.00752688172043
8.37500000000001	0.977499004380725
7.49056603773585	1.05779569892473
5.08108108108108	0.955962305183037
4.48780487804878	1.01260080645161
3.54545454545455	0.987390029325514
3.01785714285714	0.947391210218662
2.38345864661654	0.742178025709435
2.38345864661654	0.886146192116341
1.76073619631902	0.830565268065268
2.10344827586207	0.892931392931393
15.0714285714286	1.01442307692308
8.37500000000001	1.03310371220819
6.89473684210526	1.02091767881242
5.71641791044776	1.01340707223060
4	0.946612223393045
3.94505494505494	0.956996105160662
3.32692307692308	0.927452488687783
2.68852459016393	0.895521145521146
3.05660377358491	0.824735449739583
2.46774193548387	0.799537037037037
1.94520547945205	0.723416965356183
24.2941176470588	0.985449735449735
12.4375000000000	0.985449735449735
12.8709677419355	1.00497903563941
8.77272727272728	1.01077097505669
6.41379310344828	1.01077097505669
5.82539682539683	0.918874172185431
4.51282051282051	0.882863021751910
3.43298969072165	0.832244008714597
3.07766990291262	0.847457627118644
2.28125000000000	0.767292716445259
1.23404255319149	0.627623751262768
21.1052631578948	0.991864406779661
15.8000000000000	0.991864406779661
11.7272727272727	0.997726333195535
9.50000000000000	1.02196471809063
7.93617021276595	1.01123136614254
6.50000000000000	0.979544126241964
5.17647058823530	0.942830239440409
4.06024096385542	0.907748184019371
18.5238095238095	0.950307377049180
12.6666666666667	0.976089015151515
10.7142857142857	1.00078125000000
8.53488372093024	1.01149553571429
7.72340425531914	1.01149553571429
6.59259259259259	1.02164446721311
5.50793650793651	1.02164446721311
4.54054054054054	0.972161091549296
20.0526315789474	0.938967391304348
15.6666666666667	0.938967391304348
11.5000000000000	1.01165721487083
9.25641025641026	1.01165721487083
7.69565217391304	1.02807971014493
6.01754385964912	1.00554660941134
4.97014925373134	0.992897727272727
3.87804878048781	0.983267457180501
}\CdevelopmentStraub

\begin{tikzpicture}
\definecolor{color1}{rgb}{1,0.7,0.1}
\definecolor{color2}{rgb}{0.15,0.45,0.55}

 \begin{groupplot}[
     group style = {group size = 3 by 2,horizontal sep=1.4cm,vertical sep=1.4cm},
     width = 1\textwidth]

\nextgroupplot[height = 4.6cm,
     width = 4.6cm,
     xlabel={$Fr$},
    ylabel={$\langle \overline{c} \rangle_\infty$, $\langle \overline{c} \rangle_{\text{TBL} \infty}$},
     grid=both,
     xmin=1,
     xmax=100,
     ymin=0.01,
     ymax=1,
     ymode=log, 
     xmode=log,
    domain=0:0.8,samples=400, legend columns=3,
   legend style={cells={align=left},anchor = south west,at={(0,1.2)},font=\normalsize},
   clip mode=individual
   ]

\addplot [fill=orange,mark=square*,only marks,fill opacity=1,draw opacity=0, mark size=2.5,clip mode=individual]
table[row sep=crcr]{%
100 100\\
}; \addlegendentry{$\langle \overline{c} \rangle_\infty$ \citep{Straub1958} \,};

\addplot [fill=orange,mark=*,only marks,fill opacity=0.5,draw opacity=0.5, mark size=2.5,clip mode=individual]
table[row sep=crcr]{%
100 100\\
}; \addlegendentry{$\langle \overline{c} \rangle$};
   
\addplot [fill=teal,mark=*,only marks,fill opacity=0.5,draw opacity=0.5, mark size=2.5,clip mode=individual]
table[row sep=crcr]{%
100 100\\
};  \addlegendentry{$\langle \overline{c} \rangle_{\text{TWL}_\text{trap}}$};

\addplot [fill=orange,mark=square*,only marks,fill opacity=0.5,draw opacity=0, mark size=2.5,clip mode=individual]
table[row sep=crcr]{%
100 100\\
}; \addlegendentry{$\langle \overline{c} \rangle$ \citep{Straub1958} \,};

\addplot [fill=violet,mark=*,only marks,fill opacity=0.5,draw opacity=0.5, mark size=2.5,clip mode=individual]
table[row sep=crcr]{%
100 100\\
}; \addlegendentry{$\langle \overline{c} \rangle_{\text{TBL}}$};

\addplot [fill=black,mark=*,only marks,fill opacity=0.5,draw opacity=0.5, mark size=2.5,clip mode=individual]
table[row sep=crcr]{%
100 100\\
}; \addlegendentry{$\langle \overline{c} \rangle_{\text{TWL}_\text{ent}}$};

\addplot [fill=violet,mark=square*,only marks,fill opacity=0.5,draw opacity=0, mark size=2.5,clip mode=individual]
table[row sep=crcr]{%
100 100\\
}; \addlegendentry{$\langle \overline{c} \rangle_{\text{TBL} \infty}$ \citep{Straub1958} \, };

\draw [black, draw opacity=0, name path=A] plot [smooth, tension=0] coordinates {(0,0) (0,0)};;
\draw [black, draw opacity=0, name path=B] plot [smooth, tension=0] coordinates {(0,0) (0,0)};
\addplot+[green!50!black, opacity=0.1,draw opacity=1,solid] fill between[of=A and B]; \addlegendentry{GVF}

\draw [black, draw opacity=0, name path=A] plot [smooth, tension=0] coordinates {(0,0) (0,0)};;
\draw [black, draw opacity=0, name path=B] plot [smooth, tension=0] coordinates {(0,0) (0,0)};
\addplot+[cyan!60!black, opacity=0.1,draw opacity=1,solid] fill between[of=A and B]; \addlegendentry{UF}

\addplot [color=violet,mark=none,line width = 2,draw opacity=1]
table[row sep=crcr]{%
100 100\\
}; \addlegendentry{Eq. (\ref{eqCdevelopment}) \, \,};

\node[align=center,color=black] at (rel axis cs:0.1,0.9) {(\textit{a})};

\addplot [fill=orange,mark=square*,only marks,fill opacity=1,draw opacity=0, mark size=2.5,clip mode=individual]
table [y=Cmeantot,  x =Fr] {\Froude};

\addplot [fill=violet,mark=square*,only marks,fill opacity=0.5,draw opacity=0, mark size=2.5,clip mode=individual]
table [y=CmeanTBL,  x =Fr] {\Froude};

\addplot [fill=orange,mark=*,only marks,fill opacity=0.5,draw opacity=0.5, mark size=2.5,clip mode=individual]
table [y=Cmeantot,  x =Fr] {\FroudeKillen};

\addplot [fill=violet,mark=*,only marks,fill opacity=0.5,draw opacity=0.5, mark size=2.5,clip mode=individual]
table [y=CmeanTBL,  x =Fr] {\FroudeKillen};

\nextgroupplot[height = 4.6cm,
     width = 4.6cm,
     xlabel={$(x-L_i)/L_i$},
      ylabel={\footnotesize{$\langle \overline{c} \rangle / \langle \overline{c} \rangle_\infty$, $\langle \overline{c} \rangle_{\text{TBL}} /  \langle \overline{c} \rangle_\infty$}},
     grid=both,
     xmin=0,
     xmax=10,
     ymin=0,
     ymax=1.5,   legend columns=3,
   legend style={cells={align=left},anchor = south west,at={(0,1.2)},font=\normalsize},
   clip mode=individual
   ]

\draw [black, draw opacity=0, name path=A] plot [smooth, tension=0] coordinates {(0,0) (6,0)};;
\draw [black, draw opacity=0, name path=B] plot [smooth, tension=0] coordinates {(0,1.5) (6,1.5)};
\addplot+[green!50!black, opacity=0.1] fill between[of=A and B];

\draw [black, draw opacity=0, name path=A] plot [smooth, tension=0] coordinates {(6,0) (10,0)};;
\draw [black, draw opacity=0, name path=B] plot [smooth, tension=0] coordinates {(6,1.5) (10,1.5)};
\addplot+[blue!60!black, opacity=0.1] fill between[of=A and B];
   
\node[align=center,color=black] at (rel axis cs:0.1,0.9) {(\textit{b})};

\addplot [fill=orange,mark=square*,only marks,fill opacity=0.5,draw opacity=0, mark size=2.5,clip mode=individual]
table [y=Cmeantot,  x =xLi] {\CdevelopmentStraub};

\addplot[color=violet,mark=none,line width = 2,draw opacity=1] table [y=Cmean,  x =xLi] {\Cdevelopmentcalc};
   
\addplot [fill=orange,mark=*,only marks,fill opacity=0.5,draw opacity=0.5, mark size=2.5,clip mode=individual]
table [y=Cmeantot,  x =Xstar] {\Cdevelopment};
   
\addplot [fill=violet,mark=*,only marks,fill opacity=0.5,draw opacity=0.5, mark size=2.5,clip mode=individual]
table [y=CmeanTBL,  x =Xstar] {\Cdevelopment};

\nextgroupplot[height = 4.6cm,
     width = 4.6cm,
     xlabel={$(x-L_i)/L_i$},
      ylabel={\footnotesize{$\langle \overline{c} \rangle_{\text{TWL}_\text{trap/ent}} / \langle \overline{c} \rangle_\infty$}},
     grid=both,
     xmin=0,
     xmax=10,
     ymin=0,
     ymax=1, 
   ]

\draw [black, draw opacity=0, name path=A] plot [smooth, tension=0] coordinates {(0,0) (6,0)};;
\draw [black, draw opacity=0, name path=B] plot [smooth, tension=0] coordinates {(0,1.5) (6,1.5)};
\addplot+[green!50!black, opacity=0.1] fill between[of=A and B];

\draw [black, draw opacity=0, name path=A] plot [smooth, tension=0] coordinates {(6,0) (10,0)};;
\draw [black, draw opacity=0, name path=B] plot [smooth, tension=0] coordinates {(6,1.5) (10,1.5)};
\addplot+[blue!60!black, opacity=0.1] fill between[of=A and B];

\node[align=center,color=black] at (rel axis cs:0.1,0.9) {(\textit{c})}; 

\addplot [fill=teal,mark=*,only marks,fill opacity=0.5,draw opacity=0.5, mark size=2.5,clip mode=individual]
table [y=CmeanTWLtrap,  x =Xstar] {\Cdevelopment};
   
\addplot [fill=black,mark=*,only marks,fill opacity=0.5,draw opacity=0.5, mark size=2.5,clip mode=individual]
table [y=CmeanTWLent,  x =Xstar] {\Cdevelopment};

\end{groupplot}
\end{tikzpicture}

%% file: Figures/fig7.tex
\pgfplotstableread{
Cmeantot Cbottom
0.198709677419355	0.0100000000000000
0.151734265734266	0.0100000000000000
0.164856115107914	0.0200000000000000
0.162418478260870	0.0200000000000000
0.149237288135593	0.0200000000000000
0.151942675159236	0.0200000000000000
0.146686046511628	0.0200000000000000
0.153562091503268	0.0200000000000000
0.246150000000000	0.0200000000000000
0.238217821782178	0.0300000000000000
0.240756302521008	0.0400000000000000
0.244017857142857	0.0400000000000000
0.254419889502763	0.0500000000000000
0.246363636363636	0.0400000000000000
0.245333333333333	0.0400000000000000
0.230075187969925	0.0300000000000000
0.248318584070796	0.0300000000000000
0.312333333333333	0.0600000000000000
0.303024691358025	0.0700000000000000
0.327916666666667	0.0800000000000000
0.296348314606742	0.0600000000000000
0.313906250000000	0.0800000000000000
0.306090909090909	0.0600000000000000
0.293691275167785	0.0600000000000000
0.230075187969925	0.0300000000000000
0.345597014925373	0.110000000000000
0.323920454545455	0.0900000000000000
0.348243243243243	0.110000000000000
0.395625000000000	0.160000000000000
0.402910447761194	0.180000000000000
0.398157894736842	0.170000000000000
0.395228758169935	0.170000000000000
0.369178767123288	0.140000000000000
0.373228481012658	0.150000000000000
0.361706470588235	0.130000000000000
0.349253246753247	0.110000000000000
0.445357142859375	0.190000000000000
0.431750000000000	0.160000000000000
0.390645161292339	0.140000000000000
0.532142857142857	0.360000000000000
0.532142857142857	0.360000000000000
0.542688679245283	0.370000000000000
0.545816326530612	0.360000000000000
0.503518518518519	0.290000000000000
0.496192052980133	0.290000000000000
0.476746031746032	0.250000000000000
0.449411764705882	0.190000000000000
0.500000000000000	0.280000000000000
0.452702702702703	0.220000000000000
0.370298013245033	0.100000000000000
0.585200000000000	0.480000000000000
0.585200000000000	0.480000000000000
0.588658536585366	0.450000000000000
0.602959183673469	0.460000000000000
0.596626506024097	0.450000000000000
0.577931034482759	0.420000000000000
0.556269841269841	0.400000000000000
0.535571428571429	0.330000000000000
0.608196721311476	0.540000000000000
0.624696969696970	0.530000000000000
0.640500000000000	0.540000000000000
0.647357142857143	0.530000000000000
0.640500000000000	0.540000000000000
0.653852459016393	0.520000000000000
0.635454545454546	0.500000000000000
0.622183098591549	0.470000000000000
0.647887500000000	0.610000000000000
0.677978723404255	0.640000000000000
0.698043478260870	0.630000000000000
0.694259259259259	0.600000000000000
0.709375000000000	0.620000000000000
0.693827160493827	0.600000000000000
0.635454545454546	0.500000000000000
0.678454545454546	0.590000000000000
}\Straub

\pgfplotstableread{
Cmeantot Cbottom
0.195912408759124	0
0.241914893617021	0
0.250294117647059	0
0.216935483870968	0
0.330142857142857	0.0500000000000000
0.150404624277457	0
0.250294117647059	0
0.242121212121212	0.0200000000000000
0.253461538461538	0.0400000000000000
0.192535211267606	0
0.257363636363636	0.0300000000000000
0.349590163934426	0.100000000000000
0.346639344262295	0.120000000000000
0.459576271186441	0.210000000000000
0.494913793103448	0.290000000000000
0.494913793103448	0.290000000000000
0.553775510204082	0.350000000000000
}\Killen

\pgfplotstableread{
Cmeantot Cbottom
0.210304241435563	0.00100000000000000
0.248783264033264	0.00400000000000000
0.279105263157895	0.0140000000000000
0.283871391076116	0.0290000000000000
0.296059343434343	0.0260000000000000
0.330622448979592	0.0540000000000000
0.355041237113402	0.0690000000000000
0.324873936170213	0.0570000000000000
0.353172436750999	0.0960000000000000
0.206732800982801	0.00100000000000000
0.238471910112360	0.00300000000000000
0.256571514423077	0.0100000000000000
0.265983682983683	0.0160000000000000
0.291258064516129	0.0330000000000000
0.314695496083551	0.0460000000000000
0.296886882129278	0.0390000000000000
0.328058225508318	0.0740000000000000
0.329708620689655	0.0730000000000000
0.331049132947977	0.0690000000000000
0.206818440082645	0.00100000000000000
0.239573671497585	0.00400000000000000
0.271521282633371	0.0100000000000000
0.264347079037801	0.0240000000000000
0.284164688427300	0.0260000000000000
0.291487752161383	0.0400000000000000
0.310328678474114	0.0530000000000000
0.326386130136986	0.0580000000000000
0.312257142857143	0.0600000000000000
}\Bung

\pgfplotstableread{
Cmeantot Cbottom
0.0656493455833598	0
0.0956679444754437	0
0.108748754541751	0
0.122272505590358	0
0.135834773891128	9.00000000000000e-05
0.157674888784355	0.000510000000000000
0.171409974419668	0.00112000000000000
0.0708659618327915	0
0.0922321886135952	0
0.110429459852906	0
0.119224335473565	0
0.134156665552399	0
0.140477574319229	0
0.0174265433901351	0
0.0450314276993874	0
0.0737722436070054	0
0.0934419744904803	0
0.108060033589845	0
0.119799689127538	0
0.122641331413253	0
0.0189168142830060	0
0.0345513065730812	0
0.0555884284791584	0
0.0692441141051056	0
0.0918107998460343	0
0.101487075643674	0
0.111393785322428	0
0.0189054590000661	0
0.0233201647095327	0
0.0397963590963111	0
0.0579746459054042	0
0.0801942640270612	0
0.0900945361528493	0
0.0989660692476227	0
0.0637181256789170	0
0.0130979072092560	0
0.0191762398752689	0
0.0175467150711549	0
0.0248511135635384	0
0.0426464577352326	0
0.0515596310158519	0
0.0641794126773937	0
0.0734170709017645	0
0.0202045598503119	0
0.0177338027093900	0
0.0202631102870488	0
0.0216034079814842	0
0.0218987261279119	0
0.0368338286442254	0
0.0317479875541152	0
0.0494656670187585	0
0.0545257460969509	0
0.0192816702221331	0
0.0272925830656012	0
0.0223571728926497	0
0.0215575389611899	0
0.0247186853306345	0
0.0277418505962490	0
0.0254886621023398	0
0.0376957501589905	0
0.0538100976789890	0
0.0304837856172916	0
0.0316539256541281	0
0.0240504484550701	0
0.0366015817678543	0
0.0316679160954363	0
0.0224034348923729	0
0.0414604623520509	0
0.0410593006294439	0
0.0502522676731024	0
0.121474530929692	0
0.160595648519152	0
0.183001124058420	6.00000000000000e-05
0.180082564889717	2.00000000000000e-05
0.188725197700139	0
0.189353902814479	0
0.208807975127241	6.00000000000000e-05
0.0190918638429145	0
0.0965336946404320	0
0.134112459311038	0
0.165133617959325	5.00000000000000e-05
0.155809306388740	0
0.162094340656878	0
0.167096061842475	0
0.182579859717546	0
0.0117553876054615	0
0.0655654568777124	0
0.113364429427553	0
0.152130619117451	1.00000000000000e-05
0.143879390227863	4.00000000000000e-05
0.160941962338383	4.00000000000000e-05
0.153713147070928	6.00000000000000e-05
0.174125808844465	0.000100000000000000
0.0162515988132410	0
0.0409146723086772	0
0.0981931336666738	0
0.133661053449762	0
0.135406061256142	6.00000000000000e-05
0.143726169836600	0
0.145782881513426	4.00000000000000e-05
0.163547970451612	0.000210000000000000
0.0120520897357380	0
0.0412071141363554	0
0.0792733413996555	0
0.116208375484859	0
0.129746393937424	1.00000000000000e-05
0.134340981717992	0
0.142110403233152	9.00000000000000e-05
0.152651897547644	0
0.0133288204502856	0
0.0157944789932148	0
0.0407294252557150	0
0.0826885263407850	0
0.103385424328530	0
0.111528816403626	0
0.128097056646511	0
0.132158462673907	2.00000000000000e-05
0.0134155359209301	0
0.0171038406015834	0
0.0291938106876530	0
0.0624268180349160	0
0.0805458326267354	0
0.0930768961846309	0
0.103721362307013	0
0.118694494018898	2.00000000000000e-05
0.0273290813938436	0
0.0130800552458260	0
0.0258522374397171	0
0.0415299805729940	0
0.0966068567142469	0
0.0799607023801358	0
0.0928650450379275	0
0.100496067201440	0
0.0305666406790956	0
0.0214405886637052	0
0.0225841048811661	0
0.0381087443320941	0
0.0513089113402506	0
0.0561125471635787	0
0.0776072825850253	0
0.0896332511395887	0
0.0502522676731024	0
0.121474530929692	0
0.160595648519152	0
0.169874042590585	0
0.191715973737248	3.00000000000000e-05
0.215947997997926	0.000900000000000000
0.232701390071523	0.00112000000000000
0.234023058152939	0.000650000000000000
0.0190918638429145	0
0.0965336946404320	0
0.134112459311038	0
0.149246336323349	0
0.165311406685388	0
0.181174040188691	0
0.197402326953158	0.000150000000000000
0.192437806333811	0
0.0117553876054615	0
0.0655654568777124	0
0.113364429427553	0
0.137388211792916	0
0.157331821652544	0
0.165776390723812	0.000190000000000000
0.174597807671624	0.000450000000000000
0.175913228238529	0.000240000000000000
0.0162515988132410	0
0.0409146723086772	0
0.0981931336666738	0
0.129136843522393	0
0.146485428340498	0
0.161282554148827	0
0.165342578571572	8.00000000000000e-05
0.162922678509727	2.00000000000000e-05
0.0120520897357380	0
0.0412071141363554	0
0.0792733413996555	0
0.115670714488774	0
0.139852977776913	0
0.144443595556043	0.000140000000000000
0.157761613506563	0.000450000000000000
0.160808544301514	0.000120000000000000
0.0133288204502856	0
0.0157944789932148	0
0.0407294252557150	0
0.0917434314857188	0
0.113912271881554	0
0.125013773811695	1.00000000000000e-05
0.139419258950292	0
0.147817803479532	3.00000000000000e-05
0.107584868642605	0
0.158060233420982	0
0.248056291873163	0.00877000000000000
0.232724865988271	0.00224000000000000
0.209531297242976	0.000170000000000000
0.237137182153393	0.000880000000000000
0.245880709885099	0.00477000000000000
0.238421059201749	0.000570000000000000
0.0417472188083381	0
0.129249748491976	0
0.213729088909348	0.00162000000000000
0.190057080430501	0.000180000000000000
0.177845778720990	0.000250000000000000
0.195903375834485	0.000170000000000000
0.214973049440379	0.00135000000000000
0.209102524682876	0.000820000000000000
0.0199428573351500	0
0.101937860701691	0
0.192224860345317	0.000340000000000000
0.162438661489054	0.000260000000000000
0.166140847129376	0
0.172543726685841	3.00000000000000e-05
0.177655375869089	0.000200000000000000
0.180049103159074	0.000290000000000000
0.0173010451442991	0
0.0802630048342033	0
0.163425857028229	0.000170000000000000
0.148309251783553	0.000280000000000000
0.168681692837417	0.000230000000000000
0.177405891464365	0.000390000000000000
0.170420590934071	0.000530000000000000
0.172729483942069	0.000840000000000000
0.0181430705319717	0
0.0562475206276969	0
0.134542541510097	0
0.129169255539327	5.00000000000000e-05
0.161150392677605	6.00000000000000e-05
0.166383070884825	0.000190000000000000
0.160809382139447	0
0.163639579355455	0.000330000000000000
0.0172960772635413	0
0.0286891764752028	0
0.0802785533482044	0
0.107966894394670	0
0.142569685038418	5.00000000000000e-05
0.150645139871578	2.00000000000000e-05
0.151817466377344	0.000430000000000000
0.156854292794813	0.000600000000000000
0.0205582088274873	0
0.0247547794349096	0
0.0522585838705452	0
0.0777496352909004	0
0.119831381908138	0
0.136578185416372	0
0.138768936631791	1.00000000000000e-05
0.146403292769635	0
0.0309700272829757	0
0.0283634227733373	0
0.0453101263379651	0
0.0596203780751709	0
0.0957942421049988	0
0.109437629173208	0
0.120897422228723	3.00000000000000e-05
0.129725759471014	7.00000000000000e-05
0.0237945834488696	0
0.0219755981437861	0
0.0364652524720876	0
0.0483600739668466	0
0.0746273848080882	0
0.0973329539939490	0
0.110574481960720	0
0.121469626074430	3.00000000000000e-05
}\Severi

\begin{tikzpicture}

 \begin{groupplot}[
     group style = {group size = 1 by 1,horizontal sep=1.3cm,vertical sep=1.4cm},
     width = 1\textwidth]

\nextgroupplot[height = 7cm,
     width = 7cm,
     ylabel={$\overline{c}_0$},
     xlabel={$\langle \overline{c} \rangle$ },
     grid=both,
     xmin=0,
     xmax=0.8,
     xtick={0, 0.25, 0.4, 0.6, 0.8},
     ytick={0, 0.2, 0.4, 0.6},
     ymin=0,
     ymax=0.9,
    legend columns=1,
   legend style={cells={align=left},anchor = north east,at={(-0.25,1)},font=\small},
   clip mode=individual
   ]

\addplot [fill=gray,mark=square*,only marks,fill opacity=0,draw opacity=1, mark size=2,clip mode=individual] 
table[row sep=crcr]{%
0 -1	\\
};\addlegendentry{\cite{Straub1958}};

\addplot [fill=blue,mark=*,only marks,fill opacity=0.5,draw opacity=1, mark size=3,clip mode=individual]
table[row sep=crcr]{%
0 -1	\\
};\addlegendentry{\cite{Killen1968}};

\addplot [fill=orange,mark=diamond*,only marks,fill opacity=0.5,draw opacity=1, mark size=3,clip mode=individual]
table[row sep=crcr]{%
0 -1	\\
}; \addlegendentry{\cite{Bung2009}};

\addplot [fill=blue,mark=10-pointed star,only marks,fill opacity=0,draw opacity=0.3, mark size=3,clip mode=individual]
table[row sep=crcr]{%
0 -1	\\
}; \addlegendentry{\cite{Severi2018}};

 \draw [black, draw opacity=0, name path=A] plot [smooth, tension=0] coordinates {(0,0) (0.25,0)};
\draw [black, draw opacity=0, name path=B] plot [smooth, tension=0] coordinates {(0,1) (0.25,1)};
\addplot+[teal!50!black, opacity=0.1] fill between[of=A and B]; 

\draw [black, draw opacity=0, name path=A] plot [smooth, tension=0] coordinates {(0,0.675) (0.25,0.675)};
\draw [black, draw opacity=0, name path=B] plot [smooth, tension=0] coordinates {(0,0.9) (0.25,0.9)};
\addplot+[teal!50!black, opacity=0.8] fill between[of=A and B];

\draw [black, draw opacity=0, name path=C] plot [smooth, tension=0] coordinates {(0.25,0) (0.8,0)};
\draw [black, draw opacity=0, name path=D] plot [smooth, tension=0] coordinates {(0.25,1) (0.8,1)};
\addplot+[violet, opacity=0.1] fill between[of=C and D]; 

\draw [black, draw opacity=0, name path=C] plot [smooth, tension=0] coordinates {(0.25,0.675) (0.8,0.675)};
\draw [black, draw opacity=0, name path=D] plot [smooth, tension=0] coordinates {(0.25,0.9) (0.8,0.9)};
\addplot+[violet, opacity=0.8] fill between[of=C and D];

\node[align=center,color=white, opacity=1, anchor=north west ] at (0.005,0.9) {\small{Superpos.} \\ {\small{principle;}} \\
{\small{[Eq. (\ref{eq:superposition})]}}
}; 

\node[align=center,color=white, opacity=1, anchor=north west ] at (0.33,0.9) {\small{Superposition +} \\ \small{two-state principle;} \\
{\small{[Eq. (\ref{eq:voidfractionfinal1})]}}
}; 

\addplot[dashed,color=black,mark=none,line width = 2,draw opacity=1] 
table[row sep=crcr]{%
0.25 0\\
0.25 1\\
};

\addplot [fill=gray,mark=square*,only marks,fill opacity=0,draw opacity=1, mark size=2,clip mode=individual]  table [x=Cmeantot, y=Cbottom] {\Straub};

\addplot [fill=blue,mark=*,only marks,fill opacity=0.5,draw opacity=1, mark size=3,clip mode=individual]
table [x=Cmeantot, y=Cbottom] {\Killen};

\addplot [fill=orange,mark=diamond*,only marks,fill opacity=0.5,draw opacity=1, mark size=3,clip mode=individual]
table [x=Cmeantot, y=Cbottom] {\Bung};

\addplot [fill=blue,mark=10-pointed star,only marks,fill opacity=0,draw opacity=0.3, mark size=3,clip mode=individual]
table [x=Cmeantot, y=Cbottom] {\Severi};

\end{groupplot}
\end{tikzpicture}